\documentclass[pre,aps,showpacs,nofootinbib,twocolumn]{revtex4}
\usepackage{amssymb}

\usepackage{graphicx}

\input{tcilatex}

\begin{document}

\title{Bosonic Spectral Function in HTSC Cuprates: \\
Part I - Experimental Evidence for Strong Electron-Phonon Interaction}
\author{$^{1}$E. G. Maksimov, $^{2,3}$M. L. Kuli\'{c}, $^{4}$O. V. Dolgov}
\address{$^{1}$ Lebedev Physical Institute, 119991 Moscow, Russia\\
$^{2}$Goethe-Universit\"{a}t Frankfurt, Theoretische Physik, 60054 Frankfurt/Main, Germany\\
$^{3}$Max-Born-Institut f\"{u}r Nichtlineare Optik und Kurzzeitspektroskopie, \\12489 Berlin,Germany \\
$^{4}$Max-Planck-Institut f\"{u}r Festk\"{o}rperphysik,70569
Stuttgart, Germany}

\begin{abstract}
In \textit{Part I} we discuss accumulating experimental evidence related to
the structure and origin of the bosonic spectral function $\alpha
^{2}F(\omega )$ in high-temperature superconducting (HTSC) cuprates near
optimal doping. Some global properties of $\alpha ^{2}F(\omega )$, such as
number and positions of peaks, are extracted by combining optics, neutron
scattering, ARPES and tunnelling measurements. These methods give convincing
evidence for strong electron-phonon interaction (EPI) with $1<\lambda
\lesssim 3$ in cuprates near optimal doping. Here we clarify how these
results are in favor of the Eliashberg-like theory for HTSC cuprates near
optimal doping.We argue that the neglect of EPI in some previous studies of
HTSC was based on a number of deceptive prejudices related to the strength
of EPI, on some physical misconceptions and misleading interpretation of
experimental results.

In \textit{Part II} we discuss some theoretical ingredients which are
necessary to explain the experimental results related to pairing mechanism
in optimally doped cuprates. These comprise the Migdal-Eliashberg theory for
EPI in strongly correlated systems which give rise to the forward scattering
peak. The latter is due to the combined effects of the weakly screened
Madelung interaction in the ionic-metallic structure of layered cuprates and
many body effects of strong correlations. While EPI is responsible for the
strength of pairing the residual Coulomb interaction (by including spin
fluctuations) triggers the d-wave pairing.
\end{abstract}

\date{\today }
\maketitle

\section{Introduction}

In spite of an unprecedentedly intensive experimental and theoretical study
after the discovery of high-temperature superconductivity (HTSC) in cuprates
there is, even twenty-two years after, no consensus on the pairing mechanism
in these materials. At present there are two important experimental facts
which are not under dispute: (1) the critical temperature $T_{c}$ in
cuprates is high, with the maximum $T_{c}^{\max }\sim 160$ $K$; (2) the
pairing in cuprates is d-wave like, i.e. $\Delta (\mathbf{k},\omega )\approx
\Delta _{s}(k,\omega )+\Delta _{d}(\omega )(\cos k_{x}-\cos k_{y})$ with $%
\Delta _{s}<0.1\Delta _{d}$. On the contrary there is a dispute concerning
the scattering mechanism which governs normal state properties and pairing
in cuprates. To this end, we stress that in the HTSC cuprates, a number of
properties can be satisfactorily explained by assuming that the
quasi-particle dynamics is governed by some electron-boson scattering and in
the superconducting state bosonic quasi-particles are gluing electrons in
Cooper pairs. Which bosonic quasi-particles are dominating in the cuprates
is the subject which will be discussed in this work. It is known that the
electron-boson (phonon) scattering is well described by the
Migdal-Eliashberg theory if the adiabatic parameter $A_{B}\equiv $ $\lambda
(\omega _{B}/W_{b})$ fulfills the condition $A_{B}\ll 1$, where $\lambda $
is the electron-boson coupling constant, $\omega _{B}$ is the characteristic
bosonic energy and $W_{b}$ is the electronic band width. The important
characteristic of the electron-boson scattering is the Eliashberg spectral
function $\alpha ^{2}F(\mathbf{k},\mathbf{k}^{\prime },\omega )$ (or its
average $\alpha ^{2}F(\omega )$) which characterizes scattering of
quasi-particle from $\mathbf{k}$ to $\mathbf{k}^{\prime }$ by exchanging
bosonic energy $\omega $. Therefore, in systems with electron-boson
scattering the knowledge of this function is of crucial importance. There
are at least two approaches differing in assumed gluing bosons. The first
one is based on the electron-phonon interaction (EPI) \cite{Maksimov-Review}%
, \cite{Kulic-Review}, \cite{Alexandrov}, \cite{Gunnarsson-review-2008},
\cite{Falter} where mediating bosons are \textit{phonons }and where the the
average spectral function $\alpha ^{2}F(\omega )$ is similar to the phonon
density of states $F_{ph}(\omega )$. Note, $\alpha ^{2}F(\omega )$ is not
the product of two functions although sometimes one defines the function $%
\alpha ^{2}(\omega )=\alpha ^{2}F(\omega )/F(\omega )$ which should
approximate the energy dependence of the strength of the EPI coupling. There
are numerous experimental evidence in cuprates which support the dominance
of the EPI scattering mechanism with a rather large coupling constant $%
1<\lambda ^{ep}\lesssim 3$ and which will be discussed in detail below. In
the EPI approach $\alpha ^{2}F(\omega )$ is extracted from tunnelling
measurements in conjunction with IR optical measurements. We stress again
that the Migdal-Eliashberg theory is well justified framework for EPI since
in most superconductors the condition $A_{ph}\ll 1$ is fulfilled. The HTSC
are on the borderline and it is a natural question - under which condition
can high T$_{c}$ be realized in the non-adiabatic limit $A_{ph}\sim 1$? The
second approach \cite{Pines} assumes that EPI is too weak to be responsible
for high $T_{c}$ in cuprates and it is based on a phenomenological model for
spin-fluctuation interaction ($SFI$) as the dominating scattering mechanism,
i.e. it is a non-phononic mechanism. In this approach the spectral function
is proportional to the imaginary part of the spin susceptibility $\func{Im}%
\chi (\mathbf{k}-\mathbf{k}^{\prime },\omega )$, i.e. $\alpha ^{2}F(\mathbf{k%
},\mathbf{k}^{\prime },\omega )\sim \func{Im}\chi (\mathbf{k}-\mathbf{k}%
^{\prime },\omega )$. NMR spectroscopy and magnetic neutron scattering give
that in HTSC cuprates $\chi (\mathbf{q},\omega )$ is peaked at the
antiferromagnetic wave vector $Q=(\pi /a,\pi /a)$ and this property is
favorable for d-wave pairing. The $SFI$ theory roots basically on the strong
electronic repulsion on Cu atoms, which is usually studied by the Hubbard
model or its (more popular) derivative the t-J model. Regarding the
possibility to explain high T$_{c}$ solely by strong correlations, as it is
reviewed in \cite{Patrick-Lee}, we stress two facts. First, at present there
is no viable theory which can justify these (non-phononic) mechanisms of
pairing. Second, the central question in this approach is - do models based
on the Hubbard Hamiltonian show up superconductivity at sufficiently high
critical temperatures ($T_{c}\sim 100$ $K$) ? A number of numerical studies
of these models offer a negative answer. For instance, the sign-free
variational Monte Carlo algorithm in the 2D repulsive ($U>0$) Hubbard model
gives no evidence for HTSC, neither the BCS- nor
Berezinskii-Kosterlitz-Thouless (BKT)-like \cite{Imada-MC}. At the same
time, similar calculations show that there is a strong tendency to
superconductivity in the attractive ($U<0$) Hubbard model for the same
strength of $U$, i.e. at finite temperature in the 2D model with $U<0$ the
BKT superconducting transition is favored. Concerning HTSC in the $t-J$
model, various numerical calculations such as Monte Carlo calculations of
the Drude spectral weight \cite{Scalapino-Drude-weight} and high temperature
expansion \cite{Pryadko} have shown that there is no superconductivity at
temperatures characteristic for cuprates and if it exists $T_{c}$ must be
rather low - few Kelvins. These numerical results tell us that the lack of
high $T_{c}$\ (even in $2D$\ BKT phase) in the repulsive ($U>0$)\
single-band Hubbard model and in the $t-J$ model is not only due to
thermodynamical $2D$-fluctuations (which at finite T suppress and destroy
superconducting phase coherence in large systems) but it is mostly due to an
\textit{inherent ineffectiveness of strong correlations to produce solely
high }$T_{c}$ \textit{in cuprates}. These numerical results certainly mean
that the simple single-band Hubbard, as well as its derivative the t-J
model, are insufficient to explain the pairing mechanism in cuprates and
some other ingredients must be included. Having in mind these facts there is
no room at present for any kind of celebration of the victory of
non-phononic mechanisms of pairing as some prefer to do.

Since $EPI$ is as a rule strong in oxides, then it is plausible that it
should be accounted for in cuprates at least in the normal metallic state.
As it will be argued later on, the experimental support for the importance
of EPI comes from optics, tunnelling, and recent ARPES measurements \cite
{Shen-review}. It is worth mentioning that recent ARPES activity was a
strong impetus for renewed experimental and theoretical studies of EPI in
cuprates. However, in spite of accumulating experimental evidence for
importance of EPI with $\lambda ^{ep}>1$, there are occasionally reports
which doubt its importance in cuprates. This is the case with recent
interpretation of some optical measurements in terms of SFI only \cite
{Hwang-Timusk-1} and with LDA band structure calculations \cite{Bohnen-Cohen}%
, \cite{Giuistino}, where both claim that EPI is negligibly small, i.e. $%
\lambda ^{ep}<0.3$.

The paper is organized as follows. There are two parts - in \textit{Part I}
we will mainly discuss experimental results in \textit{cuprates near optimal
doping} and minimal theoretical explanations which are related to the
spectral function $\alpha ^{2}F(\omega )$ as well to the transport spectral
function $\alpha _{tr}^{2}F(\omega )$ and how these are related to EPI in
cuprates. In this work we consider mainly those direct one-particle and
two-particles probes of low energy quasi-particle excitations (by including
gap and pseudogap) and scattering rates which give informations on the
structure of the spectral functions $\alpha ^{2}F(\mathbf{k},\mathbf{k}%
^{\prime },\omega )$ and $\alpha _{tr}^{2}F(\omega )$ in systems near
optimal doping. These are angle-resolved photoemission ($ARPES$), various
arts of tunnelling spectroscopies such as superconductor/insulator/ normal
metal ($SIN$) junctions and break junctions, scanning-tunnelling microscope
spectroscopy ($STM$), infrared ($IR$) and Raman optics, inelastic neutron
scattering ($INS$) etc. We shall argue that these direct probes give
evidence for a rather strong EPI in cuprates as dominating scattering
mechanism of quasi-particles. Some other experiments on EPI are also
discussed in order to complete the arguments for the importance of EPI in
cuprates. The detailed contents of Part I is the following. In \textit{%
Section II} we discuss some prejudices related to $EPI$ and the Fermi-liquid
behavior of HTSC cuprates. We argue that any non-phononic mechanism of
pairing should have very large bare critical temperature $T_{c0}\gg T_{c}$
in the presence of the large EPI coupling constant, $\lambda ^{ep}\geq 1$,
if its spectral function is weakly momentum dependent, i.e. if $\alpha ^{2}F(%
\mathbf{k},\mathbf{k}^{\prime },\omega )\approx \alpha ^{2}F(\omega )$. The
fact that EPI is large in the normal state of cuprates and the condition
that it must conform with d-wave pairing implies inevitably that EPI in
cuprates must be \textit{strongly momentum dependent}. In \textit{Section III%
} we discuss direct and indirect experimental evidence for the importance of
EPI and for the weakness of SFI in cuprates. These are:

(\textbf{A}) \textit{Magnetic neutron scattering} \textit{measurements - }%
These measurements provide dynamic spin susceptibility $\chi (\mathbf{q}%
,\omega )$ which is in the $SFI$ phenomenological approach \cite{Pines}
related to the Eliashberg spectral function, i.e. $\alpha ^{2}F_{sf}(\mathbf{%
k},\mathbf{k}^{\prime },\omega )\sim \func{Im}\chi (\mathbf{q}=\mathbf{k}-%
\mathbf{k}^{\prime },\omega )$. We stress that such an approach can be
qualitatively justified only in the weak coupling limit, $g_{sf}\ll W_{b}$,
where $W_{b}$ is the band width. Here we discuss experimental results which
give evidence for strong rearrangement (with respect to $\omega $) of $\func{%
Im}\chi (\mathbf{Q},\omega )$ by doping toward the optimal doped HTSC \cite
{Bourges}. It turns out that in the optimally doped cuprates with $%
T_{c}=92.5 $ $K$ $\func{Im}\chi (\mathbf{Q},\omega )$ is drastically
suppressed compared to that in slightly underdoped ones with $T_{c}=91$ $K$,
and this is strong evidence for the smallness of the SFI coupling constant.

\textbf{(\textit{B})} \textit{Optical conductivity} \textit{measurements} -
From these measurements one can extract the transport relaxation rate $%
\gamma _{tr}(\omega )$ and indirectly an approximative shape of the
transport spectral function $\alpha ^{2}F_{tr}(\omega )$. In that respect we
discuss: (i) the misleading concept concerning the relation between the
optical relaxation rate $\gamma _{tr}(\omega )$ and the quasi-particle
relaxation rate $\gamma (\omega )$. This (misleading) concept has been
appearing repeatedly in the last twenty years despite the fact that this
controversy is resolved many years ago \cite{Allen}, \cite{Dolgov-Shulga},
\cite{Shulga}, \cite{Maksimov-Review}, \cite{Kulic-Review}, \cite{Kulic-AIP}%
; (ii) some methods of extraction of the optical spectral function $\alpha
_{tr}^{2}F(\omega )$ from optical reflectivity measurements. It turns out
that the width and the shape of the extracted $\alpha _{tr}^{2}F(\omega )$
favor EPI; (iii) the restricted sum-rule for the optical weight as a
function of temperature which can also be explained by strong $EPI$ \cite
{Maks-Karakoz-1}, \cite{Maks-Karakoz-2}; (iv) good agreement with
experiments of the temperature dependence of the resistivity $\rho (T)$
calculated with the extracted $\alpha _{tr}^{2}F(\omega )$. Recent
femtosecond time-resolved optical spectroscopy in $La_{2-x}Sr_{x}CuO_{4}$
gives additional evidence for importance of EPI and ineffectiveness of SFI
\cite{Kusar-2008}.

\textbf{(\textit{C})} \textit{ARPES} \textit{measurements and EPI} - From
these measurements one can extract the self-energy $\Sigma (\mathbf{k}%
,\omega )$ from which one can extract some properties of $\alpha ^{2}F(%
\mathbf{k},\mathbf{k}^{\prime },\omega )$. Here we discuss the following
items: (i) appearance of the nodal and anti-nodal kinks in optimally and
slightly underdoped cuprates, as well as the structure of the ARPES
self-energy ($\Sigma (\mathbf{k},\omega )$) and its isotope dependence,
which are all due to EPI; (ii) appearance of different slopes of $\Sigma (%
\mathbf{k},\omega )$ at low ($\omega \ll \omega _{ph}$) and high energies ($%
\omega \gg \omega _{ph}$) which can be explained with strong EPI; (iii)
formation of small polarons in the undoped HTSC which is due to strong EPI -
this gives rise to phonon side bands which are clearly seen in ARPES of
undoped HTSC.

(\textbf{D}) \textit{Tunnelling spectroscopy - }It is well known that this
method is of an immense importance in obtaining the spectral function $%
\alpha ^{2}F(\omega )$ from tunnelling conductance. In this part we discuss
the following items: (i) extraction of the Eliashberg spectral function $%
\alpha ^{2}F(\omega )$ with $\lambda =2-4$ from the tunnelling conductance
of break-junctions \cite{Tunneling-Vedeneev}-\cite{Ponomarev-Tunnel} which
gives that the maxima of $\alpha ^{2}F(\omega )$ coincide with the maxima in
the phonon density of states; (ii) the presence of the dip in dI/dV in STM
which shows the pronounced oxygen isotope effect and important role of these
phonons: (iii) the presence of fine and doping independent structure in I(V)
characteristics due to phonon emission by the Josephson current in layered
HTSC cuprates with intrinsic Josephson junctions.

\textbf{(\textit{E})} \textit{Phonon} \textit{neutron scattering} \textit{%
measurements} - From these experiments one can extract the phonon density of
state $F_{ph}(\omega )$ and strengths of the quasi-particle coupling with
various phonon modes. Here we argue, that the large softening and broadening
of the half-breathing Cu-O bond-stretching phonon, of apical oxygen phonons
and of oxygen $B_{1g}$ buckling phonons (in LSCO, BISCO,YBCO) cannot be
explained by LDA. It is curious that the magnitude of softening can be
partially obtained by LDA but the calculated widths of some important modes
are an order of magnitude smaller than the neutron scattering data show.
This remarkable fact implies \textit{the} \textit{inadequacy of LDA in
strongly correlated systems} and a more sophisticated many body theory for
EPI is needed. This problem will be discussed in more details in \textit{%
Part II } \cite{MaKuDoAk}. In \textit{Section IV} brief summary of the
\textit{Part I} is given. Since \textit{we are dealing with electron-boson
scattering in cuprate near optimal doping}, then in \textit{Section V} -
Appendix we introduce the reader briefly into the Migdal-Eliashberg theory
for superconductors (and normal metals) where the quasi-particle spectral
function $\alpha ^{2}F(\mathbf{k},\mathbf{k}^{\prime },\omega )$ and the
transport spectral function $\alpha _{tr}^{2}F(\omega )$ are defined.

Finally, at the end of the day one poses a question - do the experimental
results of the above enumerated spectroscopic methods allow a building of a
satisfactory and physically reasonable microscopic theory for basic
scattering and pairing mechanism in cuprates? The posed question is very
modest compared with a much stringent request for the \textit{theory of
everything} - which would be able to explain all properties of HTSC
materials. Such an ambitious project is not realized even in those
low-temperature conventional superconductors where it is definitely proved
that the pairing is due to EPI and many properties are well accounted for by
the Migdal-Eliashberg theory. Let us mention only two examples. First, the
experimental value for the coherence peak in the microwave response $\sigma
_{s}(T<T_{c})$ at $17$ $GHz$ in $Nb$ is much higher than the theoretical
value obtained by the Migdal-Eliashberg theory \cite{Marsiglio-1994}. So to
say, the theory explains the coherence peak at $17$ $GHz$ in $Nb$
qualitatively but not quantitatively. However, the measurements at higher
frequency $\sim 60$ $GHz$ are in agreement with the Migdal-Eliashberg theory
\cite{Klein-1994}. Second, the experimental boron (B) isotope effect in $%
MgB_{2}$ (with $T_{c}\approx 40$ $K$) is much smaller than the theoretical
value, i.e. $\alpha _{B}^{\exp }\approx 0.3<\alpha _{B}^{th}=0.5$, although
the pairing is due solely by EPI for boron vibrations \cite{MgB2-isotop}.
Since the theory of everything is impossible in the complex materials such
as HTSC cuprates in \textit{Part I} and \textit{II} we shall not discuss
those phenomena which need much more microscopic details and/or more
sophisticated many-body theory. These are selected by chance: (i)
peculiarities of the coherence peak in the microwave response $\sigma (T)$
in HTSC cuprates, which is peaked at $T$ much smaller than $T_{c}$, contrary
to the case of LTSC where it occurs near $T_{c}$; (ii) $T_{c}$ dependence on
the number of $CuO_{2}$ in the unit cell; (iii) temperature dependence of
the Hall coefficient; (iv) distribution of states in the vortex core, etc.
However, in a separate paper - \textit{Part II} \cite{MaKuDoAk} we shall
discuss some minimal theoretical concepts which can explain at least
qualitatively and semi-quantitatively results related to the above
enumerated spectroscopic methods. Due to the presence of strong correlations
and quasi-2D electronic structure some of these concepts go beyond the LDA
approach. In our opinion at this stage of the HTSC physics some \textit{%
important ingredients} of the future theory are already recognized. These
are: (1) very peculiar quasi-2D ionic-metallic structure with a rather weak
screening along the c-axis, which is a prerequisite for strong EPI; (2)
strong Coulomb interaction and correlations which are responsible for strong
magnetism in undoped cuprates and for important renormalizations of EPI.
Since both ingredients belong to the class of strong coupling problems, at
present there is no quantitative theory and therefore we must rely on
approximative and model theories. Even these approaches allow us qualitative
(and semi-quantitative) explanations of some important properties which are
due to the interplay of EPI and strong correlations. The latter two cause
the appearance of momentum dependent EPI - peaked at small transfer momenta
\cite{Kulic-Review}. Based on such an approach we are able to explain
(understand) at least qualitatively some very puzzling experimental results,
for instance: \textbf{(}\textit{a)} why is d-wave pairing realized in the
presence of strong EPI? \textbf{(}\textit{b}\textbf{)} why is the transport
coupling constant ($\lambda _{tr}$) smaller than the pairing one $\lambda $,
i.e. $\lambda _{tr}\lesssim \lambda /3$? (\textit{c}) Why is the mean-field
(one-body) LDA approach unable to give reliable values for the EPI coupling
constant in cuprates and how many-body effects help; (\textit{d}) why is
d-wave pairing robust in presence of non-magnetic impurities and defects?
(e) why the ARPES nodal and antinodal kinks are differently renormalized in
the superconducting states?

In real materials there are numerous experimental evidence for
nanoscale inhomogeneities in HTSC oxides . For instance recent STM
experiments show rather large gap dispersion at least on the
surface of BISCO crystals \cite {Davis} giving rise for a
pronounced inhomogeneity of the superconducting order parameter,
i.e. $\Delta (\mathbf{k},\mathbf{R})$ where $\mathbf{k}$ is the
relative momentum of the Cooper pair and $\mathbf{R}$ is the
center of
mass of Cooper pairs. One possible reason for the inhomogeneity of $\Delta (%
\mathbf{k},\mathbf{R})$ and disorder on the atomic scale can be due to
extremely high doping level of $\sim (10-20)$ $\%$ in HTSC cuprates which is
many orders of magnitude larger than in standard semiconductors ($10^{21}$
vs $10^{15}$ carrier concentration). There are some claims that high $T_{c}$
is exclusively due to these inhomogeneities (of an extrinsic or intrinsic
origin) which may increase EPI \cite{Phillips}, while other try to explain
high $T_{c}$ within the inhomogeneous Hubbard or t-J model. In \textit{Part
II} \cite{MaKuDoAk} we argue that the concept of an increase of T$_{c}$ by
inhomogeneity is ill-defined, since the increase of $T_{c}$ is defined with
respect to the average value $\bar{T}_{c}$. However, the latter quantity is
experimentally not well defined and an alleged increase of $T_{c}$ by the
material inhomogeneity cannot be tested at all.

\section{ EPI vs non-phononic mechanisms - facts vs prejudices}

Concerning the high $T_{c}$ in cuprates, one of the central questions is -
which interaction(s) is(are) responsible for strong quasi-particle
scattering in the normal state and for the superconducting pairing? In the
last twenty-two years, the scientific community was overwhelmed by all kinds
of (im)possible proposed pairing mechanisms, most of which are hardly
verifiable in any material, if at all. This trend is still continuing
nowadays (although with smaller slope), in spite of the fact that the
accumulated experimental results eliminate all but few.

\textit{A. Fermi vs non-Fermi liquid in cuprates}

After discovery of HTSC in cuprates there was a large amount of evidence on
strong scattering of quasi-particles which contradicts the canonical
(popular but narrow) definition of the Fermi liquid, thus giving rise to
numerous proposals of the so called non-Fermi liquids, such as Luttinger
liquid, RVB theory, marginal Fermi liquid, etc. In our opinion there is no
need for these radical approaches in explaining basic physics in cuprates at
least in optimally, slightly underdoped and overdoped metallic and
superconducting HTSC cuprates. This subject will be discussed more in
\textit{Part II} and here we give some clarifications related to the dilemma
of Fermi vs non-Fermi liquid. The definition of the \textit{canonical Fermi
liquid} (based on the Landau work) in interacting Fermi systems comprises
the following properties: (1) there are quasi-particles with charge $q=\pm e$%
, spin $s=1/2$ and low-laying energy excitations $\xi _{\mathbf{k}%
}(=\epsilon _{\mathbf{k}}-\mu )$ which are much larger than their inverse
life-times, i.e. $\xi _{\mathbf{k}}\gg 1/\tau _{\mathbf{k}}\sim \xi _{%
\mathbf{k}}^{2}/W_{b}$. Since the level width $\Gamma =2/\tau _{\mathbf{k}}$
of the quasi-particle is negligibly small, this means that the excited
states of the Fermi liquid are placed in one-to-one correspondence with the
excited states of the free Fermi gas; (2) at $T=0$ $K$ there is an energy
level with the Fermi surface at which $\xi _{\mathbf{k}_{F}}=0$ and the
Fermi quasi-particle distribution function $n_{F}(\xi _{\mathbf{k}})$ has
finite jump; (3) the number of quasi-particles under the Fermi surface is
equal to the total number of conduction particles (we omit here other
valence and core electrons) - the Luttinger theorem; (4) the interaction
between quasi-particles are characterized with a few (Landau) parameters
which describe low-temperature thermodynamics and transport properties.
Having this definition in mind one can say that if fermionic quasi-particles
interact with some bosonic excitation (for instance with phonons) and if the
coupling is sufficiently strong, then the former are not described by the
canonical Fermi liquid since at energies and temperatures of the order of
the Debay temperature $k_{B}\Theta _{D}(\equiv \hbar \omega _{D})$ (more
precisely $\sim \Theta _{D}/5$), i.e. for $\xi _{\mathbf{k}}\sim \Theta _{D}$
one has $\tau _{\mathbf{k}}^{-1}\gtrsim \xi _{\mathbf{k}}$ and the
quasi-particle picture (in the sense of the Landau definition) is broken
down. In that respect an electron-boson system can be classified as a
\textit{non-canonical Fermi liquid }for sufficiently strong electron-boson
coupling. It is nowadays well known that for instance Al, Zn are weak
coupling systems since for $\xi _{\mathbf{k}}\sim \Theta _{D}$ one has $\tau
_{\mathbf{k}}^{-1}\ll \xi _{\mathbf{k}}$ and they are well described by the
Landau theory. However, the electron-phonon system is satisfactory described
by the Migdal-Eliashberg theory and the Boltzmann theory, where
thermodynamic and transport properties depend on the spectral function $%
\alpha ^{2}F_{sf}(\mathbf{k},\mathbf{k}^{\prime },\omega )$ and its higher
momenta. Since in HTSC cuprates the electron-boson (phonon) coupling is
rather strong and $T_{c}$ is large, i.e. of the order of characteristic
boson energies ($\omega _{B}$), $T_{c}\sim \omega _{B}/4$, then it is
natural that in the normal state (at $T>$ $T_{c}$) we deal inevitably with a
strong interacting non-canonical Fermi liquid which is at least
qualitatively and semi-quantitatively described by the Migdal-Eliashberg
theory. In order to justify this statement we shall in the following
elucidate some properties in more details by studying optical, ARPES,
tunnelling and other experiments.

\textit{B}. \textit{Prejudice on the limitation of the strength of EPI}

In spite of reach experimental evidence in favor of strong EPI in HTSC
oxides there was a disproportion in the research activity (especially
theoretical) in the past, since the investigation of the SFI mechanism of
pairing prevailed in the literature. This retrograde trend was partly due to
an incorrect statement in \cite{Cohen} on the possible upper limit of T$_{c}$
in the phonon mechanism of pairing. Since in the past we have discussed this
problem thoroughly in numerous papers - for the recent one see \cite
{Maksimov-Dolgov-2007}, we shall outline here the main issue and results
only.

It is well known that in an electron-ion crystal, besides the attractive
EPI, there is also repulsive Coulomb interaction. In case of an isotropic
and homogeneous system with weak quasi-particle interaction, the effective
potential $V_{eff}(\mathbf{k},\omega )$ in the leading approximation looks
like as for two external charges ($e$) embedded in the medium with the
\textit{total longitudinal dielectric function} $\varepsilon _{tot}(\mathbf{k%
},\omega )$ ($\mathbf{k}$ is the momentum and $\omega $ is the frequency)
\cite{Kirzhnitz}, i.e.
\begin{equation}
V_{eff}(\mathbf{k},\omega )=\frac{V_{ext}(\mathbf{k})}{\varepsilon _{tot}(%
\mathbf{k},\omega )}=\frac{4\pi e^{2}}{k^{2}\varepsilon _{tot}(\mathbf{k}%
,\omega )}.  \label{Veff}
\end{equation}
In case of strong interaction between quasi-particles, the state of embedded
quasi-particles changes significantly due to interaction with other
quasi-particles, giving rise to $V_{eff}(\mathbf{k},\omega )\neq 4\pi
e^{2}/k^{2}\varepsilon _{tot}(\mathbf{k},\omega )$. In that case $V_{eff}$
depends on other (than $\varepsilon _{tot}(\mathbf{k},\omega )$) response
functions. However, in the case when Eq.(\ref{Veff}) holds, i. e. when the
weak-coupling limit is realized, $T_{c}$ is given by $T_{c}\approx \bar{%
\omega}\exp (-1/(\lambda ^{ep}-\mu ^{\ast })$ \cite{Allen-Mitrovic}, \cite
{Kirzhnitz}$)$. Here, $\lambda ^{ep}$ is the EPI coupling constant, $\bar{%
\omega}$ is an average phonon frequency and $\mu ^{\ast }$ is the Coulomb
pseudo-potential, $\mu ^{\ast }=\mu /(1+\mu \ln E_{F}/\bar{\omega})$ ($E_{F}$
is the Fermi energy). The couplings $\lambda ^{ep}$ and $\mu $ are expressed
by $\varepsilon _{tot}(\mathbf{k},\omega =0)$
\[
\mu -\lambda ^{ep}=\langle N(0)V_{eff}(\mathbf{k},\omega =0)\rangle
\]
\begin{equation}
=N(0)\int_{0}^{2k_{F}}\frac{kdk}{2k_{F}^{2}}\frac{4\pi e^{2}}{%
k^{2}\varepsilon _{tot}(\mathbf{k},\omega =0)},  \label{NVeff}
\end{equation}
where $N(0)$ is the density of states at the Fermi surface and $k_{F}$ is
the Fermi momentum - see more in \cite{Maksimov-Review}. In \cite{Cohen} it
was claimed that lattice stability of the system with respect to the charge
density wave formation implies the condition $\varepsilon _{tot}(\mathbf{k}%
,\omega =0)>1$ for all $\mathbf{k}$. If this was correct then from Eq.(\ref
{NVeff}) it follows that $\mu >\lambda ^{ep}$, which limits the maximal
value of T$_{c}$ to the value $T_{c}^{\max }\approx E_{F}\exp (-4-3/\lambda
^{ep})$. In typical metals $E_{F}<(1-10)$ $eV$ and if one accepts the
statement in \cite{Cohen}, i.e. that $\lambda ^{ep}\leq \mu \leq 0.5$, one
obtains $T_{c}\sim (1-10)$ $K$. \ The latter result, if it would be true,
means that EPI is ineffective in producing not only high-T$_{c}$
superconductivity but also low-temperature superconductivity (LTS). However,
this result is apparently in conflict first of all with experimental results
in LTSC, where in numerous systems $\mu \leq \lambda ^{ep}$ and $\lambda
^{ep}>1$. For instance, $\lambda ^{ep}\approx 2.6$ is realized in $PbBi$
alloy which is definitely much higher than $\mu (<1)$.

\begin{center}
\begin{figure}[tbp]
\resizebox{.4\textwidth}{!}
{\includegraphics*[width=6cm]{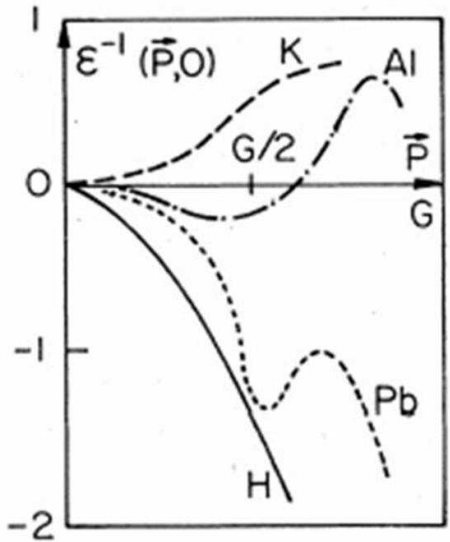}}
\caption{Inverse total static dielectric function $\protect\varepsilon^{-1}(%
\mathbf{p})$ for normal metals (K, Al, Pb and metallic H) in $\mathbf{p}%
=(1,0,0)$ direction. $\mathbf{G}$ is the reciprocal lattice vector.}
\label{Epsilon-k}
\end{figure}
\end{center}

Moreover, the basic theory tells us that $\varepsilon _{tot}(\mathbf{k}\neq
0,\omega )$ is not the response function \cite{Kirzhnitz}. Namely, if a
small external potential $\delta V_{ext}(\mathbf{k},\omega )$ is applied to
the system it induces screening by charges of the medium and the total
potential is given by $\delta V_{tot}(\mathbf{k},\omega )=\delta V_{ext}(%
\mathbf{k},\omega )/\varepsilon _{tot}(\mathbf{k},\omega )$ which means that
$1/\varepsilon _{tot}(\mathbf{k},\omega )$ is the response function. The
latter obeys the Kramers-Kronig dispersion relation which implies the
following stability condition: $1/\varepsilon _{tot}(\mathbf{k},\omega =0)<1$
for $\mathbf{k}\neq 0$, i.e. either $\varepsilon _{tot}(\mathbf{k}\neq
0,\omega =0)>1$ or $\varepsilon _{tot}(\mathbf{k}\neq 0,\omega =0)<0$. This
important theorem invalidates the above\textbf{\ }restriction on the maximal
value of T$_{c}$ in the EPI mechanism. We stress that the condition $%
\varepsilon _{tot}(\mathbf{k}\neq 0,\omega =0)<0$ is not in conflict with
the lattice instability. For instance, in inhomogeneous systems such as
crystal, the total longitudinal dielectric function is matrix in the space
of reciprocal lattice vectors ($\mathbf{Q}$), i.e. $\hat{\varepsilon}_{tot}(%
\mathbf{k+Q},\mathbf{k+Q}^{\prime },\omega )$, and $\varepsilon _{tot}(%
\mathbf{k},\omega )$ is defined by $\varepsilon _{tot}^{-1}(\mathbf{k}%
,\omega )=\hat{\varepsilon}_{tot}^{-1}(\mathbf{k+0},\mathbf{k+0},\omega )$.
It is well known that in dense metallic systems with one ion per cell (such
as metallic hydrogen ) and with the electronic dielectric function $%
\varepsilon _{el}(\mathbf{k},0)$, one has \cite{DKM}
\begin{equation}
\varepsilon _{tot}(\mathbf{k},0)=\frac{\varepsilon _{el}(\mathbf{k},0)}{%
1-1/\varepsilon _{el}(\mathbf{k},0)G_{ep}(\mathbf{k})}.  \label{eps-tot}
\end{equation}
At the same time the frequency of the longitudinal phonon $\omega _{l}(%
\mathbf{k})$ is given by
\begin{equation}
\omega _{l}^{2}(\mathbf{k})=\frac{\Omega _{p}^{2}}{\varepsilon _{el}(\mathbf{%
k},0)}[1-\varepsilon _{el}(\mathbf{k},0)G_{ep}(\mathbf{k})],  \label{ph-fr}
\end{equation}
where $\Omega _{p}^{2}$ is the ionic plasma frequency, $G_{ep}$ is the local
(electric) field correction - see Ref. \cite{DKM}. The real condition for
lattice stability requires that $\omega _{l}^{2}(\mathbf{k})>0$, which
implies that for $\varepsilon _{el}(\mathbf{k},0)>0$ one has $\varepsilon
_{el}(\mathbf{k},0)G_{ep}(\mathbf{k})<1$. The latter condition gives
automatically $\varepsilon _{tot}(\mathbf{k},0)<0$. Furthermore, the
calculations \cite{DKM} show that in the metallic hydrogen crystal, $%
\varepsilon _{tot}(\mathbf{k},0)<0$ for all $\mathbf{k\neq 0}$. Moreover,
the analyzes of crystals with more ions per unit cell \cite{DKM} gives that $%
\varepsilon _{tot}(\mathbf{k\neq 0},0)<0$ is \textit{more a rule than an
exception }- see Fig.~\ref{Epsilon-k}. The physical reason for $\varepsilon
_{tot}(\mathbf{k\neq 0},0)<0$ are local field effects described above by $%
G_{ep}(\mathbf{k})$. Whenever the local electric field $\mathbf{E}_{loc}$
acting on electrons (and ions) is different from the average electric field $%
\mathbf{E}$, i.e. $\mathbf{E}_{loc}\neq \mathbf{E}$, there are corrections
to $\varepsilon _{tot}(\mathbf{k},0)$ (and to $\varepsilon _{e}(\mathbf{k}%
,0) $) which may lead to $\varepsilon _{tot}(\mathbf{k},0)<0$.

The above analysis tells us that in real crystals $\varepsilon _{tot}(%
\mathbf{k},0)$ can be negative in the large portion of the Brillouin zone
giving rise to $\lambda ^{ep}-\mu >0$ in Eq.(\ref{NVeff}). This means that
the dielectric function $\varepsilon _{tot}$ \textit{does not limit }$T_{c}$%
\textit{\ in the phonon mechanism of pairing}. This result does not mean
that there is no limit on T$_{c}$ at all - see more in \cite
{Maksimov-Dolgov-2007} and references therein. We mention in advance that
the local field effects play important role in HTSC oxides, due to their
layered structure with very \textit{unusual ionic-metallic binding}, thus
giving rise to large $EPI$ - see more in the subsequent sections. It is
pertinent to note that one of the author of \cite{Cohen} recognizes the
possibility $\varepsilon _{tot}(\mathbf{k},0)<0$ and in \cite{Cohen2} even
makes interesting proposals for compounds with large EPI and $T_{c}>100$ $K$%
, while the other author \cite{Anderson2} still ignores rigors of scientific
arguments and negates importance of EPI in HTSC cuprates.

In conclusion we point out that there are no theoretical and experimental
arguments for ignoring EPI in HTSC cuprates. To this end it is necessary to
answer several important questions which are related to experimental
findings in HTSC cuprates (oxides): (1) if EPI is responsible for pairing in
HTSC cuprates and if superconductivity is of $d-wave$ type, how are these
two facts compatible? (2) why is the transport EPI coupling constant $%
\lambda _{tr}$ (entering resistivity) much smaller than the pairing EPI
coupling constant $\lambda ^{ep}(>1)$ (entering T$_{c}$), i.e. why one has $%
\lambda _{tr}(\approx 0.4-1.2)\ll \lambda ^{ep}(\sim 2-4)$? (3) is high T$%
_{c}$ possible for a moderate EPI coupling constant, let say for $\lambda
^{ep}\leq 1$, and under which conditions? (4) if EPI is ineffective for
pairing in HTSC oxides, inspite of $\lambda ^{ep}>1$, why it is so?

\textit{C. Is a non-phononic pairing realized in HTSC?}

Regarding EPI one can pose a question - whether it contributes significantly
to d-wave pairing in cuprates? Surprisingly, despite numerous experiments in
favor of EPI, a number of researchers still believe that EPI is irrelevant
for pairing \cite{Pines}. This belief is mainly based: (i) on the,
previously discussed, incorrect lattice stability criterion, which implies
small EPI; (ii) on the well established experimental fact that d-wave
pairing is realized in cuprates \cite{Tsui-Kirtley}, which is believed to be
incompatible with EPI. Having in mind that EPI in HTSC is strong with $%
1<\lambda ^{ep}<3$ (see below), we assume for the moment that the leading
pairing mechanism in cuprates, which gives d-wave pairing, is due to some
non-phononic mechanism, like the \textit{exitonic} one, with the high energy
gluing boson ($\Omega _{nph}\gg \omega _{ph}$) and with the bare critical
temperature $T_{c0}$ and look for the effect of EPI. If EPI is \textit{%
isotropic}, like in most LTSC materials, then it would be very detrimental
for d-wave pairing - the pair breaking effect. In the case of dominating
\textit{isotropic} EPI in the normal state and the exitonic-like pairing,
then near $T_{c}$ the linearized Eliashberg equations have an approximative
form for \textit{weak} non--phonon interaction (with the characteristic
frequency $\Omega _{nph}$)
\[
Z(\omega _{n})\Delta (\mathbf{k},\omega _{n})=\pi T_{c}\sum_{n^{\prime
}}^{\Omega _{_{nph}}}\sum_{\mathbf{q}}V_{nph}(\mathbf{k},\mathbf{q}%
,n,n_{^{\prime }})\frac{\Delta (\mathbf{q},\omega _{n^{\prime }})}{\left|
\omega _{n^{\prime }}\right| }
\]
\begin{equation}
Z(\omega _{n})\approx 1+\Gamma _{ep}/\omega _{n}.  \label{LinElia}
\end{equation}
For pure d-wave pairing one has $V_{nph}(\mathbf{k},\mathbf{q},n,n_{^{\prime
}})=V_{nph}\Theta (\Omega _{nph}-\left| \omega _{n}\right| )\Theta (\Omega
_{nph}-\left| \omega _{n^{\prime }}\right| )\times (\cos k_{x}-\cos
k_{y})(\cos q_{x}-\cos q_{y})$ and $\Delta (\mathbf{k},\omega _{n})=\Delta
_{d}\Theta (\Omega _{nph}-\left| \omega _{n}\right| )(\cos k_{x}-\cos k_{y})$
which gives the equation for T$_{c}$ - see \cite{Maksimov-Review}
\begin{equation}
\ln \frac{T_{c}}{T_{c0}}=\Psi (\frac{1}{2})-\Psi (\frac{1}{2}+\frac{\Gamma
_{ep}}{2\pi T_{c}}).  \label{gama}
\end{equation}
Here $\Psi $ is the di-gamma function. At temperatures near $T_{c}$ one has $%
\Gamma _{ep}\approx 2\pi \lambda _{ep}T_{c}$ $\ $and the solution of Eq. (%
\ref{gama}) is approximately $T_{c}\approx T_{c0}\exp \{-\lambda ^{ep}\}$,
which means that for $T_{c}^{\max }\sim 160$ $K$ and $\lambda ^{ep}>1$ the
bare $T_{c0}$ due to the non-phononic interaction must be very large, i.e. $%
T_{c0}>500$ $K$.

Concerning other non-phononic mechanisms, such as the SFI one, the effect of
the isotropic EPI in the framework of Eliashberg equations was studied
numerically in \cite{Licht}. The latter is based on Eqs.(\ref{Z-Eli}-\ref
{Fi-Eli}) in Appendix A. with kernels
\begin{equation}
\lambda _{\mathbf{kp}}^{Z}(i\nu _{n})=\lambda _{\mathbf{kp}}^{sf}(i\nu
_{n})+\lambda _{\mathbf{kp}}^{ep}(i\nu _{n})  \label{lambdaZ}
\end{equation}
\begin{equation}
\lambda _{\mathbf{kp}}^{\Delta }(i\nu _{n})=\lambda _{\mathbf{kp}}^{ep}(i\nu
_{n})-\lambda _{\mathbf{kp}}^{sf}(i\nu _{n}),  \label{lambdaD}
\end{equation}
where $\lambda _{\mathbf{kp}}^{sf}(i\nu _{n})$ is taken in the FLEX
approximation \cite{Scalapino-Review}. The calculations \cite{Licht} confirm
the very detrimental effect of the isotropic EPI on the d-wave pairing due
to SFI. For the bare SFI $T_{c0}\sim 100$ $K$ and $\lambda ^{ep}>1$ the
calculations give very small $T_{c}\ll 100$ $K$. These results tell us that
a more realistic pairing interaction must be operative in cuprates and that
EPI is \textit{strongly momentum dependent }\cite{Kulic1}. Only in that case
is strong EPI conform with d-wave pairing, either as its main cause or as a
supporter of a non-phononic mechanism. In \textit{Part II} we shall argue
that the strongly momentum dependent EPI is the main player in cuprates
providing the strength of the pairing mechanism, while the residual Coulomb
interaction and SF, although weaker, trigger it to d-wave pairing.

\section{Experimental evidence for strong EPI and weak SFI}

In the following we discuss some important experiments which give evidence
for strong EPI in cuprates. Before doing it; we shall discuss some magnetic
neutron scattering measurements in cuprates whose results are against the
SFI mechanism of pairing. The experimental results related to the pronounced
imaginary part of the susceptibility $Im\chi (\mathbf{k},k_{z},\omega )$ at
the AF wave vector $\mathbf{k}=\mathbf{Q}=(\pi ,\pi )$ were interpreted in a
number of papers as a support for the SFI mechanism for pairing \cite{Pines}%
. We briefly explain \textit{inadequacy} of such an interpretation.

\subsection{Magnetic neutron scattering and the spin fluctuation spectral
function}

\textit{A. SFI affects }$T_{c}$\textit{\ very little}

\begin{figure}[tbp]
\resizebox{.4\textwidth}{!}
{\includegraphics*[width=10cm]{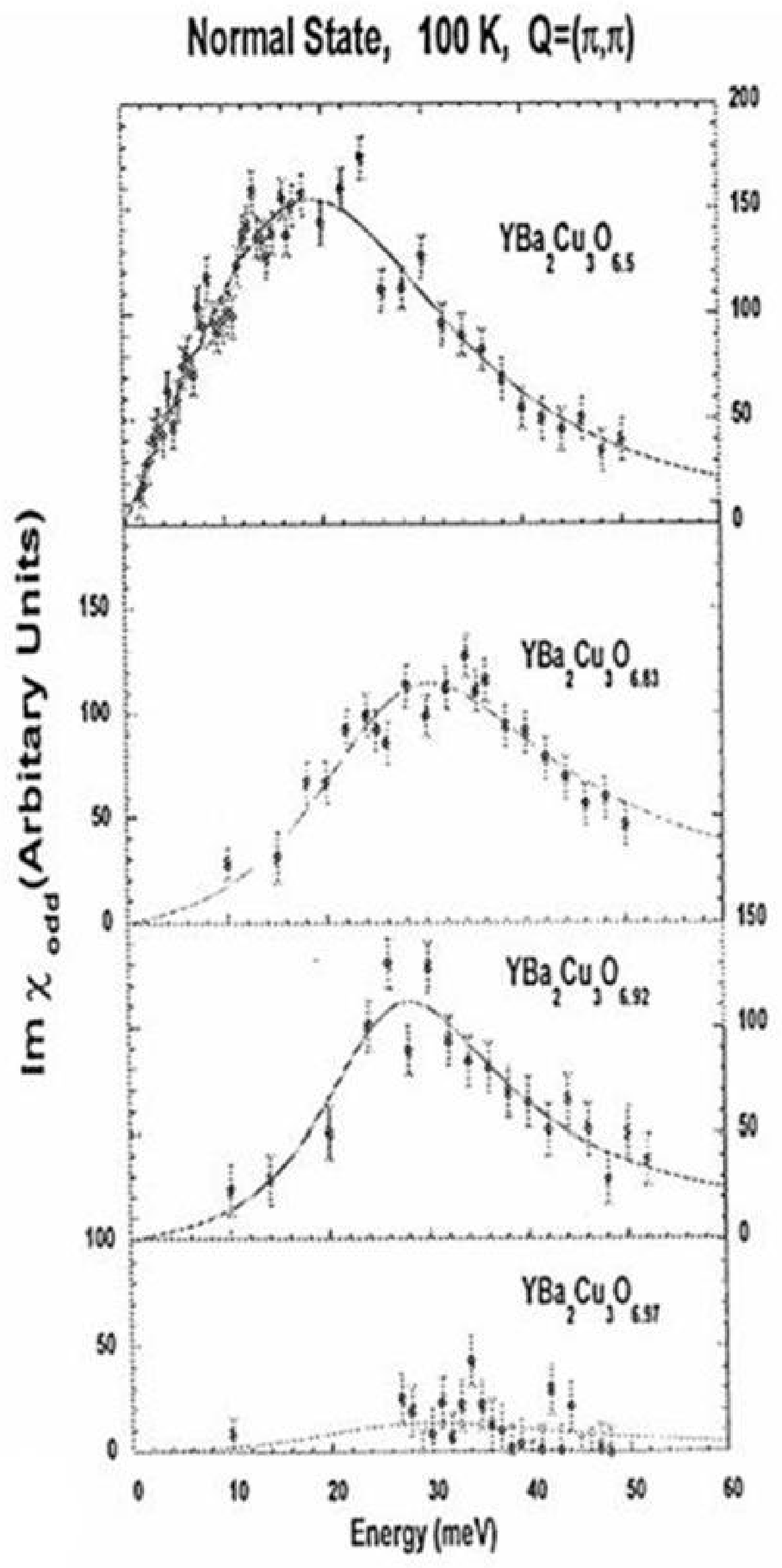}}
{\includegraphics*[width=9cm]{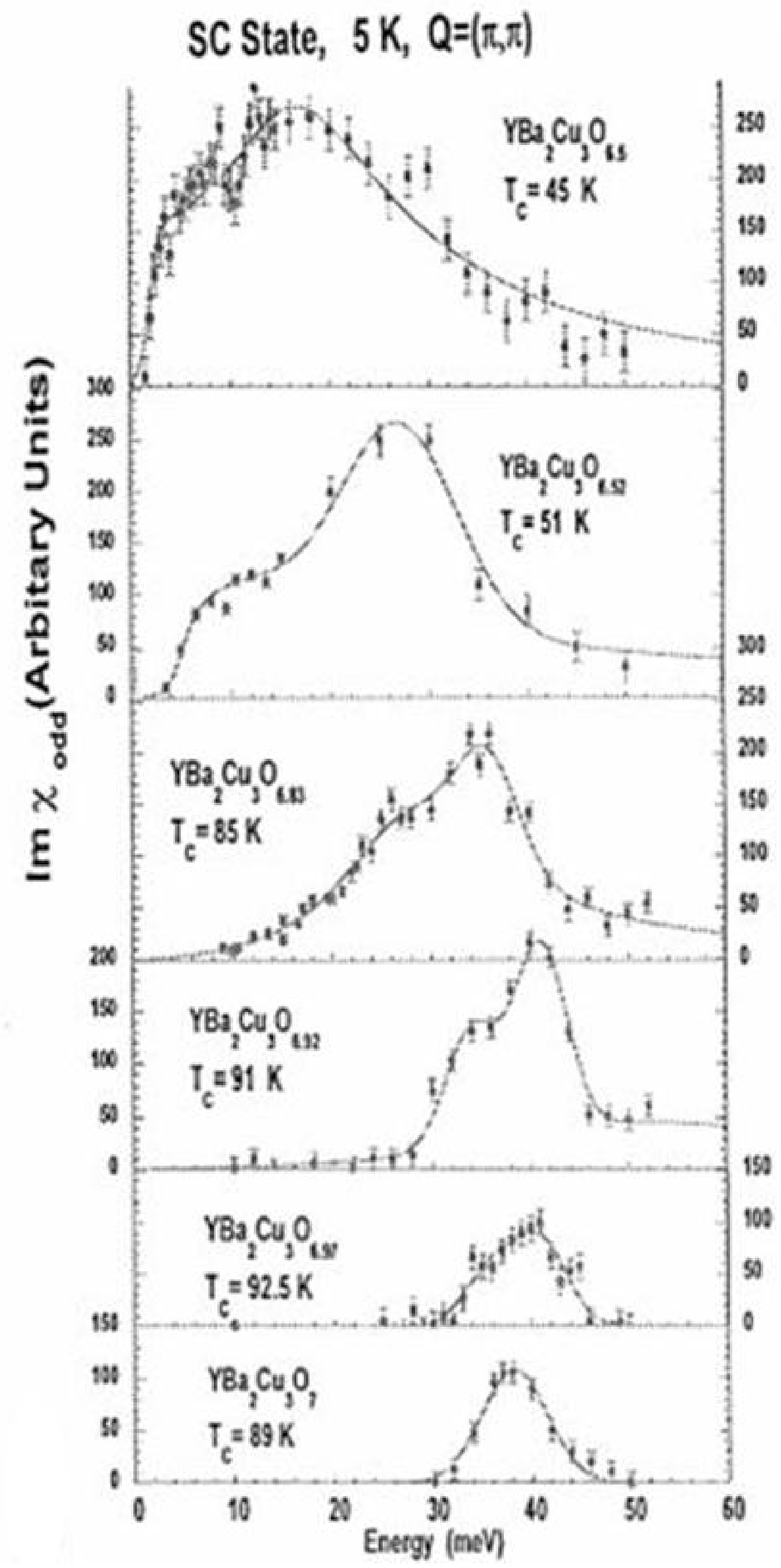}}
\caption{Magnetic spectral function $Im\protect\chi ^{(-)}(\mathbf{k},%
\protect\omega )$ in $YBa_{2}Cu_{3}O_{6+x}$. (Top) In the normal state at $%
T=100$ $K$ and at $Q=(\protect\pi ,\protect\pi )$. $100$ counts in the
vertical scale corresponds to $\protect\chi _{max}^{(-)}\approx 350\protect%
\mu _{B}^{2}/eV$. (Bottom) In the superconducting state at $T=5$ $K$ and at $%
Q=(\protect\pi ,\protect\pi )$. From Ref. \protect\cite{Bourges}.}
\label{SuscFig}
\end{figure}

Before discussing experimental results in cuprates on the imaginary part of
the spin susceptibility $Im\chi (\mathbf{k},\omega )$ we point out that in
theories based on spin fluctuations the effective pairing potential $V_{sf}(%
\mathbf{k},\omega )$, which is repulsive, \ is assumed in the form \cite
{Pines}
\begin{equation}
V_{sf}(\mathbf{q},\omega +i0^{+})=g_{sf}^{2}\int_{-\infty }^{\infty }\frac{%
d\nu }{\pi }\frac{Im\chi (\mathbf{q},\nu +i0^{+})}{\nu -\omega }.
\label{Vsf}
\end{equation}
This form of $V_{sf}$ can be theoretically justified in the weak coupling
limit ($U\ll W_{b}$) only. This mechanism of pairing could be effective in
cuprates only if the spin susceptibility (spectral function) Im$\chi (%
\mathbf{q},\omega )$ is strongly peaked at the AF wave vector $\mathbf{Q}%
=(\pi /a,\pi /a)$. What is the experimental situation? The breakthrough came
from magnetic neutron scattering experiments on $YBa_{2}Cu_{3}O_{6+x}$ by
Bourges group \cite{Bourges}. They showed that $\func{Im}\chi ^{(odd)}(%
\mathbf{q},\omega )$ (the odd part of the spin susceptibility in the bilayer
system) at $\mathbf{q}=\mathbf{Q}=(\pi ,\pi )$ is strongly dependent on the
hole doping as it is shown in Fig.~\ref{SuscFig}.

By varying doping \textit{there is a huge rearrangement} of $\func{Im}\chi
^{(odd)}(\mathbf{Q},\omega )$ in the frequency interval which is important
for superconducting pairing, let say $5$ $meV<\omega <60$ $meV$ as it is
seen in the last two curve in Fig. \ref{SuscFig}(top). At the same time
there is only a small variation of the corresponding critical temperature $%
T_{c}$! For instance, in the underdoped $YBa_{2}Cu_{3}O_{6.92}$\ crystal Im$%
\chi ^{(odd)}(\mathbf{Q},\omega )$, and $S(\mathbf{Q})=N(\mu
)g_{sf}^{2}\int_{0}^{60}d\omega \func{Im}\chi ^{(odd)}(\mathbf{Q},\omega
)\sim \lambda ^{sf}\cdot \left\langle \omega \right\rangle $, is much larger
than that in the near optimally doped $YBa_{2}Cu_{3}O_{6.97}$, i.e. $%
S_{6.92}(\mathbf{Q})\gg S_{6.97}(\mathbf{Q})$, although the difference in
the corresponding critical temperatures $T_{c}$ is very small, i.e. $%
T_{c}^{(6.92)}=91$ $K$ (in $YBa_{2}Cu_{3}O_{6.92}$) and $T_{c}^{(6.97)}=92.5$
$K$ (in $YBa_{2}Cu_{3}O_{6.97}$). This \textit{pronounced rearrangement and
decrease }of Im$\chi ^{(odd)}(\mathbf{Q},\omega )$ by doping, but a
negligible change in $T_{c}$ in YBCO is clearly seen in Fig.~\ref{SuscFig}%
(top), which is \textit{strong evidence} against the $SFI$ mechanism of
pairing. These results in fact mean that the $SFI$ coupling constant $%
\lambda ^{sf}(\sim g_{sf}^{2})$ is small, i.e. $\lambda _{sf}^{(\exp )}\ll 1$%
, and the $SFI$ pairing mechanism is ineffective in cuprates. We stress that
in the phenomenological theory of the SFI pairing \cite{Pines}, an
unrealistically large coupling $g_{sf}>0.7$ $eV$ was assumed which gives $%
\lambda ^{sf}\sim 2$. The latter value cannot be justified neither
experimentally nor theoretically. Let us add that the \textit{%
anti-correlation} between the decrease of $\func{Im}\chi (\mathbf{Q},\omega
) $ and increase of $T_{c}$ by increasing doping toward the optimal value is
also present in the NMR spectral function $I_{\mathbf{Q}}=\lim_{\omega
\rightarrow 0}\func{Im}\chi (\mathbf{Q},\omega )/\omega $ which determines
the longitudinal relaxation rate $1/T_{1}$ - see \cite{Kulic-Review}. This
result additionally disfavors the SFI model of pairing \cite{Pines}, i.e.
the strength of pairing interaction is little affected by SFI. As we shall
discuss below the role of SFI together with the stronger direct Coulomb
interaction is to trigger d-wave pairing.

A less direct argument for \textit{smallness of the SFI coupling constant,
i.e.} $g_{sf}\leq 0.2$ $eV$ and $\lambda ^{sf}\sim 0.2-0.3$ comes from other
experiments related to the magnetic resonance peak in the superconducting
state, and this will be discussed next.

\textit{B. Ineffectiveness of the magnetic resonance peak }

In the superconducting state of optimally doped YBCO and BISCO, $\func{Im}%
\chi (\mathbf{Q},\omega )$ is significantly suppressed at low frequencies
except near the resonance energy $\omega _{res}\approx 41$ $meV$ where a
pronounced narrow peak appears - the \textit{magnetic} \textit{resonance peak%
}. We stress that there is no magnetic resonance peak in LSCO sand
consequently one can question the importance of the resonance peak in the
scattering processes. The relative intensity of this peak (compared to the
total one) is small, i.e. $I_{0}\sim (1-5)\%$ - see Fig~\ref{SuscFig}
(bottom). In underdoped cuprates this peak is present also in the normal
state as it is seen in Fig~\ref{SuscFig} (top). After the discovery of the
resonance peak there were attempts to relate it: (\textit{i}) to the origin
of the superconducting condensation energy and (\textit{ii}) to the kink in
the energy dispersion or the peak-dimp structure in the ARPES spectral
function. In order that the property (i) holds it is necessary that the peak
intensity $I_{0}$ is small \cite{Kivelson}. $I_{0}$ is obtained by equating
the condensation energy $E_{con}$ with the change of the magnetic energy in
the superconducting state, i.e. $\delta E_{mag}\approx 4I_{0}E_{mag}$, where
\[
E_{con}\approx N(0)\Delta ^{2}/2
\]
\begin{equation}
E_{mag}=J\iint \frac{d\omega d^{2}k}{(2\pi )^{3}}(1-\cos k_{x}-\cos k_{y})S(%
\mathbf{k},\omega ).  \label{Emag}
\end{equation}
By taking $2\Delta \approx 4T_{c}$ and the realistic value $N(0)\sim
1/(10J)\sim 1$ $states/eV\cdot spin$, one obtains $I_{0}\sim
10^{-1}(T_{c}/J)^{2}\sim 10^{-3}$. However, such a small intensity cannot be
responsible for the anomalies in ARPES and optical spectra since it gives
rise to small coupling constant $\lambda ^{res}$ for the interaction of
holes with the resonance peak, i.e. $\lambda ^{res}\approx
(2I_{0}N(0)g_{sf}^{2}/\omega _{res})\ll 1$. Such a small coupling does not
affect superconductivity. Moreover, by studying the width of the resonance
peak one can extract the SFI coupling constant $g_{sf}$. Thus, the magnetic
resonance disappears in the normal state of the optimally doped YBCO, which
can be qualitatively understood by assuming that its broadening scales with
the resonance energy $\omega _{res}$, i.e. $\gamma ^{res}<\omega _{res}$,
where the line-width is given by $\gamma ^{res}=4\pi (N(0)g_{sf})^{2}\omega
_{res}$ \cite{Kivelson}. This limits to $g_{sf}<0.2$ $eV$. We stress that
the obtained $g_{sf}$ is much smaller than the one assumed in the
phenomenological spin-fluctuation theory \cite{Pines} where $g_{sf}\sim
0.6-0.7$ $eV$, but much larger than in \cite{Kivelson} (where $g_{sf}<0.02$ $%
eV$). The smallness of $g_{sf}$ comes out also from the analysis of the
antiferromagnetic state in underdoped metals of LSCO and YBCO \cite
{Kulic-Kulic}, where the small magnetic moment $\mu (<0.1$ $\mu _{B})$
points to an itinerant antiferromagnetism with small coupling constant $%
g_{sf}\leq 0.2$ $eV$. The conclusion is that the magnetic resonance in the
optimally doped YBCO is a consequence of the onset of superconductivity and
not its cause.

There is also a principal reason against the pairing due to the resonance
peak at least in optimally doped cuprates. Since its intensity near $T_{c}$
is vanishingly small, though not affecting pairing at the second order phase
transition at $T_{c}$, then if it would be the origin for superconductivity
the phase transition at T$_{c}$ would be \textit{first order}, contrary to
experiments. Recent ARPES experiments give evidence that the magnetic
resonance cannot be related to the kinks in ARPES spectra \cite{Lanzara},
\cite{Valla} - see the discussion below.

We shall argue below that despite its smallness, spin fluctuations can,
together with other contributions of the residual Coulomb interaction,
\textit{trigger} d-wave pairing, while the \textit{strength of pairing is
due to EPI }which is peaked at small transfer momenta - see more \ below and
in \cite{Kulic-Review}, \cite{Kulic-AIP}.

\subsection{Optical conductivity and EPI}

Optical spectroscopy gives information on \textit{optical conductivity} $%
\sigma (\omega )$ and on two-particle excitations, from which one can
indirectly extract the transport spectral function $\alpha _{tr}^{2}F(\omega
)$. Since this method probes bulk sample (on the skin depth), contrary to
ARPES\ and tunnelling methods which probe tiny regions ($10-15$ \AA ) near
the sample surface, this method is very indispensable. However, $\sigma
(\omega )$ \textit{is not a directly} \textit{measured quantity} but it is
derived from the reflectivity $R(\omega )=\left| (\sqrt{\varepsilon
_{ii}(\omega )}-1)/(\sqrt{\varepsilon _{ii}(\omega )}+1)\right| ^{2}$ with
the transversal dielectric tensor $\varepsilon _{ii}(\omega )=\varepsilon
_{ii,\infty }+\varepsilon _{ii,latt}+4\pi i\sigma _{ii}(\omega )/\omega $.
Here, $\varepsilon _{ii,\infty }$ is the high frequency dielectric function,
$\varepsilon _{ii,latt}$ describes the contribution of the lattice
vibrations and $\sigma _{ii}(\omega )$ describes the optical (dynamical)
conductivity of conduction carriers. $R(\omega )$ was usually measured in
the limited frequency interval $\omega _{\min }<\omega <\omega _{\max }$.
Therefore, some physical modelling for $R(\omega )$ is needed in order to
guess it outside this range - see more in reviews \cite{Maksimov-Review},
\cite{Kulic-Review}. This was the reason for numerous inadequate
interpretations of optic measurements in cuprates, as well as the
misconceptions and misinterpretations that will be uncover below. An
illustrative example for \ this claim is large dispersion in the reported
value of $\omega _{pl}$ - from $0.06$ to $25$ $eV$, i.e. almost three orders
of magnitude - see discussion in \cite{Bozovic-Plasma}. This tells us also
that in some periods science suffers from a lack of rigorousness and
objectiveness. However, it turns out that $IR$ measurements of $R(\omega )$
in conjunction with elipsometric measurements of $\varepsilon _{ii}(\omega )$
at high frequencies allows reliable determination of $\sigma (\omega )$.

1. \textit{Transport and quasiparticle relaxation rates }

The widespread misconception in studying the quasi-particle
scattering in cuprates was an ad hoc assumption that the
\textit{transport relaxation rate}
$\gamma _{tr}(\omega )$ is equal to the \textit{quasi-particle relaxation rate%
} $\gamma (\omega )$, in spite of the well known fact that $\gamma
_{tr}(\omega )\neq \gamma (\omega )$ \cite{Allen}. This incorrect assumption
led to the abandoning of EPI as relevant scattering mechanism in cuprates.
Although we have discussed this problem several times before, we want to do
it again, since the correct understanding of the scattering mechanism in
cuprates will take us forward in understanding of the pairing mechanism.

\begin{figure}[tbp]
\resizebox{.5\textwidth}{!}
{\includegraphics*[width=9cm]{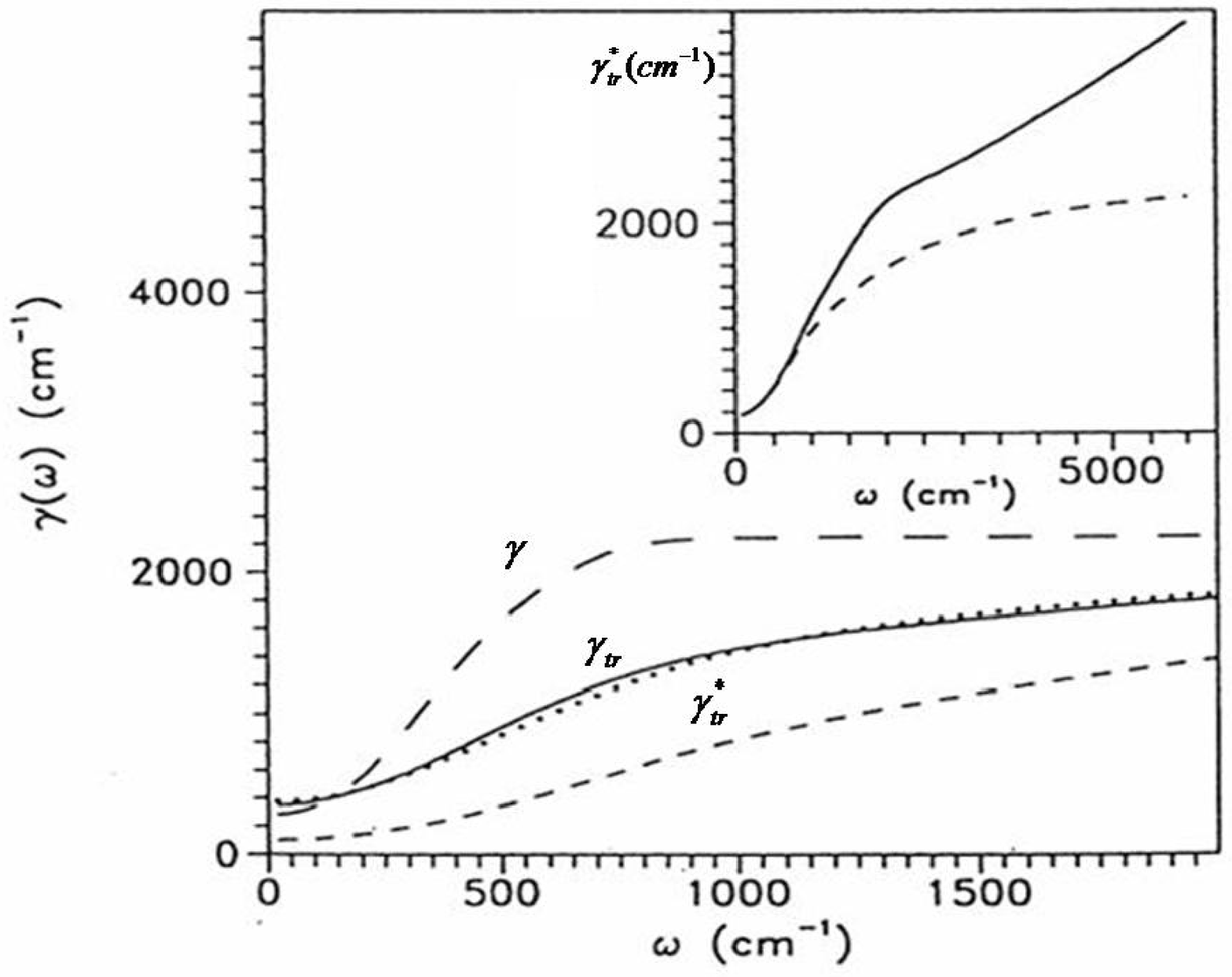}}
{\includegraphics*[width=6cm]{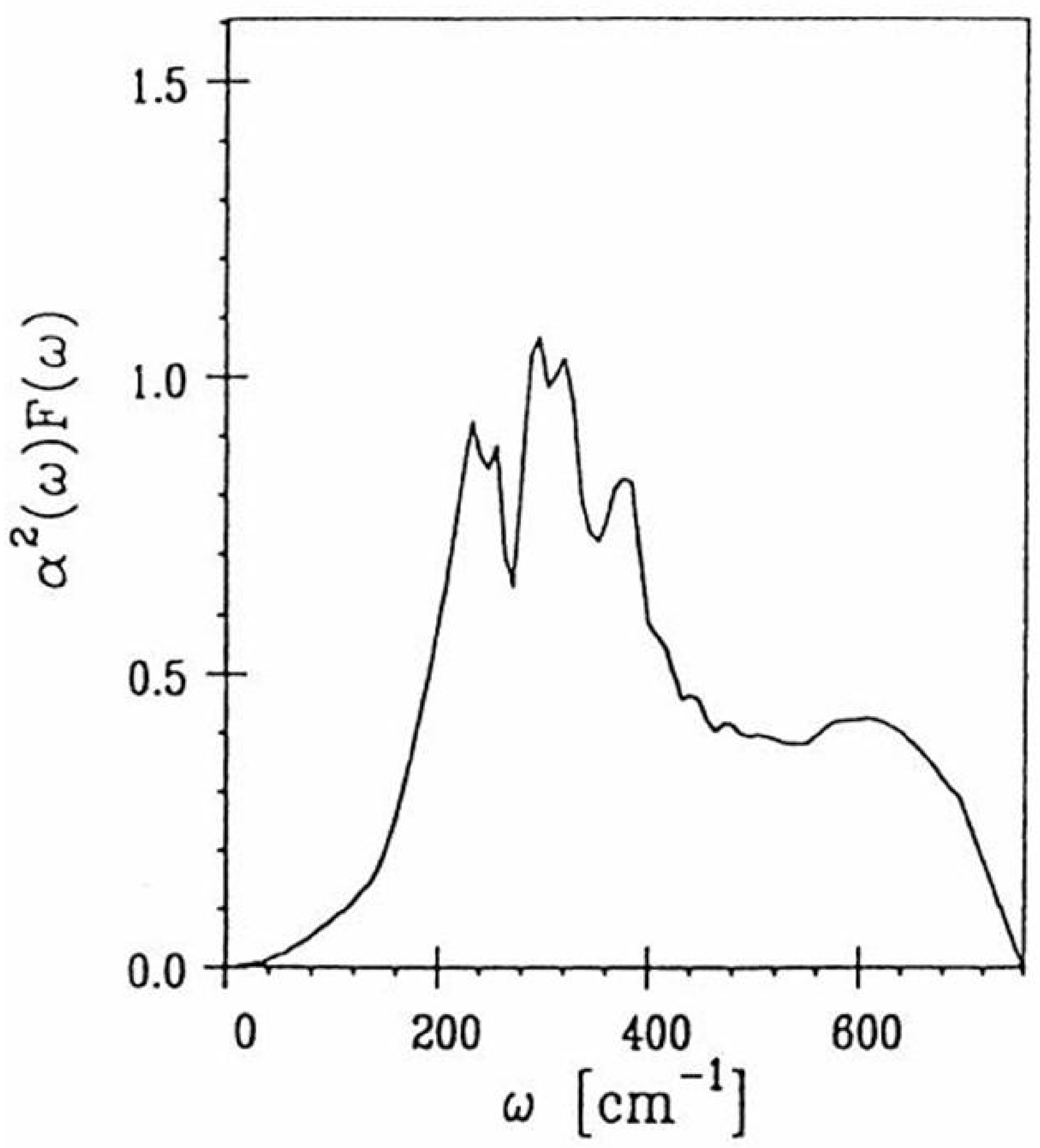}}
\caption{(a) Scattering rates $\protect\gamma (\protect\omega ,T)$, $\protect%
\gamma _{tr}(\protect\omega ,T)$ and $\protect\gamma _{tr}^{\ast }$ - from
top to bottom, for the Eliashberg function in (b). From \protect\cite
{Dolgov-Shulga}. (b)Eliashberg spectral function $\protect\alpha _{ep}^{2}(%
\protect\omega )F_{ph}(\protect\omega )$ obtained from tunnelling
experiments on break junctions \protect\cite{Tunneling-Vedeneev}. Inset
shows $\protect\gamma _{tr}^{\ast }$ with (full line) and without (dashed
line) interband transitions protect\protect\cite{Maksimov-Review}.}
\label{Rates}
\end{figure}

The dynamical conductivity $\sigma (\omega )$ consists of two parts, i.e. $%
\sigma (\omega )=\sigma ^{inter}(\omega )+\sigma ^{intra}(\omega )$ where $%
\sigma ^{inter}(\omega )$ describes \textit{interband} \textit{transitions}
which contribute at higher frequencies, while $\sigma ^{intra}(\omega )$ is
due to \textit{intraband} transitions which are relevant at low frequencies $%
\omega <1$ $eV$. (In $IR$ measurements the frequency is usually given in $%
cm^{-1}$, where the following conversion holds: $1cm^{-1}=29.98$ $%
GHz=0.123985$ $meV=1.44$ $K$.) The experimental data for $\sigma (\omega
)=\sigma _{1}+i\sigma _{2}$ in cuprates are usually processed by the
generalized (extended) Drude formula \cite{Allen}, \cite{Schlesinger}, \cite
{Dolgov-Shulga}, \cite{Shulga}
\begin{equation}
\sigma (\omega )=\frac{\omega _{p}^{2}}{4\pi }\frac{1}{\gamma _{tr}(\omega
)-i\omega m_{tr}(\omega )/m_{\infty }}\equiv \frac{1}{\tilde{\omega}%
_{tr}(\omega )},  \label{Drude}
\end{equation}
which is a useful representation for systems with single band electron-boson
scattering which is justified in HTSC cuprates - see the discussion below.
(The usefulness of introducing the optic relaxation $\tilde{\omega}%
_{tr}(\omega )$ will be discussed in Appendix B.) Here, $i=a,b$ enumerates
the plane axis, $\omega _{p}$, $\gamma _{tr}(\omega ,T)$ and $m_{op}(\omega
) $ are the electronic plasma frequency, the transport (optical) scattering
rate and the optical mass, respectively. Very frequently, the quantity $%
\gamma _{tr}^{\ast }(\omega ,T)=\gamma _{tr}(\omega ,T)(m_{\infty
}/m_{tr}(\omega ))=\func{Im}\sigma (\omega )/\omega \func{Re}\sigma (\omega
) $ \cite{Schlesinger}, which is determined from the half-width of the
Drude-like expression for $\sigma (\omega )$, was analyzed since it is
independent of $\omega _{p}^{2}$. In the weak coupling limit $\lambda
^{ep}<1 $, the formula for conductivity given in Eqs. (\ref{Sigma-tr}-\ref
{K2}) can be written in the form of Eq.(\ref{Drude}) where $\gamma _{tr}$
reads \cite{Dolgov-Shulga}-\cite{Shulga}
\[
\gamma _{tr}(\omega ,T)=\pi \sum_{l}\int_{0}^{\infty }d\nu \alpha
_{tr,l}^{2}F_{l}(\nu )[2(1+2n_{B}(\nu ))
\]
\begin{equation}
-2\frac{\nu }{\omega }-\frac{\omega +\nu }{\omega }n_{B}(\omega +\nu )+\frac{%
\omega -\nu }{\omega }n_{B}(\omega -\nu )].  \label{Gamma-tr}
\end{equation}
Here $n_{B}(\omega )$ is the Bose distribution function. (For the explicit
form of the transport mass $m_{tr}(\omega )$ see \cite{Allen}, \cite
{Dolgov-Shulga}, \cite{Shulga}, \cite{Maksimov-Review}, \cite{Kulic-Review}%
.) In the presence of impurity scattering one should add $\gamma _{tr}^{imp}$
to $\gamma _{tr}$. It turns out that Eq.(\ref{Gamma-tr}) holds within a few
percents also for large $\lambda ^{ep}(>1)$. Note, that $\alpha
_{tr,l}^{2}F_{l}(\nu )\neq \alpha _{l}^{2}F_{l}(\nu )$ and the index $l$
enumerates all scattering bosons - phonons, spin fluctuations, etc. For
comparison, we give the quasi-particle scattering rate $\gamma (\omega ,T)$

\[
\gamma (\omega ,T)=2\pi \int\limits_{0}^{\infty }d\nu \alpha ^{2}F(\nu
)\{2n_{B}(\nu )
\]
\begin{equation}
+n_{F}(\nu +\omega )+n_{F}(\nu -\omega )\}+\gamma ^{imp},  \label{gamma-n}
\end{equation}
where $n_{F}$ is the Fermi distribution function. By comparing Eq.(\ref
{gamma-n}) and Eq.(\ref{Gamma-tr}), it is seen that $\gamma _{tr}$ and $%
\gamma $ are different quantities, $\gamma _{tr}\neq \gamma $, i.e. the
former describes the relaxation of Bose particles (electron-hole pairs)
while the latter one the relaxation of Fermi particles. This difference
persists also at $T=0$ $K$ \ where one has (due to simplicity we omit in the
following summation over $l$ ) \cite{Allen}
\begin{equation}
\gamma _{tr}(\omega )=\frac{2\pi }{\omega }\int_{0}^{\omega }d\nu (\omega
-\nu )\alpha _{tr}^{2}(\nu )F(\nu )  \label{Gamma-tr-0}
\end{equation}
and
\begin{equation}
\gamma (\omega )=2\pi \int_{0}^{\omega }d\nu \alpha ^{2}(\nu )F(\nu ).
\label{Gamma-qp-0}
\end{equation}
In the case of EPI, the above equations give that $\gamma ^{ep}(\omega
)=const$ for $\omega >\omega _{ph}^{\max }$ while $\gamma _{tr}^{ep}(\omega
) $ (as well as $\gamma _{tr}^{\ast }$) is monotonic growing for $\omega
>\omega _{ph}^{\max }$, where $\omega _{ph}^{\max }$ is the maximal phonon
frequency. This is clearly seen by comparing $\gamma (\omega ,T)$, $\gamma
_{tr}(\omega ,T)$ and $\gamma _{tr}^{\ast }$ which are calculated for the
EPI spectral function $\alpha _{ep}^{2}(\omega )F_{ph}(\omega )$ extracted
from tunnelling experiments in YBCO (with $\omega _{ph}^{\max }\sim 80$ $meV$%
) \cite{Tunneling-Vedeneev} - see Fig.~\ref{Rates}.

The results shown in Fig.~\ref{Rates} clearly demonstrate the physical
difference between two scattering rates $\gamma ^{ep}$ and $\gamma
_{tr}^{ep} $. It is also seen that $\gamma _{tr}^{\ast }(\omega ,T)$ is more
a linear function of $\omega $ than $\gamma _{tr}(\omega ,T)$. From these
calculations one concludes that the quasi-linearity of $\gamma _{tr}(\omega
,T)$ (and $\gamma _{tr}^{\ast }$) is not in contradiction with the EPI
scattering mechanism but it is in fact a natural consequence of EPI. We
stress that such behavior of $\gamma ^{ep}$ and $\gamma _{tr}^{ep}$, shown
in Fig.~\ref{Rates}, is in fact not exceptional for HTSC cuprates but it is
generic for many metallic systems, for instance 3D metallic oxides, low
temperature superconductors such as $Al$, $Pb$, etc. - see more in \cite
{Maksimov-Review}, \cite{Kulic-Review}.

\begin{figure}[tbp]
\resizebox{.4\textwidth}{!}
{\includegraphics*[width=8cm]{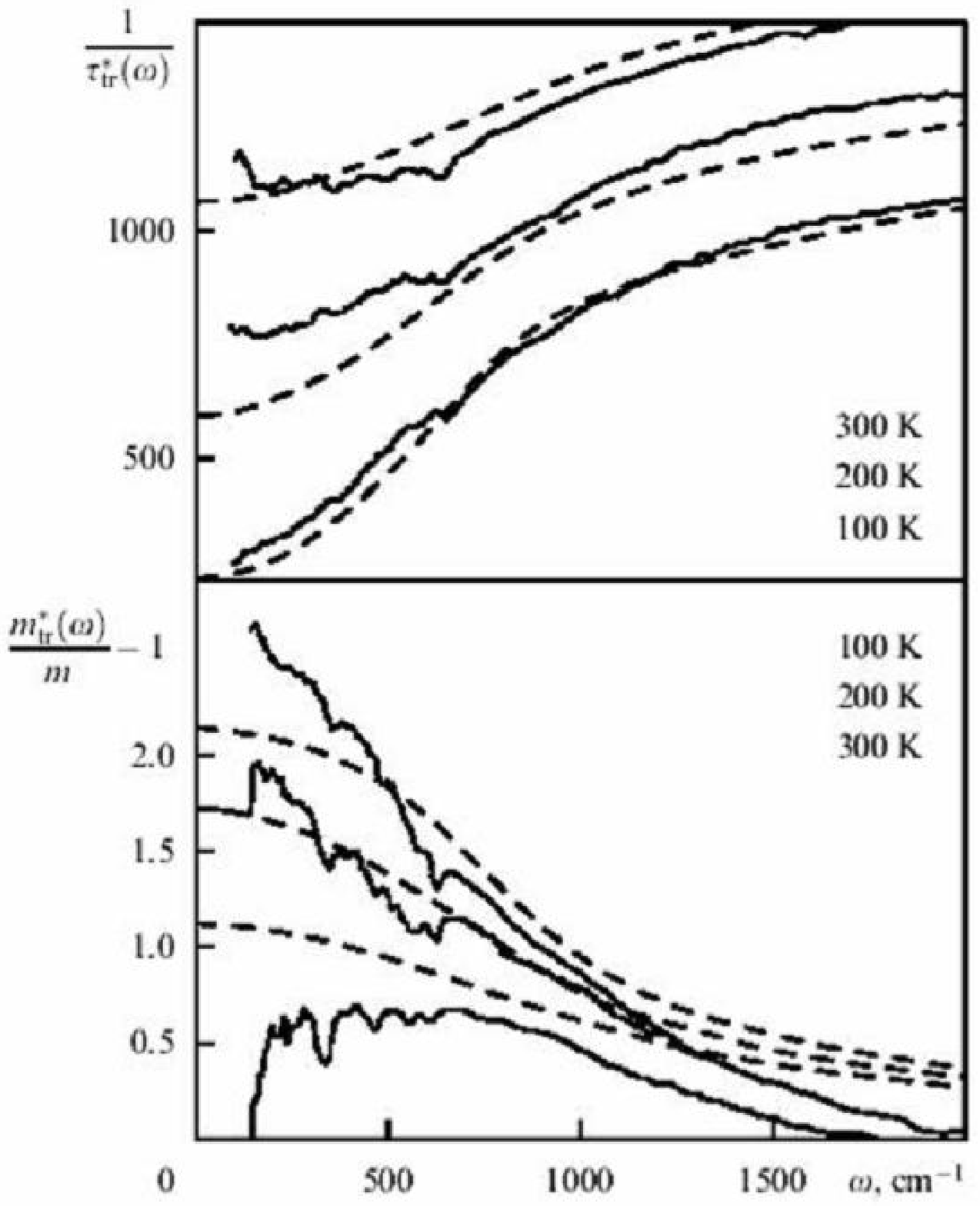}}
{\includegraphics*[width=8cm]{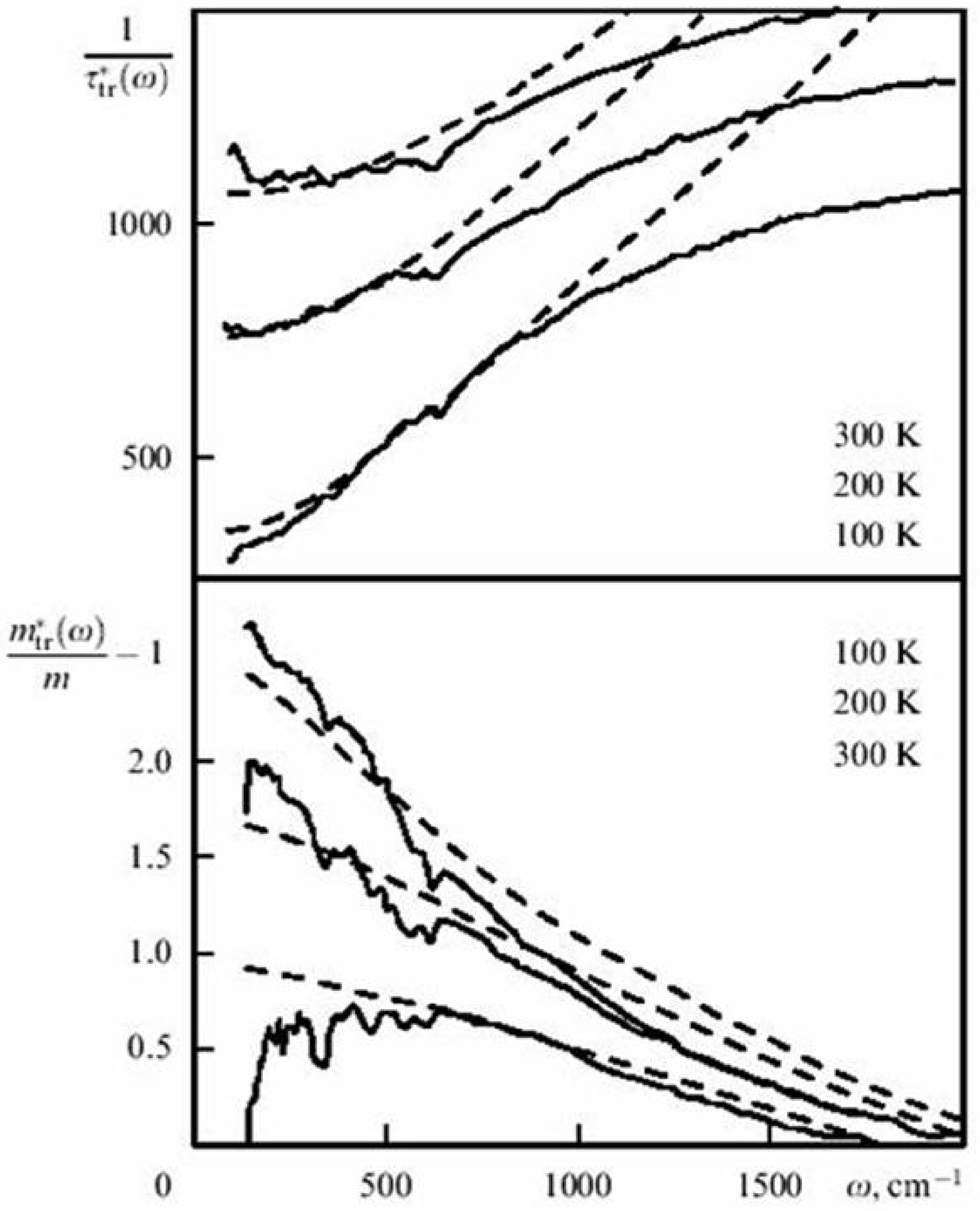}}
\caption{(Top) Experimental transport scattering rate $\protect\gamma %
_{tr}^{\ast }$ (solid lines) for BISCO and the theoretical curve by using
Eq. (\ref{Sigma-tr}) and transport mass $m _{tr}^{\ast }$ with $\protect%
\alpha^{2}F(\protect\omega)$ described in text (dashed lines). (Bottom)
Comparison with the marginal Fermi liquid theory - dashed lines. From
\protect\cite{Maksimov-Review}}
\label{MaksRev15-16}
\end{figure}

Let us discuss briefly the experimental results for $R(\omega )$ and $\gamma
_{tr}^{\ast }(\omega ,T)$ and compare these with theoretical predictions
obtained by using a single band model and $\alpha _{ep}^{2}(\omega
)F_{ph}(\omega )$ from tunnelling data with the EPI coupling $\lambda =2$
\cite{Tunneling-Vedeneev}. In the case of YBCO the agreement between
measured and calculated $R(\omega )$ is very good up to frequencies $\omega
<6000$ $cm^{-1}$ which confirms the importance of EPI in scattering
processes. For higher frequencies, where a mead infrared peak appears, it is
necessary to account for interband transitions \cite{Maksimov-Review}. In
optimally doped $Bi_{2}Sr_{2}CaCu_{2}O_{6}$ \cite{Romero92} the experimental
results for $\gamma _{tr}^{\ast }(\omega ,T)$ are explained theoretically by
assuming that the EPI spectral function $\alpha _{ep}^{2}(\omega )F(\omega
)\sim F_{ph}(\omega )$, where $F_{ph}(\omega )$ is the phononic DOS in BISCO
while $\alpha _{ep}^{2}(\omega )\sim \omega ^{1.6}$, $\lambda =1.9$ and $%
\gamma _{im}\approx 320$ $cm^{-1}$ - see Fig. ~\ref{MaksRev15-16}(a). The
agreement is rather good. At the same time the fit of $\gamma _{tr}^{\ast
}(\omega ,T)$ by the marginal Fermi liquid fails as it is evident in Fig.~%
\ref{MaksRev15-16}(b).

\begin{figure}[tbp]
\resizebox{.5\textwidth}{!}
{\includegraphics*[width=7cm]{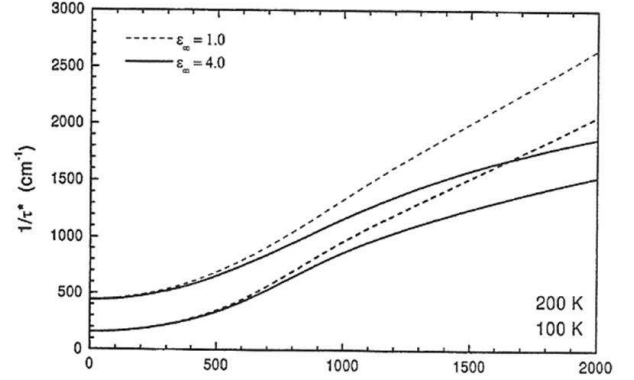}}
\caption{Dependence of $\protect\gamma _{tr}^{\ast }(\protect\omega ,T)$ on $%
\protect\varepsilon _{\infty }$ in $Bi_{2}Sr_{2}CaCu_{2}O_{8}$ on different
temperatures for $\protect\varepsilon _{\infty }=4$ (solid lines) and $%
\protect\varepsilon _{\infty }=1$ (dashed lines). From \protect\cite
{Kaufmann}.}
\label{GammaEpsilon}
\end{figure}

Now we will comment on the so called pronounced linear behavior of $\gamma
_{tr}(\omega ,T)$ (and $\gamma _{tr}^{\ast }(\omega ,T)$) which served in
the past for numerous inadequate conclusions. We stress that the measured
quantity is reflectivity $R(\omega )$ and derived ones are $\sigma (\omega )$%
, $\gamma _{tr}(\omega ,T)$ and $m_{tr}(\omega )$, which are very sensitive
to the value of the dielectric constant $\varepsilon _{\infty }$. This is
clearly demonstrated in Fig.~\ref{GammaEpsilon} for Bi2212 where it is seen
that $\gamma _{tr}(\omega ,T)$ (and $\gamma _{tr}^{\ast }(\omega ,T)$) for $%
\varepsilon _{\infty }=1$ is linear up to much higher $\omega $ than in the
case $\varepsilon _{\infty }>1$.

In some experiments \cite{Puchkov}, \cite{Timusk-old} $\gamma _{tr}(\omega
,T)$ (and $\gamma _{tr}^{\ast }(\omega ,T)$) is linear up to very high $%
\omega $ which means that the ion background and interband transitions
(contained in $\varepsilon _{\infty }$) are not properly taken into account
since it is assumed too small $\varepsilon _{\infty }$. The recent
elipsometric measurements on YBCO \cite{BorisMPI} give the reliable value
for $\varepsilon _{\infty }\approx 4-6$. The latter gives rise to a much
less spectacular linearity in the relaxation rates than it was the case
immediately after the discovery of HTSC cuprates.

Furthermore, we would like to comment on two points concerning $\sigma $, $%
\gamma _{tr}$, $\gamma $ and their interrelations. First, the
parametrization of $\sigma (\omega )$ with the generalized Drude formula in
Eq.(\ref{Drude}) and its relation to the transport scattering rate $\gamma
_{tr}(\omega ,T)$ and the transport mass $m_{tr}(\omega ,T)$ is useful if we
deal with electron-boson scattering in a single band problem. In \cite
{Shulga} it is shown that $\sigma (\omega )$ of a two-band model with only
elastic impurity scattering can be represented by the generalized (extended)
Drude formula with $\omega $ and $T$ dependence of effective parameters $%
\gamma _{tr}^{eff}(\omega ,T)$, $m_{tr}^{eff}(\omega ,T)$ despite the fact
that the inelastic electron-boson scattering is absent. To this end we
stress that the single-band approach is fully justified for a number of HTSC
cuprates such as LSCO, BISCO etc. Second, at the beginning we said that $%
\gamma _{tr}(\omega ,T)$ and $\gamma (\omega ,T)$ are physically different
quantities and it holds $\gamma _{tr}(\omega ,T)\neq \gamma (\omega ,T)$. In
order to give the physical picture and qualitative explanation we assume
that $\alpha _{tr}^{2}F(\nu )\approx \alpha ^{2}F(\nu )$. In that case the
renormalized frequencies, the quasi-particle one $\tilde{\omega}(\omega
)=Z(\omega )\omega =\omega -\Sigma (\omega )$ (for the definition of $%
Z(\omega )$ see Appendix A.) and the transport one $\tilde{\omega}%
_{tr}(\omega )$ - defined above, are related and at $T=0$, they are given by
\cite{Allen}, \cite{Shulga}
\begin{equation}
\tilde{\omega}_{tr}(\omega )=\frac{1}{\omega }\int_{0}^{\omega }d\omega
^{\prime }2\tilde{\omega}(\omega ^{\prime }).  \label{tr-qp}
\end{equation}
It gives the relation between $\gamma _{tr}(\omega )$ and $\gamma (\omega )$%
, $m_{tr}(\omega )$ and $m^{\ast }(\omega )$ respectively
\begin{equation}
\gamma _{tr}(\omega )=\frac{1}{\omega }\int_{0}^{\omega }d\omega ^{\prime
}\gamma (\omega ^{\prime })  \label{gammatr-gamma}
\end{equation}
\begin{equation}
\omega m_{tr}(\omega )=\frac{1}{\omega }\int_{0}^{\omega }d\omega ^{\prime
}2\omega ^{\prime }m^{\ast }(\omega ^{\prime }).  \label{mtr-m}
\end{equation}
The physical meaning of Eq.(\ref{tr-qp}) is the following: in optical
measurements one photon with the energy $\omega $ is absorbed and two
excited particles (electron and hole) are created above and below the Fermi
surface. If the electron has energy $\omega ^{\prime }$ and the hole $\omega
-\omega ^{\prime }$, then they relax as quasi-particles with the
renormalized $\tilde{\omega}$. Since $\omega ^{\prime }$ takes values $%
0<\omega ^{\prime }<\omega $ then the optical relaxation $\tilde{\omega}%
_{tr}(\omega )$ is the energy-averaged $\tilde{\omega}(\omega )$ according
to Eq.(\ref{tr-qp}). The factor 2 is due to the two quasi-particles,
electron+hole. At finite $T$, the generalization reads \cite{Allen}, \cite
{Shulga}
\begin{equation}
\tilde{\omega}_{tr}(\omega )=\frac{1}{\omega }\int_{0}^{\infty }d\omega
^{\prime }[1-n_{F}(\omega ^{\prime })-n_{F}(\omega -\omega ^{\prime })]2%
\tilde{\omega}(\omega ^{\prime }).  \label{omtr-T}
\end{equation}

2. \textit{Inversion of the optical data and }$\alpha _{tr}^{2}(\omega
)F(\omega )$\textit{\ }

\begin{figure}[tbp]
\resizebox{.4\textwidth}{!}
{\includegraphics*[width=7cm]{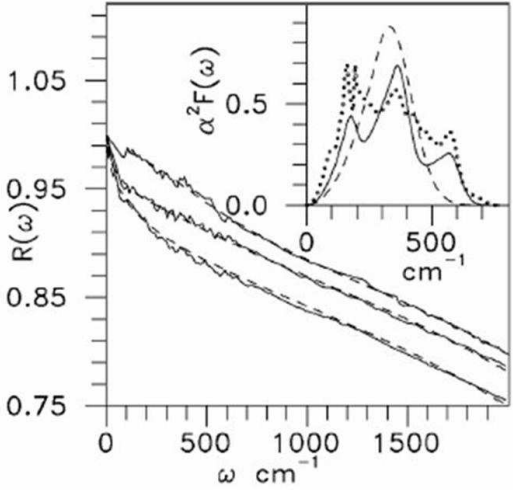}}
\caption{Experimental (solid lines) and calculated (dashed lines) of $R(%
\protect\omega)$ in optimally doped YBCO \protect\cite{Schutzmann} at T=100,
200, 300 K (from top to bottom). Inset: the two reconstructed $\protect%
\alpha _{tr}^{2}(\protect\omega )F(\protect\omega )$ at T=100 K. The phonon
density of states $F(\protect\omega )$ - dotted line. From \protect\cite
{Dolgov-Shulga}}
\label{ShulgaFig1}
\end{figure}

\begin{figure}[tbp]
\resizebox{.4\textwidth}{!}
{\includegraphics*[width=7cm]{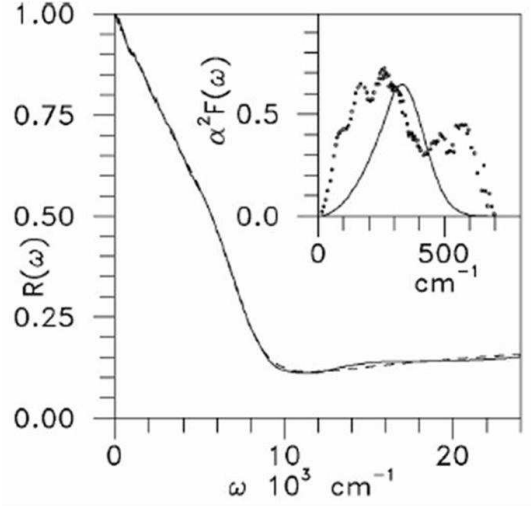}}
\caption{Experimental (solid line) and calculated (dashed line) of $R(%
\protect\omega )$ in optimally doped BISCO \protect\cite{Kamaras} at T=100
K. Inset: the reconstructed $\protect\alpha _{tr}^{2}(\protect\omega )F(%
\protect\omega )$ - solid line. The phonon density of states $F(\protect%
\omega )$ - dotted line. From \protect\cite{Dolgov-Shulga}}
\label{ShulgaFig2}
\end{figure}

In principle, the transport spectral function $\alpha _{tr}^{2}(\omega
)F(\omega )$ can be precisely extracted from $\sigma (\omega )$, i.e. $%
\gamma _{tr}(\omega )$, only at $T=0$ $K$, which follows from Eq.( \ref
{Gamma-tr-0})
\[
\alpha _{tr}^{2}(\omega )F(\omega )=\frac{1}{2\pi }\frac{\partial ^{2}}{%
\partial \omega ^{2}}(\omega \gamma _{tr}(\omega )
\]
\begin{equation}
=\frac{\omega _{p}^{2}}{8\pi ^{2}}\frac{\partial ^{2}}{\partial \omega ^{2}}[%
\omega Re\frac{1}{\sigma (\omega )}]\mid _{T=0}.  \label{spect-func}
\end{equation}
However, real measurements are performed at finite $T$ (and also at $T>T_{c}$%
) and the inversion procedure is in principle an ill-posed problem since $%
\alpha _{tr}^{2}(\omega )F(\omega )$ is the deconvolution of the
inhomogeneous Fredholm integral equation of the first kind with the
temperature dependent Kernel $K_{2}(\omega ,\nu ,T)$ in Eq.(\ref{Gamma-tr}).
An ill-posed mathematical problem, like this one, is very sensitive to input
since experimental data contain less information than one needs. This can
cause the fine structure of $\alpha _{tr}^{2}(\omega )F(\omega )$ gets
blurred in the extraction procedures and it can be temperature dependent
even when the true $\alpha _{tr}^{2}(\omega )F(\omega )$ is $T$ independent.
In the context of HTSC cuprates, this problem was first studied in \cite
{Dolgov-Shulga}, \cite{Shulga} with the following results: (\textit{1}) the
extracted shape of $\alpha _{tr}^{2}(\omega )F(\omega )$ in $%
YBa_{2}Cu_{3}O_{7-x}$\ is not unique and it is temperature dependent, i.e.
at higher $T>T_{c}$ the peak structure is smeared and only a single peak
(slightly shifted to higher $\omega $) is present. For instance, the
experimental data of $R(\omega )$ in YBCO were reproduced by two different
spectral functions $\alpha _{tr}^{2}(\omega )F(\omega )$, one with single
peak and the other with three peaks structure as it is shown in Fig.~\ref
{ShulgaFig1} The similar situation is realized in optimally doped BISCO as
it is seen in Fig.~\ref{ShulgaFig2}. It is important to stress that the
width of the extracted $\alpha _{tr}^{2}(\omega )F(\omega )$ in both
compounds coincide with the width of the phonon density of states $%
F_{ph}(\omega )$ \cite{Dolgov-Shulga}, \cite{Shulga}, \cite{Kaufmann}; (%
\textit{2}) the upper energy bound for $\alpha _{tr}^{2}(\omega )F(\omega )$
can be extracted with certainty and it coincides approximately with the
maximal phonon frequency in cuprates $\omega _{ph}^{\max }\lesssim 80$ $meV$
as it is seen in Figs. \ref{ShulgaFig1}-\ref{ShulgaFig2}.

These results undoubtedly demonstrate the importance of EPI in cuprates \cite
{Dolgov-Shulga}, \cite{Shulga}, \cite{Maksimov-Review}. We point out that
the width of $\alpha _{tr}^{2}(\omega )F(\omega )$ which is extracted from
the optical measurements \cite{Dolgov-Shulga}, \cite{Shulga}, \cite
{Maksimov-Review} coincides with the width of the quasi-particle spectral
function $\alpha ^{2}(\omega )F(\omega )$ obtained in tunneling and ARPES
spectra (which we shall discuss below), i.e. both functions are spread over
the energy interval $0<\omega <\omega _{ph}^{\max }(\lesssim 80$ $meV)$.
Since in cuprates this interval coincides with the width in the phononic
density of states $F(\omega )$ and since the maxima of $\alpha ^{2}(\omega
)F(\omega )$ and $F(\omega )$ almost coincide, this is further strong
evidence for the dominance of EPI.

To this end, we would like to comment on two important points. \textit{First}%
, in some reports \cite{Carbotte},\cite{Hwang-Timusk-1}, \cite
{Hwang-Timusk-2} it was assumed that $\alpha _{tr}^{2}(\omega )F(\omega )$
of cuprates can be extracted also in the superconducting state by using Eq. (%
\ref{spect-func}). However, Eq. (\ref{spect-func}) holds exclusively in the
normal state (at T=0) since $\sigma (\omega )$ can be described by the
generalized (extended) Drude formula in Eq. (\ref{Drude}) only in the normal
state. Such an approach apparently does not hold in the superconducting
state since the dynamical conductivity depends not only on the
electron-boson scattering but also on coherence factors and on the momentum
and energy dependent order parameter $\Delta (\mathbf{k},\omega )$. In such
a case it is unjustified to extract $\alpha _{tr}^{2}(\omega )F(\omega )$
from Eq. (\ref{spect-func}). \textit{Second}, if $R(\omega )$ (and $\sigma
(\omega )$) in cuprates are due to some other bosonic scattering which is
pronounced up to much higher energies $\omega _{c}\gg \omega _{ph}^{\max }$,
this should be seen in the extracted spectral function $\alpha
_{tr}^{2}(\omega )F(\omega )$. Such an assumption is made, for instance, in
the phenomenological spin-fluctuation approach \cite{Pines} where it is
assumed that $\alpha ^{2}(\omega )F(\omega )=g_{sf}^{2}$Im$\chi (\omega )$
where Im$\chi (\omega )$ is extended up to the large energy cutoff $\omega
_{c}\approx 400$ $meV$. This assumption is apparently in conflict with the
above theoretical and experimental analysis which shows that solely EPI can
describe $R(\omega )$ very well and that the contribution from higher
energies $\omega \gg \omega _{ph}^{\max }$ must be small and therefore
irrelevant for pairing \cite{Dolgov-Shulga}, \cite{Shulga}, \cite{Kaufmann}.
This is also confirmed by tunnelling measurements - see below.

Despite the experimentally established facts for the importance of EPI, some
reports appeared recently claiming that SFI dominates and that $\alpha
_{tr}^{2}(\omega )F(\omega )\approx g_{sf}^{2}\func{Im}\chi (\omega )$ where
$\func{Im}\chi (\omega )=\int d^{2}k\chi (\mathbf{k},\omega )$ \cite
{Hwang-Timusk-1}, \cite{Hwang-Timusk-2}. This claim is based on reanalyzing
some old IR measurements \cite{Hwang-Timusk-1}, \cite{Hwang-Timusk-2}. The
transport spectral function $\alpha _{tr}^{2}(\omega )F(\omega )$ is
extracted in \cite{Hwang-Timusk-1} by using the maximum entropy method in
solving the Fredholm equation. However, in order to exclude negative values
in the extracted $\alpha _{tr}^{2}(\omega )F(\omega )$ they imply a biased
condition that $\alpha _{tr}^{2}(\omega )F(\omega )$ has a rather large tail
at large energies - up to 400 meV. Then it is not surprising at all that
their extracted $\alpha _{tr}^{2}(\omega )F(\omega )$ at large $\omega $
resembles qualitatively $\func{Im}\chi (\omega )$ obtained by the magnetic
neutron scattering on $La_{2-x}Sr_{x}CuO_{4}$ \cite{Vignolle}. In other
words one obtains as output what is assumed in the input. It turns out that
even such a biased assumption in \cite{Hwang-Timusk-1} by extracting $\alpha
_{tr}^{2}(\omega )F(\omega )$ does not reproduce the experimental curve $%
\func{Im}\chi (\omega )$ \cite{Vignolle} in some important respects. (%
\textit{1}) The relative heights of the two peaks in the extracted spectral
function $\alpha _{tr}^{2}(\omega )F(\omega )$ at lower temperatures are
opposite to that in $\func{Im}\chi (\omega )$ \cite{Vignolle} - see Fig. 1
in \cite{Hwang-Timusk-1}. (\textit{2}) The strong temperature dependence of
the extracted $\alpha _{tr}^{2}(\omega )F(\omega )$, found in \cite
{Hwang-Timusk-1}, \cite{Hwang-Timusk-2} is in fact not the intrinsic
property of the spectral function but it is due to the high sensitivity of
the extraction procedure on temperature. As we already explained before,
this is due to the ill-posed problem of solving the Fredholm integral
equation of the first kind with strong $T$-dependent kernel. (\textit{3})
The extracted spectral weight $\alpha _{tr}^{2}(\omega )F(\omega )$ in \cite
{Hwang-Timusk-1} has much smaller values at larger frequencies ($\omega >100$
$meV$) than it is the case for the measured Im$\chi (\omega )$, i.e. $%
(I(\omega >100$ $meV)/I(\omega _{\max }))\ll $Im$\chi (\omega >100$ $meV)/$Im%
$\chi (\omega _{\max })$ - see Fig. 1 in \cite{Hwang-Timusk-1}. In spite of
the fact that the main weight of the extracted $\alpha _{tr}^{2}(\omega
)F(\omega )$ \cite{Hwang-Timusk-1} lies in the range of phononic
frequencies, $0<\omega <\omega _{\max }^{ph}$ it is not in agreement with
that obtained in tunnelling and ARPES measurements. (\textit{4}) To this end
it is suspicious that the transport coupling constant $\lambda _{tr}$
extracted\ in \cite{Hwang-Timusk-1} is so large, i.e. $\lambda _{tr}>3$
contrary to the previous findings that $\lambda _{tr}<1.5$ \cite
{Dolgov-Shulga}, \cite{Shulga}, \cite{Kaufmann}. Since in HTSC one has $%
\lambda >\lambda _{tr}$ this would probably give $\lambda \approx 6$ that is
not confirmed by other experiments. It is necessary to stress that the
estimated $\lambda _{tr}$ depends strongly on the value of plasma frequency,
i.e. on $\omega _{pl}^{2}$ - see Fig. \ref{lam-plas} below, and it might be
that for the latter the larger value is assumed in \cite{Hwang-Timusk-1}. (%
\textit{5}) The interpretation of $\alpha _{tr}^{2}(\omega )F(\omega )$ in
LSCO and BISCO solely in terms of Im$\chi (\omega )$ is in contradiction
with the magnetic neutron scattering in the optimally doped and slightly
underdoped YBCO \cite{Bourges} - that was discussed above, where in the
former Im$\chi (\mathbf{Q},\omega )$ is small in the normal state - it is
even below the experimental noise. This means that if the assumption that $%
\alpha _{tr}^{2}(\omega )F(\omega )\approx g_{sf}^{2}$Im$\chi (\omega )$
were correct then the contribution to Im$\chi (\omega )$ from the momenta $%
0<k<Q$ would be dominant and very detrimental for d-wave superconductivity
and T$_{c}$ would be rather low. Since the results of magnetic neutron
scattering in YBCO \cite{Bourges} are very convincing and trustful, then the
conclusion is that the SFI coupling constant $\lambda ^{sf}(\sim g_{sf}^{2})$
must be small, i.e. $g_{sf}<0.2$ $eV$ \ and $\lambda ^{sf}<0.2$. The latter
is in accordance with other independent estimates (discussed also above) of $%
\lambda ^{sf}(\ll 2)$.

Finally, we point out that very similar (to cuprates) properties, of $\sigma
(\omega )$, $R(\omega )$ (and $\rho (T)$ and electronic Raman spectra) were
observed in 3D isotropic metallic oxides $La_{0.5}Sr_{0.5}CoO_{3}$ and $%
Ca_{0.5}Sr_{0.5}RuO_{3}$ which are non-superconducting \cite{Bozovic} and in
$Ba_{1-x}K_{x}BiO_{3}$ which superconducts at $T_{c}\simeq 30$ $K$ at $x=0.4$%
. This means that in all of them, the scattering mechanism might be of
similar origin. Since in these compounds there are no signs of
antiferromagnetic fluctuations (which are present in cuprates), then EPI
plays important role.

3. \textit{Restricted optical sum-rule }

The \textit{restricted} \textit{optical sum-rule }was studied intensively in
HTSC cuprates. It shows peculiarities not present in low-temperature
superconductors. It turns out that the restricted spectral weight $W(\Omega
_{c},T)$ is strongly temperature dependent in the normal and superconducting
state, that was interpreted either to be due to EPI \cite{Maks-Karakoz-1},
\cite{Maks-Karakoz-2} or to some non-phononic mechanisms \cite{Hirsch}. In
the following we demonstrate that the temperature dependence of $W(\Omega
_{c},T)=W(0)-\beta T^{2}$ in the normal state can be explained in a natural
way by the $T$-dependence of the EPI transport relaxation rate $\gamma
_{tr}^{ep}(\omega ,T)$ \cite{Maks-Karakoz-1}, \cite{Maks-Karakoz-2}. Since
the problem of the restricted sum-rule attracted much interest, it will be
considered here in some details. In fact there are two kinds of sum rules
related to $\sigma (\omega )$. The first one is the \textit{total sum rule}
which in the normal state reads

\begin{equation}
\int_{0}^{\infty }\sigma _{1}^{N}(\omega )d\omega =\frac{\omega _{pl}^{2}}{8}%
=\frac{\pi ne^{2}}{2m},  \label{TSR}
\end{equation}
while in the \textit{superconducting state} it is given by the
Tinkham-Ferrell-Glover (TFG) sum-rule

\begin{equation}
\int_{0}^{\infty }\sigma _{1}^{S}(\omega )d\omega =\frac{c^{2}}{8\lambda
_{L}^{2}}+\int_{+0}^{\infty }\sigma _{1}^{S}(\omega )d\omega =\frac{\omega
_{pl}^{2}}{8}.  \label{TSR-SC}
\end{equation}
Here,$\ n$ - the total electron density, $e$ - the electron charge, $m$ -
the bare electron mass and $\lambda _{L}$ - the London penetration depth.
The first (singular) term $c^{2}/8\lambda _{L}^{2}$ is due to the
superconducting condensate which contributes $\sigma _{1,cond}^{S}(\omega
)=(c^{2}/4\lambda _{L}^{2})\delta (\omega )$. The total sum rule represents
the fundamental property of matter - the conservation of the electron
number, and to calculate it one should use the total Hamiltonian $\hat{H}%
_{tot}=\hat{T}_{e}+\hat{H}_{int}$ where all electrons, bands and their
interactions $\hat{H}_{int}$ (Coulomb, EPI, with impurities, etc.) are
accounted for. Here, $T_{e}$ is the \textit{kinetic energy of bare electrons}
\begin{equation}
\hat{T}_{e}=\sum_{\sigma }\int d^{3}x\hat{\psi}_{\sigma }^{\dagger }(x)\frac{%
\mathbf{\hat{p}}^{2}}{2m}\hat{\psi}_{\sigma }(x)=\sum_{\mathbf{p},\sigma }%
\frac{\mathbf{p}^{2}}{2m_{e}}\hat{c}_{\mathbf{p}\sigma }^{\dagger }\hat{c}_{%
\mathbf{p}\sigma }.  \label{Kin-energy}
\end{equation}

The \textit{partial sum rule} is related to the energetics in the \textit{%
conduction (valence) band\ }which is described by the Hamiltonian of the
valence (band) electrons
\begin{equation}
\hat{H}_{v}=\sum_{\mathbf{p},\sigma }\epsilon _{\mathbf{p}}\hat{c}_{v,%
\mathbf{p}\sigma }^{\dagger }\hat{c}_{v,\mathbf{p}\sigma }+\hat{V}_{v,Coul}.
\label{Hval}
\end{equation}
It contains the band-energy with the dispersion $\epsilon _{\mathbf{p}}$ and
the effective Coulomb interaction of the valence electrons $\hat{V}_{v,Coul}$%
. In this case the \textit{partial sum-rule} in the normal state reads \cite
{Maldague} (for general form of $\epsilon _{\mathbf{p}}$)
\begin{equation}
\int_{0}^{\infty }\sigma _{1,v}^{N}(\omega )d\omega =\frac{\pi e^{2}}{2V}%
\sum_{\mathbf{p}}\frac{\langle \hat{n}_{v,\mathbf{p}}\rangle _{H_{v}}}{m_{%
\mathbf{p}}}  \label{Total}
\end{equation}
where the number operator $\hat{n}_{v,\mathbf{p}}=\sum_{\sigma }\hat{c}_{%
\mathbf{p}\sigma }^{\dagger }\hat{c}_{\mathbf{p}\sigma }$; $1/m_{\mathbf{p}%
}=\partial ^{2}\epsilon _{\mathbf{p}}/\partial p_{x}^{2}$ is the reciprocal
mass and $V$ is volume . In practice measurements are performed up to finite
frequency and the integration over $\omega $ goes up to some cutoff
frequency $\Omega _{c}$ (of the order of the band plasma frequency). In this
case the restricted sum-rule has the form

\[
W(\Omega _{c},T)=\int_{0}^{\Omega _{c}}\sigma _{1,v}^{N}(\omega )d\omega
\]
\begin{equation}
=\frac{\pi }{2}\left[ K^{d}+\Pi (0)\right] -\int_{0}^{\Omega _{c}}\frac{%
Im\Pi (\omega )}{\omega }d\omega .  \label{rest-sum}
\end{equation}
where $K^{d}$ is the diamagnetic Kernel and $\Pi (\omega )$ is the
paramagnetic (current-current) response function - see more in \cite
{Maks-Karakoz-1}, \cite{Maks-Karakoz-2}. In the case when the interband gap $%
E_{g}$ is the largest scale in the problem, i.e. when $W_{b}<\Omega
_{c}<E_{g}$, in this region one has approximately Im$\Pi (\omega )\approx 0$
and the limit $\Omega _{c}\rightarrow \infty $ in Eq.(\ref{rest-sum}) is
justified. In that case one has $\Pi (0)\approx \int_{0}^{\Omega _{c}}($Im$%
\Pi (\omega )/\omega )d\omega $ which gives the approximate formula for $%
W(\Omega _{c},T)$
\[
W(\Omega _{c},T)=\int_{0}^{\Omega _{c}}\sigma _{1,v}^{N}(\omega )d\omega
\approx \frac{\pi }{2}K^{d}
\]
\begin{equation}
=e^{2}\pi \sum_{\mathbf{p}}\frac{\partial ^{2}\epsilon _{\mathbf{p}}}{%
\partial \mathbf{p}^{2}}n_{\mathbf{p}},  \label{app-rest-sum}
\end{equation}
where $\epsilon _{\mathbf{p}}$ is the band-energy and $n_{\mathbf{p}%
}=\left\langle \hat{n}_{v,\mathbf{p}}\right\rangle $ is the
quas-iparticle distribution function in the interacting system.
Note that the right hand side of Eq.(\ref{app-rest-sum}) does not
depend on the cutoff energy $\Omega _{c}$. So one should be
careful not to interpret blindly the experimental result in
cuprates by this formula and for that reason the best way is to
calculate $W(\Omega _{c},T)$ by using the exact result in Eq.(\ref{rest-sum}%
) which apparently depends on $\Omega _{c}$. However, Eq.(\ref{app-rest-sum}%
) is useful for appropriately chosen $\Omega _{c}$, since it allows us to
get semi-quantitative and qualitative results. In most papers\ related to
the restricted sum-rule in HTSC, it was assumed, due to simplicity, the
\textit{tight-binding model with nearest neighbors} (n.n.) with the energy $%
\epsilon _{\mathbf{p}}=-2t(\cos p_{x}a+\cos p_{y}a)$ and $1/m_{\mathbf{p}%
}=-2ta^{2}\cos p_{x}a$. It is straightforward to show that in this case one
has
\[
W(\Omega _{c},T)=\int_{0}^{\Omega _{c}}\sigma _{1,v}^{N}(\omega )d\omega
\]
\begin{equation}
\approx \frac{\pi e^{2}a^{2}}{2V}\langle -T_{v}\rangle ,  \label{Partial}
\end{equation}
where $\langle T_{v}\rangle _{H_{v}}=\sum_{\mathbf{p}}\epsilon _{\mathbf{p}%
}\langle n_{v}\rangle _{H_{v}}$ is the averaged kinetic energy of the band
electrons and $\omega _{pl,v}$ is the (band) plasma frequency. In this
approximation $W(\Omega _{c},T)$ is a direct measure of the averaged kinetic
energy. In\textit{\ }the\textit{\ superconducting state} the partial
sum-rule reads
\[
W_{s}(\Omega _{c},T)=\frac{c^{2}}{8\lambda _{L}^{2}}+\int_{+0}^{\Omega
_{c}}\sigma _{1,v}^{S}(\omega )d\omega
\]
\begin{equation}
=\frac{\pi e^{2}a^{2}}{2}\langle -T_{v}\rangle _{s}.  \label{Partial-sc}
\end{equation}
In order to introduce the reader to the complexity of the problem of
T-dependence of $W(\Omega _{c},T)$, let us consider the electronic system in
the normal state and in absence of quasi-particle interaction. In that case
one has $n_{\mathbf{p}}=f_{\mathbf{p}}$ ($f_{\mathbf{p}}$ is the Fermi
distribution function) and $W_{n}(\Omega _{c},T)$ increases with the
decrease temperature, i.e. $W_{n}(\Omega _{c},T)=W_{n}(0)-\beta _{b}T^{2}$
where $\beta _{b}\sim 1/W_{b}$. To this end, let us mention in advance that
the experimental value $\beta _{\exp }$ is much larger than $\beta _{b}$,
i.e. $\beta _{\exp }\gg \beta _{b}$ thus telling us that the simple
Sommerfeld-like smearing of $f_{\mathbf{p}}$ by the temperature effects
cannot explain the T-dependence of $W(\Omega _{c},T)$ as it was put forward
in some papers. We stress that the smearing of $f_{\mathbf{p}}$ by
temperature lowers the spectral weight compared to that at $T=0$ $K$, i.e. $%
W_{n}(\Omega _{c},T)<W_{n}(\Omega _{c},0)$. In that respect it is not
surprising at all that there is a lowering of $W_{s}(\Omega _{c},T)$ in the
BCS superconducting state, $W_{s}^{BCS}(\Omega _{c},T=0)<W_{n}(\Omega
_{c},0) $ since $f_{\mathbf{p}}$ is smeared due to the appearance of the
superconducting gap, $2f_{\mathbf{p}}=1-(\xi _{\mathbf{p}}/E_{\mathbf{p}%
})th(E_{\mathbf{p}}/2T)$, $E_{\mathbf{p}}=\sqrt{\xi _{\mathbf{p}}^{2}+\Delta
^{2}}$, $\xi _{\mathbf{p}}=\epsilon _{\mathbf{p}}-\mu $, and the maximal
decrease of $W_{s}(\Omega _{c},T)$ is at $T=0$.

\begin{figure}[tbp]
\resizebox{.5\textwidth}{!}
{\includegraphics*[width=8cm]{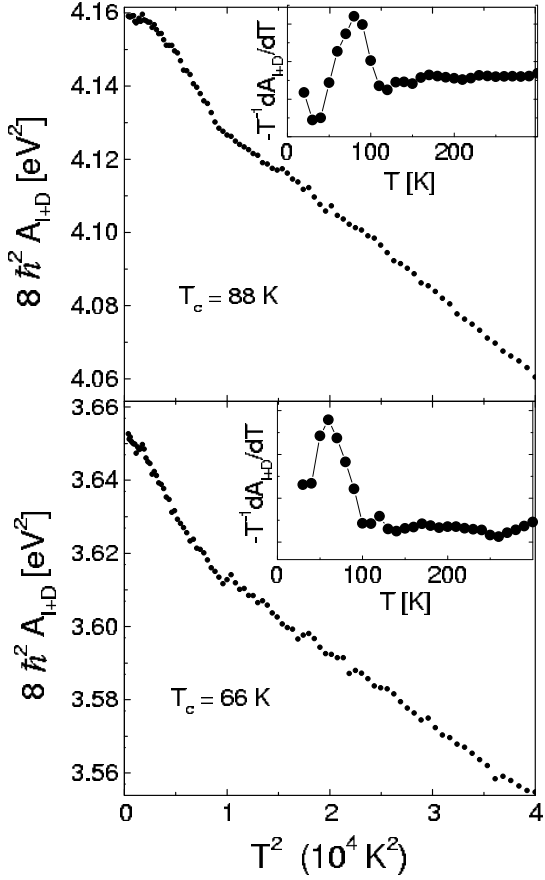}}
\caption{Measured spectral weight $W_{s}(\Omega _{c},T)$ for $\protect\omega %
_{c} \approx 1.25 eV$ in two underdoped $Bi2212$ (with $T_{c}=88$ $K$ and $%
T_{c}=66$ $K$). From \protect\cite{Molegraaf}.}
\label{Weight-NonBCS}
\end{figure}

Let us enumerate and analyze the \textit{main experimental results in
cuprates}. \textbf{1}. In the \textit{normal state} ($T>T_{c}$) of most
cuprates, one has $W_{n}(\Omega _{c},T)=W_{n}(0)-\beta _{ex}T^{2}$ with $%
\beta _{\exp }\gg \beta _{b}$, i.e. $W_{n}(\Omega _{c},T)$ is increasing by
decreasing $T$, even at T below the opening of the pseudogap. The change of $%
W_{n}(\Omega _{c},T)$ from room temperature down to $T_{c}$ is no more than $%
5$ $\%$.\ \textbf{2}. In the \textit{superconducting state} ($T<T_{c}$) of
some underdoped and optimally doped Bi-2212 compounds \cite{Molegraaf}, \cite
{Carbone} (and underdoped Bi-2212 films \cite{Santander-2003}) there is
\textit{an effective increase} of $W_{s}(\Omega _{c},T)$ with respect to
that in the normal state, i.e. $W_{s}(\Omega _{c},T)>W_{n}(\Omega _{c},T)$
for $T<T_{c}$. This is non-BCS behavior shown in Fig. ~\ref{Weight-NonBCS}.

\begin{figure}[tbp]
\resizebox{.5\textwidth}{!}
{\includegraphics*[width=8cm]{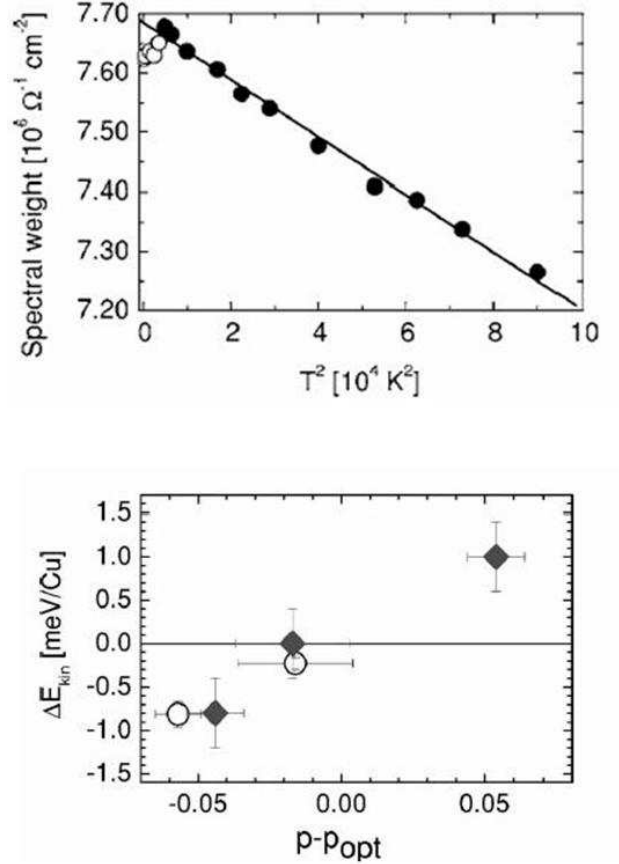}}
\caption{(Top) Spectral weight $W_{n}(\Omega _{c},T)$ of the overdoped $%
Bi2212 $ for $\Omega _{c}=1 eV$. Closed symbols - normal state. Open symbols
- superconducting state. (Bottom) Change of the kinetic energy $\Delta
E_{kin}=E_{kin,S}-E_{kin,N}$ in $meV$ per Cu site vs the charge $p$ per Cu
with respect to the optimal value $p_{opt}$. From \protect\cite
{Deutscher-optics}.}
\label{Weight-BCS}
\end{figure}

In some optimally doped and in most overdoped cuprates, there is decreasing
of $W_{s}(\Omega _{c},T)$ at $T<T_{c}$ which is the BCS-like behavior \cite
{Deutscher-optics} as it is seen in Fig.~\ref{Weight-BCS}

We stress that the non-BCS behavior of $W_{s}(\Omega _{c},T)$ for underdoped
and optimally doped systems was obtained by assuming that $\Omega
_{c}\approx (1-1.2)$ $eV$. However, in Ref. \cite{BorisMPI} these results
have been questioned and the conventional BCS-like behavior was observed ($%
W_{s}(\Omega _{c},T)<W_{n}(\Omega _{c},T)$) in the optimally doped YBCO and
slightly underdoped Bi-2212 by using larger cutoff energy $\Omega _{c}=1.5$ $%
eV$. Although the results obtained in \cite{BorisMPI} looks very trustfully,
it is fair to say that the issue of the reduced spectral weight in the
superconducting state of cuprates is still unsettled and under dispute. In
overdoped Bi-2212 films, the BCS-like behavior $W_{s}(\Omega
_{c},T)<W_{n}(\Omega _{c},T)$ was observed, while in LSCO it was found that $%
W_{s}(\Omega _{c},T)\approx const$, i.e. $W_{s}(\Omega _{c},T<T_{c})\approx
W_{n}(\Omega _{c},T_{c})$.

\begin{figure}[tbp]
\resizebox{.4 \textwidth}{!}
{\includegraphics*[width=8cm]{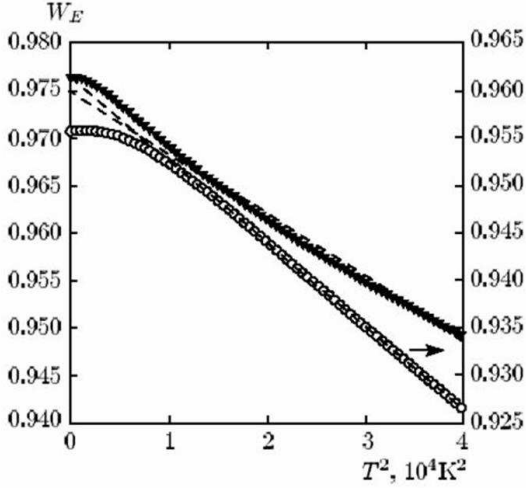}}
\caption{Spectral weight $W(\Omega _{c},T)$ for Einstein phonons with $%
\Omega _{E}=200$ $K$ (full triangles) and $\Omega _{E}=400$ $K$ (open
circles, left axis). Dashed lines is $T^{2}$ asymptotic. From \protect\cite
{Maks-Karakoz-2}.}
\label{MaksKarakoz4}
\end{figure}

How to explain the strong temperature dependence of $W(\Omega _{c},T)$ in
the normal and superconducting state? In \cite{Maks-Karakoz-1}, \cite
{Maks-Karakoz-2} it was shown that the EPI relaxation $\gamma ^{ep}(T)$
plays the main role in the $T$-dependence of $W(\Omega _{c},T)$. The main
theoretical results of \cite{Maks-Karakoz-1}, \cite{Maks-Karakoz-2} are the
following. (1) The calculations based on the exact formula in Eq.(\ref
{app-rest-sum}) give that for $\Omega _{c}\gg \Omega _{D}$, the difference
in spectral weights of the normal and superconducting state is small, i.e. $%
W_{n}(\Omega _{c},T)\approx W_{s}(\Omega _{c},T)$ (in the following we call
it $W$) since $W_{n}(\Omega _{c},T)-W_{s}(\Omega _{c},T)\sim \Delta
^{2}/\Omega _{c}^{2}$. In the case of large $\Omega _{c}$ based on the
approximate formula Eq.(\ref{app-rest-sum}), one obtains
\begin{equation}
W(\Omega _{c},T)\approx \frac{\omega _{pl}^{2}}{8}\left[ 1-\frac{\gamma (T)}{%
W_{b}}-\frac{\pi ^{2}}{2}\frac{T^{2}}{W_{b}^{2}}\right] .  \label{Wapp}
\end{equation}
In the case of EPI, one has $\gamma =\gamma ^{ep}(T)+\gamma ^{imp}$ where $%
\gamma ^{ep}(T)=\int_{0}^{\infty }dz\alpha ^{2}(z)F(z)\coth (z/2T)$. It
turns out that for $\alpha ^{2}(\omega )F(\omega )$ shown in Fig.~\ref{Rates}%
, one obtains: (i) $\gamma ^{ep}(T)\sim T^{2}$ in the temperature interval $%
100$ $K<T<200$ $K$ as it is seEn in Fig.~\ref{MaksKarakoz4}, \cite
{Maks-Karakoz-1}, \cite{Maks-Karakoz-2}; (ii) the second term in Eq.(\ref
{Wapp}) is much larger than the last one (the Sommerfeld-like term). For the
EPI coupling constant $\lambda _{tr}^{ep}=1.5$ one obtains rather good
agreement with experiments. At lower temperatures, $\gamma ^{ep}(T)$
deviates from the $T^{2}$ behavior and the deviation depends on the
structure of the spectrum in $\alpha ^{2}(\omega )F(\omega )$. It is seen in
Fig.~\ref{MaksKarakoz4} that for a softer Einstein spectrum (with $\Omega
_{E}=200$ $K$), $W(\Omega _{c},T)$ lies above the curve with the $T^{2}$
asymptotic, while the one with a harder phononic spectrum (with $\Omega
_{E}=400$ $K$) lies below it.
This result means that different behavior of $W(\Omega _{c},T)$ in the
superconducting state of cuprates for different doping might be simply
related to different contributions of low and high frequency phonons. We
stress that such a behavior of $W(\Omega _{c},T)$ was observed in
experiments \cite{Molegraaf}, \cite{Carbone}, \cite{BorisMPI} and the above
analysis tells us that the theory based on EPI explains in a consistent way
the strange temperature behavior of $W(\Omega _{c},T)$ above and below $%
T_{c} $ and that there is no need to invoke exotic scattering mechanisms.%
\newline

4. \textit{Resistivity }$\rho (T)$

\begin{figure}[tbp]
\resizebox{.4\textwidth}{!}
{\includegraphics*[width=7cm]{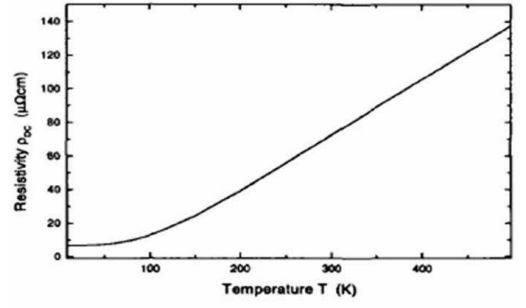}}
{\includegraphics*[width=8cm]{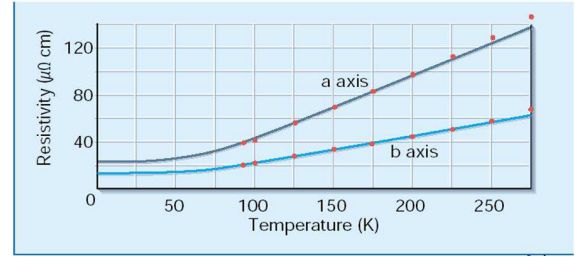}}
\caption{(a) Calculated resistivity $\protect\rho (T)$ for the EPI spectral
function $\protect\alpha _{tr}^{2}(\protect\omega )F(\protect\omega )$ in
\protect\cite{KMS}. (b) Measured resistivity in a(x)- and b(y)-crystal
direction of YBCO \protect\cite{Friedman} and calculated Bloch-Gr\"{u}neisen
curve for $\protect\lambda ^{ep}=1$, \protect\cite{Allen-kinky}.}
\label{Res-T}
\end{figure}

The temperature dependence of the in-plane resistivity $\rho _{ab}(T)$ in
cuprates is a direct consequence of the quasi-$2D$ motion of quasi-particles
and of the inelastic scattering which they experience. At present, there is
no consensus on the origin of the linear temperature dependence of the
in-plane resistivity $\rho _{ab}(T)$ in the normal state and there is rather
widespread believe that it can not be due to EPI. The inadequacy of this
belief was already demonstrated by analyzing the dynamic conductivity $%
\sigma (\omega )$ which is successfully explained by EPI. Since $\rho
(T)=1/\sigma (\omega =0)$
\begin{equation}
\rho (T)=\frac{4\pi }{\omega _{p}^{2}}\gamma _{tr}(T)+\rho _{imp}
\label{inv-sig}
\end{equation}
\begin{equation}
\gamma _{tr}(T)=\frac{\pi }{T}\int_{0}^{\infty }d\omega \frac{\omega }{\sin
^{2}(\omega /2T)}\alpha _{tr}^{2}(\omega )F(\omega ).  \label{g-tr-T}
\end{equation}
It is quite natural that in some temperature region, $\rho (T)$ in cuprates
can be explained by EPI as it is shown in Fig. \ref{Res-T}. It turns out
that $\gamma _{tr}(T)\sim T$ for $T>\alpha \Theta _{D}$, $\alpha <1$
depending on the shape of $\alpha _{tr}^{2}(\omega )F(\omega )$. In case of
the Debye spectrum, it is realized for $T>\Theta _{D}/5$ i.e.
\begin{equation}
\rho (T)\simeq 8\pi ^{2}\lambda _{tr}^{ep}\frac{k_{B}T}{\hbar \omega _{p}^{2}%
}=\rho ^{\prime }T.  \label{rho}
\end{equation}

There is an experimental constraint on $\lambda _{tr}$, i.e.
\begin{equation}
\lambda _{tr}\approx 0.25\omega _{pl}^{2}(eV)\rho ^{\prime }(\mu \Omega
\mathrm{cm}/K),  \label{lambdatr}
\end{equation}
which imposes a limit on it. For instance, for $\omega _{pl}\approx (2-3)$ $%
eV$ \cite{Bozovic} and $\rho ^{\prime }\approx 0.6$ in the oriented YBCO
films and $\rho ^{\prime }\approx 0.3-0.4$ in single crystals of BSCO, one
obtains $\lambda _{tr}\approx 0.4-1.2$. In case of YBCO single crystals,
there is a pronounced anisotropy in $\rho _{a,b}(T)$ \cite{Friedman} which
gives $\rho _{x}^{\prime }(T)=0.6\mu \Omega \mathrm{cm}/K$ and $\rho
_{y}^{\prime }(T)=0.25\mu \Omega \mathrm{cm}/K$. According to Eq.(\ref
{lambdatr}), one obtains $\lambda _{tr}(\omega _{pl})$ which is shown in Fig.%
\ref{lam-plas}, where the plasma frequency $\omega _{pl}$ which enters Eqs.(%
\ref{inv-sig}-\ref{rho}) can be calculated by LDA and also extracted from
the width ($\sim $ $\omega _{pl}^{\ast })$ of the Drude peak at small
frequencies, where $\omega _{pl}=\sqrt{\varepsilon _{\infty }}\omega
_{pl}^{\ast }$.

\begin{figure}[tbp]
\resizebox{.5\textwidth}{!}
{\includegraphics*[width=7cm]{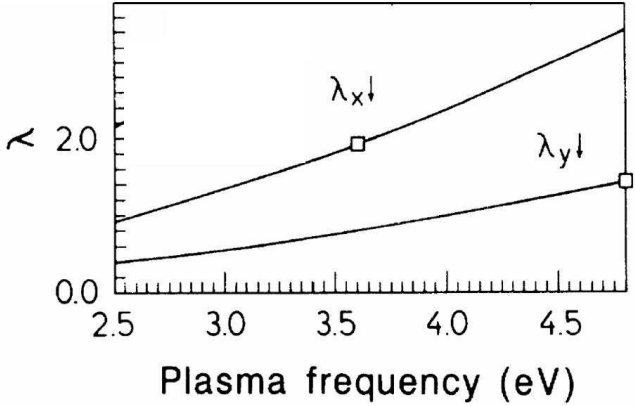}}
\caption{Transport EPI spectral function coupling constant in YBCO as a
function of plasma frequency $\protect\omega _{p}$ as derived from the
experimental slope of resistivity $\protect\rho ^{\prime }(T)$ in Eq.(\ref
{lambdatr}). $\protect\lambda _{x}$ for $\protect\rho _{x}^{\prime }(T)=0.6%
\protect\mu \Omega \mathrm{cm}/K$ and $\protect\lambda _{y}$ for $\protect%
\rho _{y}^{\prime }(T)=0.25\protect\mu \Omega \mathrm{cm}/K$ \protect\cite
{Friedman}. Squares are LDA values \protect\cite{Mazin-Dolgov}.}
\label{lam-plas}
\end{figure}
We shall argue below that from tunnelling experiments \cite
{Tunneling-Vedeneev}-\cite{Tsuda} one obtains in the framework of the
Eliashberg theory that the EPI coupling constant is large $\lambda
^{ep}\approx 2-3$ which implies that $\lambda _{tr}\sim (\lambda /3)$, i.e.
EPI is reduced in transport properties due to some reasons that shall be
discussed in \textit{Part II}. Such a large reduction of $\lambda _{tr}$
cannot be obtained within the LDA band structure calculations which means
that $\lambda ^{ep}$ and $\lambda _{tr}$ contain renormalization which do
not enter in the $LDA$ theory. In \textit{Part II} we shall argue that the
strong suppression of $\lambda _{tr}$ may have its origin in strong
electronic correlations and the long-range Madelung energy \cite{Kulic1},
\cite{Kulic2}.

4. \textit{Femtosecond time-resolved optical spectroscopy}

The \textit{femtosecond time-resolved optical spectroscopy }(FTROS) has been
developed intensively in the last couple of years and applied successfully
to HTSC cuprates. In this method a femtosecond ($1fs=10^{-15}\sec $) laser
pump excites in materials electron-hole pairs via interband transitions.
These hot carriers release their energy via electron-electron (with the
relaxation time $\tau _{ee}$) and electron-phonon scattering reaching states
near the Fermi energy within $10-100$ $fs$ - see \cite{Mihailovic-Kabanov}.
The typical energy density of the laser pump pulses with the wavelength $%
\lambda \approx 810$ nm ($1.5$ $eV$) was around $F\sim 1\mu J/cm^{2}$ (the
\textit{excitation fluenc} $F$) which produces approximately $3\times
10^{10} $ carriers per puls (by assuming that each photon produces $\hbar
\omega /\Delta $ carriers, $\Delta $ is the superconducting gap). By
measuring photoinduced changes of the reflectivity in time, i.e. $\Delta
R(t)/R_{0}$, one can extract information on the further relaxation dynamics
of the low-laying electronic excitations. Since $\Delta R(t)$ relax to
equilibrium the fit with exponential functions is used
\begin{equation}
\frac{\Delta R(t)}{R_{0}}=f(t)\left[ Ae^{-\frac{t}{\tau _{A}}}+Be^{-\frac{t}{%
\tau _{B}}}+...\right] ,  \label{ftr}
\end{equation}
where $f(t)=H(t)[1-\exp \{-t/\tau _{ee}\}]$ ($H(t)$ is the Heavyside
function) describes the finite rise-time. The parameters $A$, $B$ depends on
the fluenc $F$. This method was used in studying the superconucting phase of
$La_{2-x}Sr_{x}CuO_{4}$, with $x=0.1$, $0.15$ and $T_{c}=30$ $K$ and $38$ $K$
respectively \cite{Kusar-2008}. In that case the signal $A\neq 0$ for $T$%
\TEXTsymbol{<}$T_{c}$ and $A=0$ for $T>T_{c}$, while the signal $B$ was
present also at $T>T_{c}$. It turns out that the signal $A$ is related to
the quasi-particle recombination across the superconducting gap $\Delta (T)$
and has a relaxation time of the order $\tau _{A}>10$ $ps$ at $T=4.5$ $K$.
At the so called threshold fluenc ($F_{T}=4.2\pm 1.7$ $\mu J/cm^{2}$ for $%
x=0.1$ and $F_{T}=5.8\pm 2.3$ $\mu J/cm^{2}$ for $x=0.15$) occurs the
vaporization (destroying) of the superconducting phase, where the parameter $%
A$ saturates. This vaporization process takes place at the time scala $\tau
_{r}\approx 0.8$ $ps$. The external fluenc is distributed in the sample over
the \textit{excitation volume} which is proportional to the optical
penetration depth $\lambda _{op}$($\approx 150$ $nm$ at $\lambda \approx 810$
$nm$) of the pump. The energy densities stored in the excitation volume at
the vaporization threshold for $x=0.1$ and $x=0.15$ are $U_{p}=F_{T}/\lambda
_{op}=2.0\pm 0.8$ $K/Cu$ and $2.6\pm 1.0$ $K/Cu$, respectively. The
important fact is that $U_{p}$ is much larger than the superconducting
condensation energy which is $U_{cond}\approx 0.12$ $K/Cu$ for $x=0.1$ and $%
U_{cond}\approx 0.3$ $K/Cu$ for $x=0.15$, i.e. $U_{p}\gg U_{cond}$. This
means that the energy difference $U_{p}-U_{cond}$ must be stored elsewhere
on the time scale $\tau _{r}$. The only present reservoir which can absorb
the difference in energy are the bosonic baths of phonons and spin
fluctuations. The energy required to heat the spin reservoir from $T=4.5K$
to $T_{c}$ is $U_{sf}=\int_{T}^{T_{c}}C_{sf}(T)dT$. The measured $C_{sf}(T)$
in $La_{2}CuO_{4}$ \cite{Kusar-2008} gives very small value $U_{sf}\approx
0.01$ $K$. In the case of the phonon reservoir on obtains $%
U_{ph}=\int_{T}^{T_{c}}C_{ph}(T)dT=9$ $K/Cu$ for $x=0.1$ and $28$
$K/Cu$ for $x=0.15$. Since $U_{sf}\ll U_{p}-U_{cond}$ the spin
reservoir cannot absorb the rest energy $U_{p}-U_{cond}$. The
situation is opposite with phonons since $U_{ph}\gg
U_{p}-U_{cond}$ \ and phonon can absorb the rest energy in the
excitation volume. The complete vaporization dynamics can be
described in the framework of the Rothwarf-Taylor model which
describes approaching of electrons and phonons to
quasi-equilibrium on the time scale of 1 ps \cite {Kabanov-PRL}.
We shall not go into details but only summarize, that only
phonon-mediated vaporization is consistent with the experiments,
thus ruling out spin-mediated quasi-particle recombination and
pairing in HTSC cuprates. This is additional proof for the
ineffectivness of the SFI scattering in cuprates.

\textit{In conclusion}, optics and resistivity measurements in normal state
of cuprates are much more in favor of EPI than against it. However, some
intriguing questions still remain to be answered: (\textbf{i}) what are the
values of $\lambda _{tr}$ and $\omega _{pl}$; (\textbf{ii}) what is the
reason that $\lambda _{tr}\ll \lambda $ is realized in cuprates; (\textbf{iii%
}) what is the role of Coulomb scattering in $\sigma (\omega )$ and $\rho
(T) $. Later on we shall argue that ARPES measurements in cuprates give
evidence for a contribution of Coulomb scattering at higher frequencies,
where $\gamma (\omega )\approx \gamma _{0}+\lambda _{c}\omega $ for $\omega
>\omega _{\max }^{ph}$ with $\lambda _{c}\approx 0.4$. So, despite the fact
that EPI is suppressed in transport properties it can be sufficiently strong
in the self-energy in some frequency and temperature range.

\subsection{ARPES and the EPI self-energy}

ARPES is nowadays a leading spectroscopy method in the solid state physics
\cite{Shen-review}. It provides direct information to the one-electron
removal spectrum in a complex many system. The method involves shining light
(photons) with energies between $5-1000$ $eV$ on the sample and by detecting
momentum ($\mathbf{k}$) - and energy($\omega $)-distribution of the outgoing
electrons. The resolution of ARPES has been significantly increased in the
last decade with the energy resolution of $\Delta E\approx 1-2$ $meV$ (for
photon energies $\sim 20$ $eV$) and angular resolution of $\Delta \theta
\lesssim 0.2{{}^{\circ }}$. The ARPES method is surface sensitive technique,
since the average escape depth ($l_{esc}$)\ of the outgoing electrons is of
the order of $l_{esc}\sim 10$ \AA , depending on the energy of incoming
photons. Therefore, very good surfaces are needed in order that the results
be representative for bulk samples. The most reliable studies were done on
the bilayer $Bi_{2}Sr_{2}CaCu_{2}O_{8}$ ($Bi2212$) and its single layer
counterpart $Bi_{2}Sr_{2}CuO_{6}$ ($Bi2201$), since these materials contain
weakly coupled $BiO$ planes with the longest inter-plane separation in the
cuprates. This results in a \textit{natural cleavage} plane making these
materials superior to others in ARPES experiments. After a drastic
improvement of sample quality in other families of HTSC materials, the ARPES
technique has became a central method in theoretical considerations.
Potentially, it gives valuable information on the quasi-particle Green's
function, i.e. on the quasi-particle spectrum and life-time effects. The
ARPES can indirectly give information on the momentum and energy dependence
of the pairing potential. Furthermore, the electronic spectrum of the (above
mentioned) cuprates is highly \textit{quasi-2D} which allows an unambiguous
determination of the initial state momentum from the measured final state
momentum, since the component parallel to the surface is conserved in
photoemission. In this case, the ARPES probes (under some favorable
conditions) directly the single particle spectral function $A(\mathbf{k}%
,\omega )$. In the following we discuss only those ARPES experiments which
give evidence for the importance of the EPI in cuprates - see more in \cite
{Shen-review}.

The \textit{photoemission} measures a nonlinear response function of the
electron system, and under some conditions it is analyzed in the so-called
\textit{three-step model}, where the total photoemission intensity $I_{tot}(%
\mathbf{k},\omega )\approx I\cdot I_{2}\cdot I_{3}$ is the product of three
independent terms: (\textbf{1}) $I$ - describes optical excitation of the
electron in the bulk; (\textbf{2}) $I_{2}$ - describes the scattering
probability of the travelling electrons; (\textbf{2}) $I_{3}$ - the
transmission probability through the surface potential barrier. The central
quantity in the three-step model is $I(\mathbf{k},\omega )$ and it turns out
that it can be written in the form (for $\mathbf{k=k}_{\parallel }$) \cite
{Shen-review} $I(\mathbf{k},\omega )\simeq I_{0}(\mathbf{k},\upsilon
)f(\omega )A(\mathbf{k},\omega )$ with $I_{0}(\mathbf{k},\upsilon )\sim \mid
\langle \psi _{f}\mid \mathbf{pA\mid }\psi _{i}\rangle \mid ^{2}$ and the
quasi-particle spectral function $A(\mathbf{k},\omega )=-$\textrm{Im}$G(%
\mathbf{k},\omega )/\pi $%
\begin{equation}
A(\mathbf{k},\omega )=-\frac{1}{\pi }\frac{Im\Sigma (\mathbf{k},\omega )}{%
[\omega -\xi (\mathbf{k})-\func{Re}\Sigma (\mathbf{k},\omega )]^{2}+Im\Sigma
^{2}(\mathbf{k},\omega )}.  \label{A}
\end{equation}
Here, $\langle \psi _{f}\mid \mathbf{pA\mid }\psi _{i}\rangle $ is the
dipole matrix element which depends on $\mathbf{k}$, polarization and energy
$\upsilon $ of the incoming photons. The knowledge of the matrix element is
of a great importance and its calculation from first principles was done
carefully in \cite{Bansil}. $f(\omega )$ is the Fermi function, $G$ and $%
\Sigma =\func{Re}\Sigma +i\func{Im}\Sigma $ are the quasi-particle Green's
function and the self-energy, respectively.

We summarize and comment here on some important ARPES results
which were obtained recently and which confirm the existence of
the Fermi surface and importance of EPI in quasi-particle
scattering \cite{Shen-review}.

\textit{ARPES in the normal state }

\begin{figure}[tbp]
\resizebox{.5\textwidth}{!} {
\includegraphics*[width=6cm]{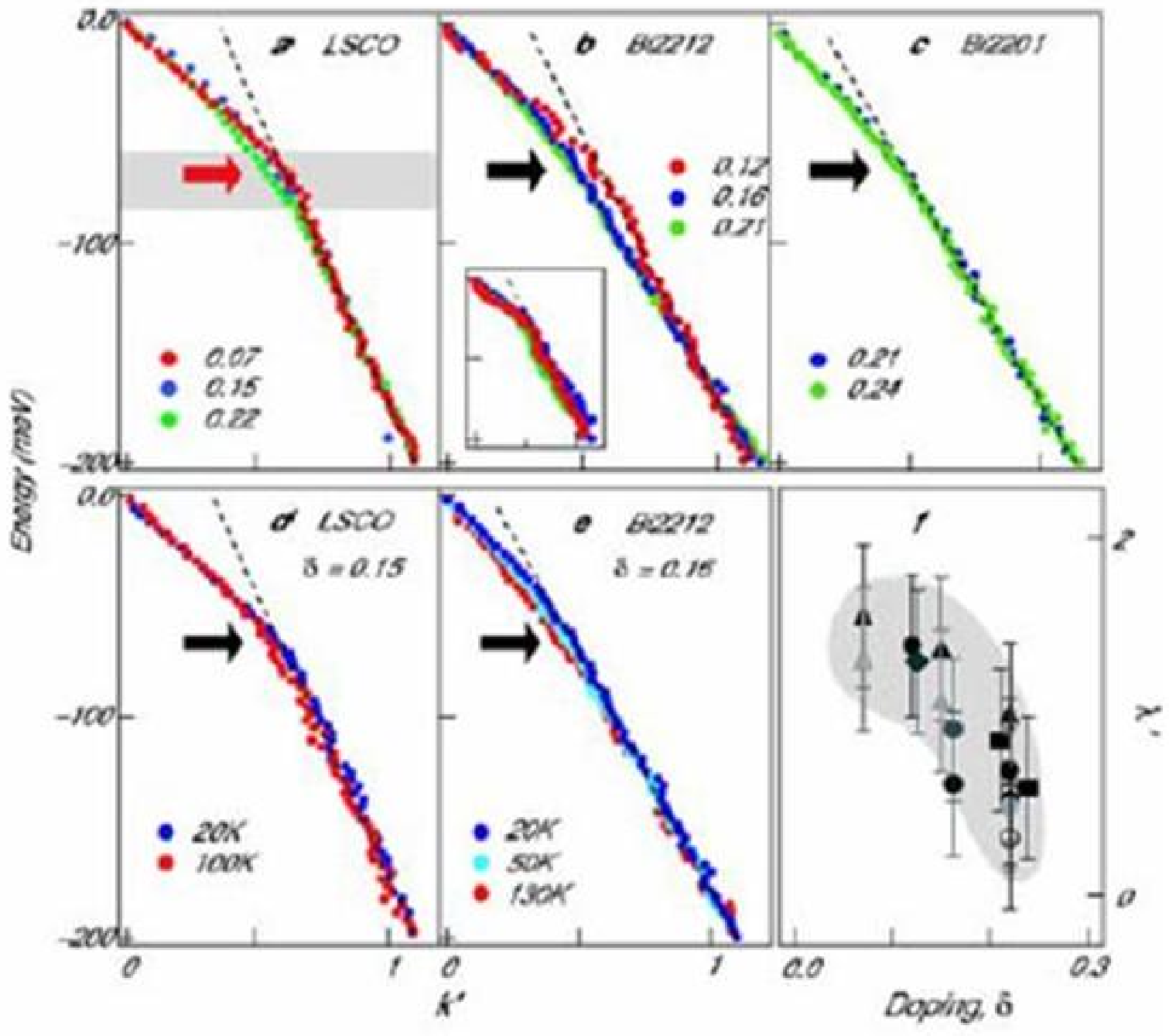}}
{\includegraphics*[width=9cm]{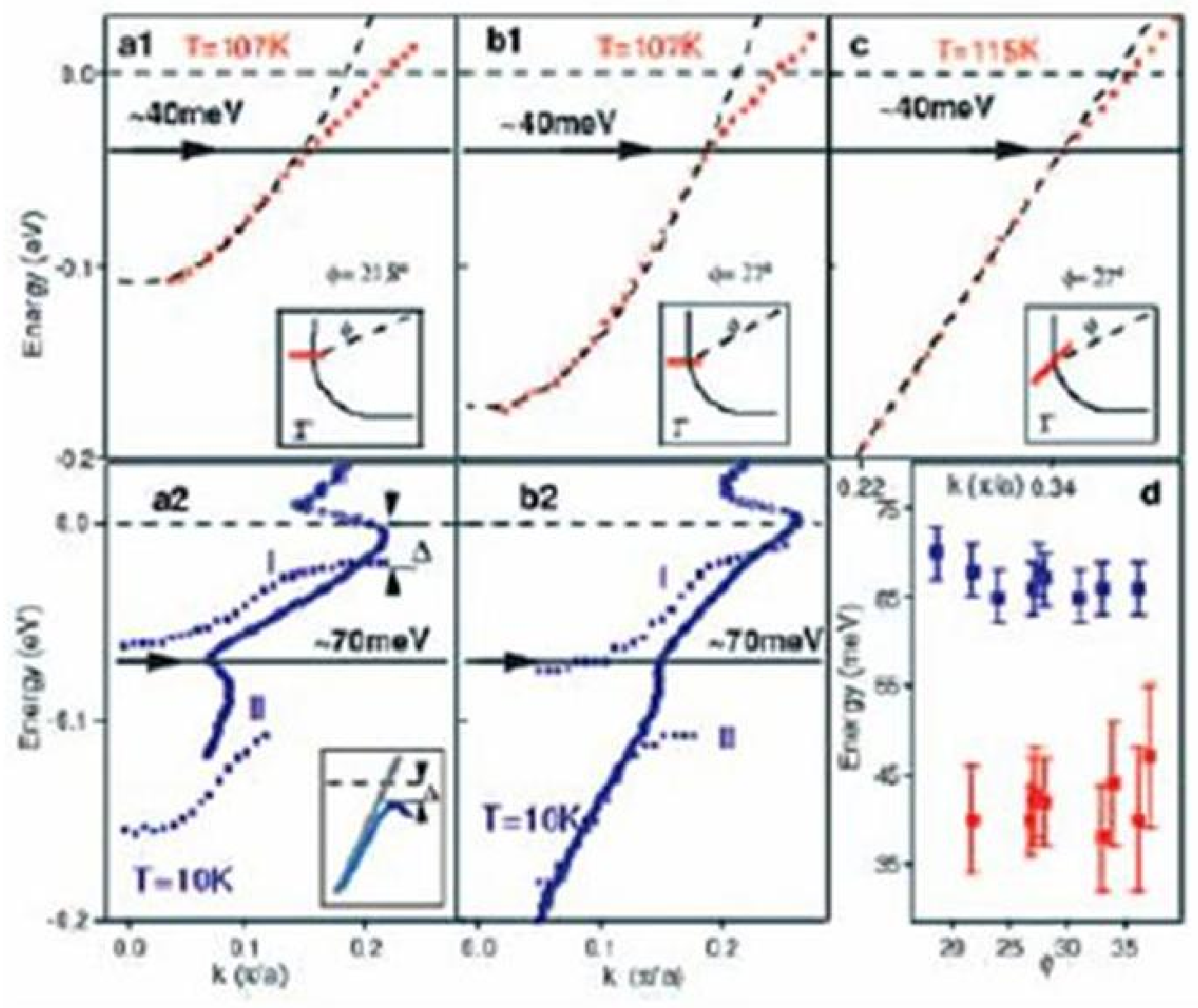}}
\caption{(Top) Quasi-particle dispersion of $Bi2212$, $Bi2201$ and $LSCO$
along the \textit{nodal} direction, plotted vs the momentum $k$ for $(a)-(c)$
different doping, and $(d)-(e)$ different $T$; black arrows indicate the
kink energy; the red arrow indicates the energy of the $q=(\protect\pi ,0)$
oxygen stretching phonon mode; inset of $(e)$- T-dependent $\Sigma ^{\prime
} $ for optimally doped $Bi2212$; $(f)$ - doping dependence of the effective
coupling constant $\protect\lambda ^{\prime }$ along $(0,0)-(\protect\pi ,%
\protect\pi )$ for the different HTSC oxides. From Ref. \protect\cite
{Lanzara}. ( Bottom) Quasi-particle dispersion $E(k)$ in the normal state
(a1, b1, c), at 107 K and 115 K, along various directions $\protect\phi $
around the \textit{anti-nodal} point. The kink at $E=40meV$ is shown by the
horizontal arrow. (a2 and b2) is $E(k)$ in the superconducting state at 10 K
with the shifted kink to $70meV$. (d) kink positions as a function of $%
\protect\phi $ in the anti-nodal region. From Ref. \protect\cite{Cuk}.}
\label{ARPESLanzFig}
\end{figure}

($N1$) There is a well defined Fermi surface in the metallic state with the
topology predicted by the LDA. However, the bands are narrower than LDA
predicts which points to a strong quasi-particle renormalization. ($N2$) The
spectral lines are broad with $\mid $Im$\Sigma (\mathbf{k},\omega )\mid \sim
\omega $ (or $\sim T$ for $T>\omega $) which tells us that the
quasi-particle liquid is a non-canonical Fermi liquid. ($N3$) There is a
bilayer band splitting in $Bi2212$ (at least in the over-doped state). The
previous experiments did not show this splitting and served for various
speculations on some exotic non-Fermi liquid scenarios. ($N4$) At
temperatures $T_{c}<T<T^{\ast }$ and in the under-doped cuprates there is a
d-wave like pseudogap $\Delta _{pg}(\mathbf{k})\sim \Delta _{pg,0}(\cos
k_{x}-\cos k_{y})$ in the quasi-particle spectrum where $\Delta _{pg,0}$
increases by lowering doping. We stress that the pseudogap phenomenon is not
well understood at present and we shall discuss this problem in Part II. Its
origin can be due to a precursor superconductivity or due to a competing
order, such as spin- or charge-density wave or something similar. ($N5$) The
ARPES self-energy gives clear evidence that EPI interaction is rather
strong. For instance, at $T>T_{c}$ there are \textit{kinks} in the
quasi-particle dispersion $\omega (\xi _{\mathbf{k}})$ in the \textit{nodal}
direction (along the $(0,0)-(\pi ,\pi )$ line) at the characteristic phonon
energy $\omega _{ph}^{(70)}\sim (60-70)$ $meV$ \cite{Lanzara}, see Fig.~\ref
{ARPESLanzFig}, and near the \textit{anti-nodal point} $(\pi ,0)$ at $40$ $%
meV$ \cite{Cuk} - see Fig.~\ref{ARPESLanzFig}.

\begin{figure}[tbp]
\resizebox{.5\textwidth}{!}
{\includegraphics*[width=8cm]{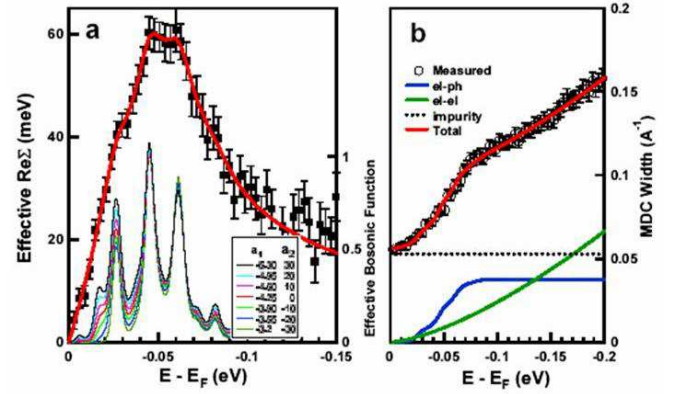}}
\caption{(a) Effective real self-energy for the non-superconducting $%
La_{2-x}Sr_{x}CuO_{4}$, $x=0.03$. Extracted $\protect\alpha _{eff}^{2}(%
\protect\omega )F(\protect\omega )$ is in the inset. (b) Top: the total MDC
width - open circles. Bottom: the EPI contribution shows saturation,
impurity contribution - dotted black line. The residual part is growing $%
\sim \protect\omega ^{1.3}$. From \protect\cite{Zhou-PRL}.}
\label{Phonons-Sigma}
\end{figure}

That these kinks exist also above $T_{c}$ excludes the scenario with the
magnetic resonance peak in $Im\chi _{s}(\mathbf{Q},\omega )$. Since the
magnetic neutron scattering give small SFI coupling constant $\lambda
^{sf}<0.3$. The kinks cannot be due to SFI as we already discussed above. ($%
N6$) The position of the nodal kink is practically doping independent which
points towards phonons as the scattering (gluing) boson. ($N7$) The
quasi-particles (holes) couple practically to the whole spectrum of phonons
since at least three group of phonons were extracted from the ARPES
effective self-energy in $La_{2-x}Sr_{x}CuO_{4}$ \cite{Zhou-PRL} - Fig.~\ref
{Phonons-Sigma}.

\begin{figure}[tbp]
\begin{center}
\resizebox{.5 \textwidth}{!} {
\includegraphics*[width=8cm]{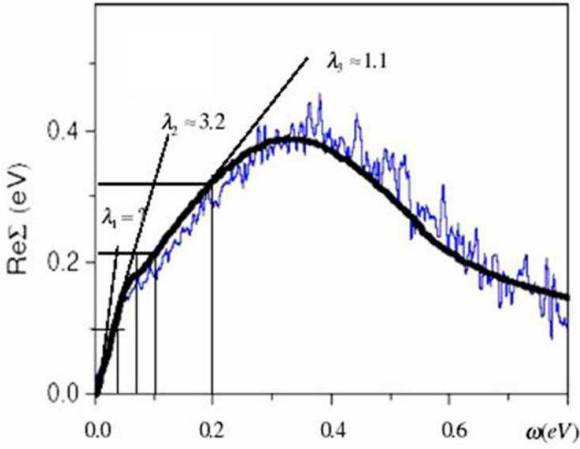}}
\end{center}
\caption{ Fig.4b from \protect\cite{Valla-2006}: $Re\Sigma (\protect\omega )
$ measured in Bi2212 (thin line) and model $Re\Sigma (\protect\omega )$
(bold line) obtained in \protect\cite{Valla-2006}. The three thin lines ($%
\protect\lambda _{1},\protect\lambda _{2},\protect\lambda _{3}$) are the
slopes of $Re\Sigma (\protect\omega )$ in different energy regions - see the
text.}
\label{Valla-Sigma}
\end{figure}

This result is in a qualitative agreement with numerous tunnelling
measurements \cite{Tunneling-Vedeneev}-\cite{Tsuda} which apparently
demonstrate that the broad spectrum of phonons couples with holes without
preferring any particular phonons - see discussion below. ($N8$) Recent
ARPES measurements in B2212 \cite{Valla} show very different slope $d\omega
/d\xi _{\mathbf{k}}$ of the quasi-particle energy $\omega (\xi _{\mathbf{k}%
}) $ at very small $\mid \xi _{\mathbf{k}}\mid \ll \omega _{ph}$ and large
energies $\mid \xi _{\mathbf{k}}\mid \gg \omega _{ph}$ - see Fig.~\ref
{Valla-Sigma}. The theoretical analysis \cite{Kulic-Dolgov-lambda} of these
results gives the total coupling constant $\lambda >3$, the EPI one $\lambda
^{ep}>2$ while the Coulomb scattering (SFI is a part of it) is $\lambda
^{c}\approx 1$ \cite{Kulic-Dolgov-lambda} - see Fig.~\ref{Valla-Sigma}. To
this end let us mention some confusion related to the value of the EPI
coupling constant extracted from ARPES. Namely, in \cite{Shen-review}, \cite
{Shen-Cuk-review}, \cite{Lanzara-isotope} the EPI self-energy was obtained
by subtracting the high energy slope of the quasi-particle spectrum $\omega
(\xi _{k})$ at $\omega \sim 0.3$ $eV$. The latter is apparently due to the
Coulomb interaction. Although the position of the low-energy kink is not
affected by this procedure (if $\omega _{ph}^{\max }\ll \omega _{c}$), the
above (subtraction) procedure gives in fact an \textit{effective EPI
self-energy }$\Sigma _{eff}^{ep}(\mathbf{k},\omega )$ and\textit{\ the
coupling constant} $\lambda _{z,eff}^{ep}(\mathbf{k})$ only. The latter is
smaller than the real EPI coupling constant $\lambda _{z}^{ep}(\mathbf{k})$.
The total self-energy is $\Sigma (\mathbf{k},\omega )=\Sigma ^{ep}(\mathbf{k}%
,\omega )+\Sigma ^{c}(\mathbf{k},\omega )$ where $\Sigma ^{c}$ is the
contribution due to the Coulomb interaction. At very low energies $\omega
\ll \omega _{c}$ one has usually $\Sigma ^{c}(\mathbf{k},\omega )=-\lambda
_{z}^{c}(\mathbf{k})\omega $, where $\omega _{c}(\sim 1$ $eV)$ is the
characteristic Coulomb energies and $\lambda _{z}^{c}$ the Coulomb coupling
constant. The quasi-particle spectrum $\omega (\mathbf{k})$ is determined
from the condition
\begin{equation}
\omega -\xi (\mathbf{k})-\func{Re}[\Sigma ^{ep}(\mathbf{k},\omega )+\Sigma
^{c}(\mathbf{k},\omega )]=0,  \label{Dis}
\end{equation}
where $\xi (\mathbf{k})$ is the bare band structure energy. At low energies $%
\omega <\omega _{ph}^{\max }\ll \omega _{c}$, Eq.(\ref{Dis}) can be
rewritten in the form
\begin{equation}
\omega -\xi ^{ren}(\mathbf{k})-\func{Re}\Sigma _{eff}^{ep}(\mathbf{k},\omega
)=0,  \label{ren-spec}
\end{equation}
\begin{equation}
\xi ^{ren}(\mathbf{k})=[1+\lambda _{z}^{c}(\mathbf{k})]^{-1}\xi (\mathbf{k})
\label{ren-en}
\end{equation}
\begin{equation}
Re\Sigma _{eff}^{ep}(\mathbf{k},\omega )=\frac{Re\Sigma _{eff}^{ep}(\mathbf{k%
},\omega )}{1+\lambda _{z}^{c}(\mathbf{k})}.  \label{eff-SE}
\end{equation}
Since at very low energies $\omega \ll \omega _{ph}^{\max }$, one has $\func{%
Re}\Sigma ^{ep}(\mathbf{k},\omega )=-\lambda _{z}^{ep}(\mathbf{k})\omega $
and $\func{Re}\Sigma _{eff}^{ep}(\mathbf{k},\omega )=-\lambda _{z,eff}^{ep}(%
\mathbf{k})\omega $, then the real coupling constant is related to the
effective one by
\[
\lambda _{z}^{ep}(\mathbf{k})=[1+\lambda _{z}^{c}(\mathbf{k})]\lambda
_{z,eff}^{ep}(\mathbf{k})>\lambda _{z,eff}^{ep}(\mathbf{k}).
\]
At higher energies $\omega _{ph}^{\max }<\omega <\omega _{c}$, which are
less important for pairing, the EPI effects are suppressed and $\Sigma ^{ep}(%
\mathbf{k},\omega )$ stops growing, one has $\func{Re}\Sigma (\mathbf{k}%
,\omega )\approx \func{Re}\Sigma ^{ep}(\mathbf{k},\omega )-\lambda _{z}^{c}(%
\mathbf{k})\omega $. The measured $\func{Re}\Sigma ^{\exp }(\mathbf{k}%
,\omega )$ at $T=10$ $K$ near and slightly away from the \textit{nodal point}
in the optimally doped Bi2212 with $T_{c}=91$ $K$ \cite{Valla-2006} is shown
in Fig.~\ref{Valla-Sigma}.

\begin{figure}[tbp]
\resizebox{.5 \textwidth}{!} {
\includegraphics*[width=8cm]{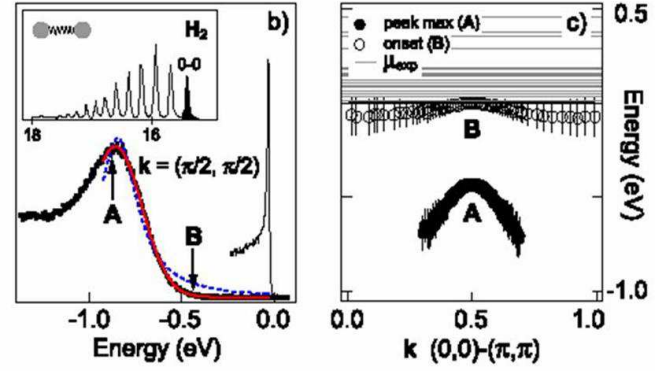}}
\caption{ (Left) The ARPES spectrum of undoped $Ca_{2}CuO_{2}Cl_{2}$ at $%
\mathbf{k}=(\protect\pi/2,\protect\pi/2)$. Gaussian shape - solid line,
Lorentzian shape - dashed line. (Right) Dispersion of the polaronic band - A
and of the quasi-particle band - B along the nodal direction. Horizontal
lines are the chemical potentials for a large number of samples. From
\protect\cite{Shen-polarons}}
\label{ARPES-polaronE}
\end{figure}

It is seen that $Re\Sigma ^{\exp }(\mathbf{k},\omega )$ has \textit{two kinks%
} - the first one at \textit{low energy} $\omega _{1}\approx \omega
_{ph}^{high}\approx 50-70$ $meV$ which is most probably of the phononic
origin \cite{Shen-review}, \cite{Shen-Cuk-review}, \cite{Lanzara-isotope},
while the second kink at \textit{higher energy} $\omega _{2}\approx \omega
_{c}\approx 350$ $meV$ is probably due to the Coulomb interaction. However,
the important results in Ref. \cite{Valla-2006} is that the slopes of $\func{%
Re}\Sigma ^{\exp }(\mathbf{k},\omega )$ at low ($\omega <\omega _{ph}^{high}$%
) and high energies ($\omega _{ph}^{high}<\omega <\omega _{c}$) are \textit{%
different}. The low-energy and high-energy slope \emph{near the nodal point}
are depicted and shown in Fig.~\ref{Valla-Sigma} schematically (thin lines).
From Fig.~\ref{Valla-Sigma} it is obvious that EPI prevails at low energies $%
\omega <\omega _{ph}^{high}$. More precisely digitalization of $\func{Re}%
\Sigma ^{\exp }(\mathbf{k},\omega )$ in the interval $\omega
_{ph}^{high}<\omega <0.4$ $eV$ gives the Coulomb coupling $\lambda
_{z}^{c}\approx 1.1$ while the same procedure at $20$ $meV\approx \omega
_{ph}^{low}<\omega <\omega _{ph}^{high}\approx 50-70meV$ gives the total
coupling constant $(\lambda _{2}\equiv )\lambda _{z}=\lambda
_{z}^{ep}+\lambda _{z}^{c}\approx 3.2$ and the EPI coupling constant $%
\lambda _{z}^{ep}(\equiv \lambda _{z,high}^{ep})\approx 2.1>2\lambda
_{z,eff}^{ep}(\mathbf{k})$, i.e. the EPI coupling is at least twice larger
than in the previous analysis of ARPES results. This estimation tells us
that at (and near) the nodal point, \textit{the EPI interaction dominates }%
in the quasi-particle scattering at low energies since $\lambda
_{z}^{ep}(\approx 2.1)\approx 2\lambda _{z}^{c}>2\lambda _{z}^{sf}$, while
at large energies (compared to $\omega _{ph}$), the Coulomb interaction with
$\lambda _{z}^{c}\approx 1.1$ dominates. We point out that EPI near the
anti-nodal point can be even larger than in the nodal point, mostly due to
the higher density of states near the anti-nodal point. ($N8$) Recent ARPES
spectra in the optimally doped \ $Bi2212$ near the nodal and anti-nodal
point \cite{Lanzara-isotope} show a pronounced isotope effect in $\func{Re}%
\Sigma ^{\exp }(\mathbf{k},\omega )$, thus pointing to the important role of
EPI - see more in the part related to the isotope effect. The isotope effect
in $\func{Re}\Sigma (\mathbf{k},\omega )$ can be well described in the
framework of the Migdal-Eliashberg theory for EPI \cite{Ma-Ku-Do} as it will
be discussed in Part II. ($N9$) ARPES experiments on $Ca_{2}CuO_{2}Cl_{2}$
give strong evidence for the formation of \textit{small polarons in undoped
cuprates }which can be only due to phonons and strong EPI, while by doping
quasi-particles appear and there are no small polarons \cite{Shen-polarons}.
Thus in \cite{Shen-polarons}, it a broad peak (around $-0.8$ $eV$) is
observed at the top of the band ($\mathbf{k}=(\pi /2,\pi /2)$) with the
dispersion similar to that predicted by the $t-J$ model - see Fig. \ref
{ARPES-polaronE}.

However, the peak in Fig.~\ref{ARPES-polaronE} is of Gaussian shape and can
be described only by coupling to bosons, i.e. this peak is a boson side band
- see more in \cite{Gunnarsson-review-2008} and references therein. The
theory based on the t-J model (in the antiferromagnetic state of the undoped
compound) by including coupling to several (half-breathing, apical oxygen,
low-lying) phonons, which is given in \cite{Gunnar-Nagaosa-Ciuchi}, explains
successfully this broad peak of the boson side band by the formation of
small polarons due to the EPI coupling ($\lambda ^{ep}\approx 1.2$). Note
that this $\lambda ^{ep}$ is for the polaron at the bottom of the band while
in the case whwn the Fermi surface exists this coupling is even larger \cite
{Gunnar-Nagaosa-Ciuchi}. In \cite{Gunnar-Nagaosa-Ciuchi} it was stressed
that even when the electron-magnon interaction is stronger than the EPI one,
the polarons are formed due to EPI. The latter involves excitation of many
phonons at the lattice site (where the hole is seating), while it is
possible to excite only one magnon at the given site.

\begin{figure}[tbp]
\begin{center}
\resizebox{.5 \textwidth}{!} {
\includegraphics*[width=8cm]{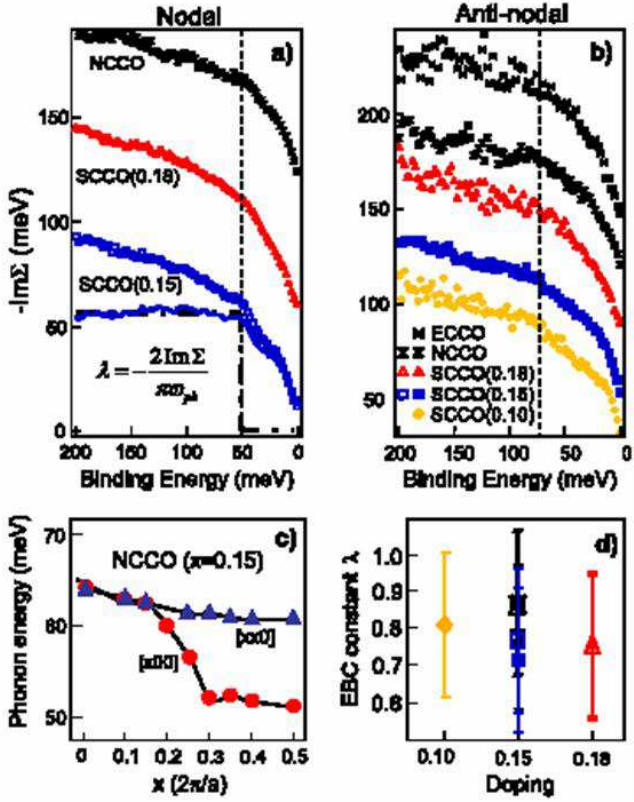}}
\end{center}
\caption{NCCO electron-doped: (a) $Im\Sigma (\protect\omega )$ measured in
the nodal point. Curves are offsets by 50 meV for clarity. The change of the
slope in the last bottom curve is at the phonon energy. (b) $Im\Sigma (%
\protect\omega )$ for the antinodal direction with $30$ $meV$ offset. (c)
Experimental phonon dispersion of the bond stretching modes. (d) Estimated $%
\protect\lambda _{eff}^{ep}$ from $Im\Sigma (\protect\omega )$. From
\protect\cite{Eisaki}.}
\label{Eisaki-kink-el}
\end{figure}

($N10$) Recent soft x-ray ARPES measurements on the \textit{electron doped}
HTSC $Nd_{1.85}Ce_{0.15}CuO_{4}$ \cite{Tsunekawa}, and $%
Sm_{(2-x)}Ce_{x}CuO_{4}$ ($x=0.1,$ $0.15,$ $0.18$), $%
Nd_{1.85}Ce_{0.15}CuO_{4}$, $Eu_{1.85}Ce_{0.15}CuO_{4}$ \cite{Eisaki} show
kink at energies $50-70$ $meV$ in the quasi-particle dispersion relation
along both nodal and antinodal directions as it is shown in Fig. \ref
{Eisaki-kink-el}.

It is seen from this figure that the effective EPI coupling constant $%
\lambda _{eff}^{ep}(<\lambda ^{ep})$ is isotropic and $\lambda
_{eff}^{ep}\approx 0.8-1$. The kink in the electron-doped cuprates is due
solely to EPI and in that respect the situation is similar to the one in the
hole-doped cuparates.

\textit{ARPES results in the superconducting state}

\begin{figure}[tbp]
\begin{center}
\resizebox{.5 \textwidth}{!}
{\includegraphics*[width=8cm]{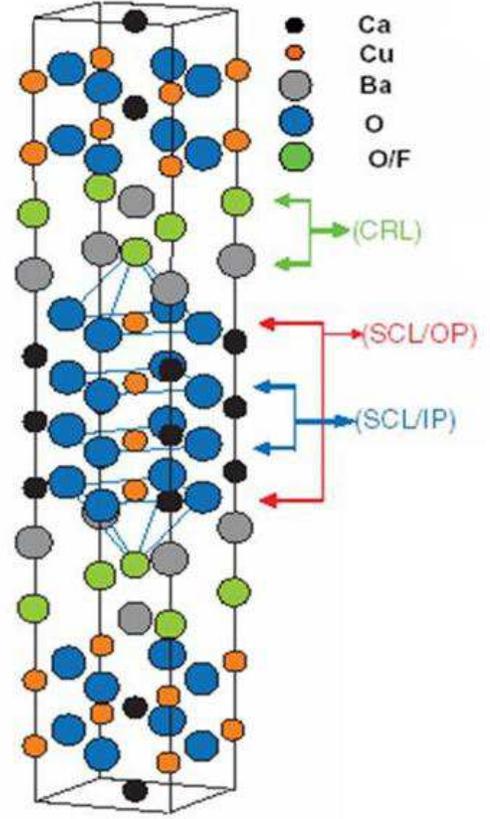}}
\end{center}
\caption{Crystal structure of $Ba_{2}Ca_{3}Cu_{4}O_{8}(O_{\protect\delta
}F_{1-\protect\delta })_{2}$. There are four $CuO_{2}$ layers in a unit cell
with the outer having apical F atoms. From \protect\cite{Chen-Shen-4-layered}%
.}
\label{4lay}
\end{figure}

($S1$) There is an anisotropic superconducting gap in most HTSC compounds
\cite{Shen-review}, which is predominately d-wave like, i.e. $\Delta _{sc}(%
\mathbf{k})\sim \Delta _{0}(\cos k_{x}-\cos k_{y})$ with $2\Delta
_{0}/T_{c}\approx 5-6$. ($S2$) The kink at $(60-70)$ $meV$ in the
quasi-particle energy around the nodal point is \textit{not-shifted} in the
superconducting state while the antinodal kink at $\omega _{ph}^{(40)}\sim
40 $ $meV$ is shifted in the superconducting state by $\Delta
_{0}(=(25-30)meV)$, i.e. $\omega _{ph}^{(40)}\rightarrow \omega
_{ph}^{(40)}+\Delta _{0}=(65-70)meV$ \cite{Shen-review}. To remind the
reader, in the standard Eliashberg theory the kink in the normal state at $%
\omega =\omega _{ph}$ should be shifted in the superconducting state to $%
\omega _{ph}+\Delta _{0}$ at any point at the Fermi surface. This puzzling
result might be a smoking gun result since it makes a constraint on the
quasi-particle interaction in cuprates. Until now there is only one
plausible explanation \cite{Kulic-Dolgov-shift} of this \textit{%
shift-non-shift puzzle} which is based on an assumption of the forward
scattering peak (FSP) in EPI - see more in Part II. The FSP in EPI means
that electrons scatter into a narrow region around the starting point in the
k-space, so that at the most part of the Fermi surface, there is weak (or
no) mixing of states with different signs of the order parameter $\Delta (%
\mathbf{k})$. ($S3$) The recent ARPES spectra \cite{Chen-Shen-4-layered} on
an undoped single crystalline 4-layered cuprate with apical fluorine (F), $%
Ba_{2}Ca_{3}Cu_{4}O_{8}F_{2}$ (F0234) gives strong evidence against SFI -
see Fig.~\ref{4lay}.

\begin{figure}[tbp]
\begin{center}
\resizebox{.5 \textwidth}{!}
{\includegraphics*[width=7cm]{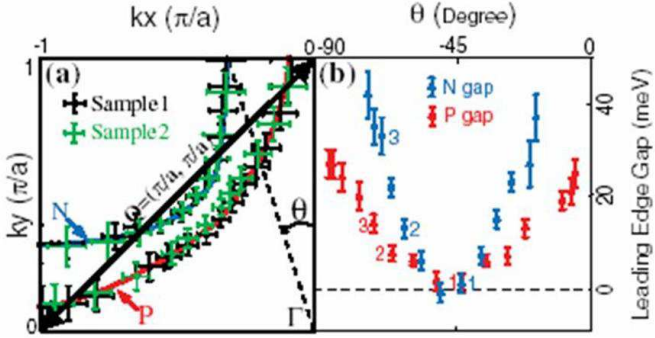}}
\end{center}
\caption{(a) Fermi surface (FS) contours from two samples of F0234. $n$ -
electron-like; $h$ - hole-like. Bold arrow is $(\protect\pi ,\protect\pi )$
scattering vector. Angle $\protect\theta $ defines the horizontal axis in
(b). (b) Leading gap edge along k-space angle from the two FS contours. From
\protect\cite{Chen-Shen-4-layered}.}
\label{Fermi4lay}
\end{figure}

Namely, F0234 is not a Mott insulator - as expected from valence charge
counting which puts $Cu$ valence as $2^{+}$, but it is \ a superconductor
with $T_{c}=60$ $K$. Moreover, the ARPES data \cite{Chen-Shen-4-layered}
reveal at least two metallic Fermi-surface sheets with corresponding volumes
equally below and above half-filling - see Fig.~\ref{Fermi4lay}.

One of the Fermi-surfaces is due to the electron-like ($N$) band (with $%
20\pm 6$\% electron-doping) and the other one due to the hole-like ($P$)
band (with $20\pm 8$ $\%$ hole-doping) and their split along the nodal
direction is significant and cannot be explained by the LDA (or DFA) method
\cite{Shen-Anders-4layer}. This electron and hole self-doping of inner and
outer layers is in an appreciable contrast to other multilayered cuprates
where there is only hole self-doping. For instance, in HgBa$_{2}$Ca$_{n}$Cu$%
_{n+1}$O$_{2n+2}$ ($n=2,3$) and (Cu,C)Ba$_{2}$Ca$_{n}$Cu$_{n+1}$O$_{3n+2}$ ($%
n=2,3,4$), the inner CuO$_{2}$ layers are less hole-doped than outer layers.
It turns out, unexpectedly, that the superconducting gap on the $N$-band
Fermi-surface is significantly larger than on the $P$-one, i.e. their ratio
is anomalous $(\Delta _{N}/\Delta _{P})\approx 2$ and $\Delta _{N}$ is an
order of magnitude larger than in the electron-doped cuprate $%
Nd_{2-x}Ce_{x}CuO_{4}$. Furthermore, the $N$-band Fermi-surface is rather
far from the antinodal point at ($\pi /2,0$). This is an extremely important
result which means that the antiferromagnetic spin fluctuations with the AF
wave-vector $\mathbf{Q}=(\pi /2,\pi /2)$, as well as the van Hove
singularity, are not important for the pairing in the $N$-band. To remind
the reader, the SFI scenario assumes that the pairing is due to spin
fluctuations with the wave-vector $\mathbf{Q}$ which connects two anti-nodal
points which are near the van Hove singularity at the hole-surface (at ($\pi
/2,0$) and ($0,\pi /2$)) giving rise to large density of states. This is
apparently not the case for the $N$-band Fermi-surface - see Fig.~\ref
{Fermi4lay}. The ARPES data give further that there is a kink at $\sim 85$ $%
meV$ in the quasi-particle dispersion of both bands, while the kink in the $%
N $-band is stronger than that in the $P$-band. This result, together with
the anomalous ratio $(\Delta _{N}/\Delta _{P})\approx 2$, strongly disfavors
SFI as a pairing mechanism.

($S4$) Despite the presence of significant elastic scattering in optimally
doped Bi-2212, there are dramatic sharpening of the spectral function near
the anti-nodal point $(\pi ,0)$ at $T<T_{c}$ \cite{Zhu}. This can be
explained by assuming that the small q-scattering (the forward scattering
peak) dominates in the elastic impurity scattering \cite{KuOudo}, \cite
{Kulic-Dolgov-imp}. As a result, one finds that the impurity scattering rate
in the superconducting state is almost zero, i.e. $\gamma _{imp}(\mathbf{k}%
,\omega )=\gamma _{n}(\mathbf{k},\omega )+\gamma _{a}(\mathbf{k},\omega )=0$
for $\mid \omega \mid <\Delta _{0}$ for any kind of pairing (s- p- d-wave
etc.) since the normal ($\gamma _{n}$)\ and the anomalous ($\gamma _{a}$)
scattering rates compensate each other - the \textit{collapse of the elastic
scattering rate}. This result is a consequence of the Anderson-like theorem
for unconventional superconductors which is due to the dominance of the
small q-scattering \cite{KuOudo}-\cite{Kulic-Dolgov-imp}. In such a case
d-wave pairing is weakly unaffected by impurities - there is small reduction
in $T_{c}$ \cite{Kulic-Dolgov-imp}, \cite{Kee-Tc}. The physics behind this
result is rather simple. The small q-scattering (forward scattering) means
that electrons scatter into a small region in the k-space, so that at the
most part of the Fermi surface there is no mixing of states with different
signs of the order parameter $\Delta (\mathbf{k})$, and the detrimental
effect of impurities is reduced. For states near the nodal points, there is
mixing but since $\Delta (\mathbf{k})$ is small in this region, there is
only small reduction in $T_{c}$ \cite{Licht}. This result points to the
importance of strong correlations in the renormalization of the impurity
scattering - see discussion in Part II.

In conclusion, in order to explain ARPES results in cuprates it is necessary
to take into account: (1) EPI interaction since it dominates in the
quasiparticle scattering in the energy region responsible for pairing; (2)
effects of elastic nonmagnetic impurities with FSP; (3) the Coulomb
interaction which dominates at higher energies $\omega >\omega _{ph}$. In
this respect, the presence of ARPES kinks and the knee-like shape of the
spectral width are serious constraints for the pairing theory.

\subsection{Tunnelling spectroscopy and spectral function $\protect\alpha %
^{2}F(\protect\omega )$}

By measuring current-voltage $I-V$ characteristics in $NIS$ (normal
metal-insulator-superconductor) tunnelling junctions with large tunnelling
barrier - see more below, one obtains from tunnelling conductance $%
G_{NS}(V)=dI/dV$ the so called tunnelling density of states in
superconductors $N_{T}(\omega )$. Moreover, by measuring of $G_{NS}(V)$ at
voltages $eV>\Delta $ it is possible to determine the Eliashberg spectral
function $\alpha ^{2}F(\omega )$ and finally to confirm (definitely) the
phonon mechanism of pairing in $LTSC$ materials, except heavy fermions. Four
tunnelling techniques were used in the study of $HTSC$ cuprates: $\mathbf{(1)%
}$ vacuum tunneling by using the $STM$ technique - scanning tunnelling
microscope; $\mathbf{(2)}$ point-contact tunnelling; $\mathbf{(3)}$
break-junction tunnelling; $\mathbf{(4)}$ planar-junction tunnelling. Each
of these techniques has some advantages although in principle the most
potential one is the STM technique \cite{Kirtley}. It should be stressed
that there are still difficulties in understanding tunnelling experiments in
$HTSC$ cuprates because of non-ideal tunnelling behavior of contacts \cite
{Kirtley}. Since tunnelling measurements probe a surface region of the order
of superconducting coherence length $\xi _{0}$, then this kind of
measurements in $HTSC$ materials with small coherence length $\xi _{0}$ ($%
\xi _{ab}\sim 20$ \AA\ in the $a-b$ plane and $\xi _{c}\sim 1-3$ \AA\ along
the $c-axis$) depends strongly on the surface quality and sample
preparation. Nowadays, many of these material problems in $HTSC$ cuprates
are understood and as a result consistent picture of tunnelling features is
starting to emerge.

From tunnelling experiments one obtains the energy gap in the
superconducting state. Since we have already discussed this problem in \cite
{Kulic-Review}, we will only briefly mention some important result, that in
most cases, $G_{NS}(V)$ has V-shape in all families of HTSC hole and
electron doped cuprates. This is due to $d-wave$ pairing with the gapless
spectrum, which is definitely confirmed in the interference experiments on
hole and electron doped cuprates \cite{Tsui-Kirtley}. Some experiments give
the U-shape of $G_{NS}(V)$ which resembles s-wave pairing. This controversy
is explained to be the property of the tunneling matrix element which
filters out states with the maximal gap.

\begin{figure}[tbp]
\begin{center}
\resizebox{.5 \textwidth}{!}
{\includegraphics*[width=8cm]{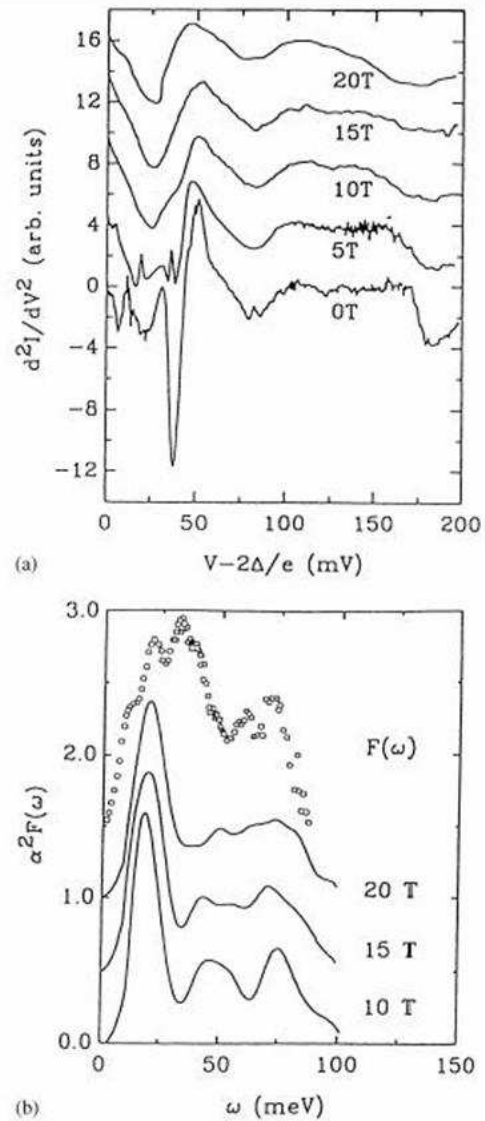}}
\end{center}
\caption{(a) Second derivative of $I(V)$ for a $Bi2212$ break junction in
various magnetic fields. The structure of minima in $d^{2}I/dV^{2}$ can be
compared with the phonon density of states $F(\protect\omega )$; (b) the
spectral functions $\protect\alpha ^{2}F(\protect\omega )$ in various
magnetic fields. From \protect\cite{Tunneling-Vedeneev}}
\label{Veden-Break-JJ}
\end{figure}

Here we are interested in the electron-boson spectral function $\alpha
^{2}F(\omega )$ in HTSC cuprates which can be extracted by using tunnelling
spectroscopy. We inform the reader in advance, that the shape and the energy
width of $\alpha ^{2}F(\omega )$, which are extracted from the second
derivative $d^{2}I/dV^{2}$ at voltages above the superconducting gap, in
most HTSC cuprates resembles the phonon density of states $F(\omega )$. This
result is strong evidence for the importance of $EPI$ in the pairing
potential of $HTSC$ cuprates. For instance, plenty of break-junctions made
from $Bi2212$ single crystals \cite{Tunneling-Vedeneev} show that the
negative peaks in $d^{2}I/dV^{2}$coincide with the peaks in the generalized
phonon density of states $F_{ph}(\omega )$ measured by neutron scattering -
see Fig.~\ref{Veden-Break-JJ}.

\begin{figure}[tbp]
\begin{center}
\resizebox{.4 \textwidth}{!}
{\includegraphics*[width=7cm]{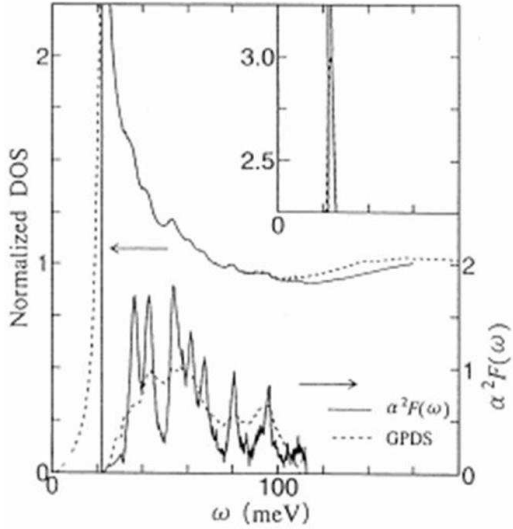}}
\end{center}
\caption{The spectral functions $\protect\alpha ^{2}F(\protect\omega )$ and
the calculated density of states at $0K$ (upper solid line) obtained from
the conductance measurements the $Bi(2212)-Au$ planar junctions. From
\protect\cite{Tun2}}
\label{Tun-Shimada}
\end{figure}

The tunnelling spectra in Bi2212 break junctions \cite{Tunneling-Vedeneev},
which are shown in Fig.~\ref{Veden-Break-JJ}, indicates that the spectral
function $\alpha ^{2}F(\omega )$ is unchanged in magnetic field which
disfavors SFI since in the latter case, this function should be sensitive to
the magnetic field. The reported broadening of the peaks in $\alpha
^{2}F(\omega )$ are partly due to the gapless spectrum of $d-wave$ pairing
in $HTSC$ cuprates. Additionally, the tunnelling density of states $%
N_{T}(\omega )$ at very low $T$ show a pronounced gap structure and it was
found that $2\bar{\Delta}/T_{c}=6.2-6.5$, where $T_{c}=74-85$ $K$ $\ $and $%
\bar{\Delta}$ is some average value of the gap. In order to obtain $\alpha
^{2}F(\omega )$ the inverse procedure was\TEXTsymbol{<}used by assuming $%
s-wave$ superconductivity and the effective Coulomb parameter $\mu ^{\ast
}\approx 0.1$ \cite{Tunneling-Vedeneev}. The obtained $\alpha ^{2}F(\omega )$
gives large $EPI$ coupling constant $\lambda ^{ep}\approx 2.3$. Although
this analysis \cite{Tunneling-Vedeneev} was done in terms of $s-wave$
pairing, it mimics qualitatively the case of $d-wave$ pairing, since one
expects that $d-wave$ pairing does not change significantly the global
structure of $d^{2}I/dV^{2}$ at $eV>\Delta $ albeit introducing a broadening
in it - see the physical meaning in Appendix A. We point out that the
results obtained in \cite{Tunneling-Vedeneev} were reproducible on more than
30 junctions. In that respect very important results on slightly overdoped $%
Bi2212-GaAs$ and on $Bi2212-Au$ planar tunnelling junctions are obtained in
\cite{Tun2} - see Fig.~\ref{Tun-Shimada}.

\begin{figure}[tbp]
\begin{center}
\resizebox{.4 \textwidth}{!}
{\includegraphics*[width=7cm]{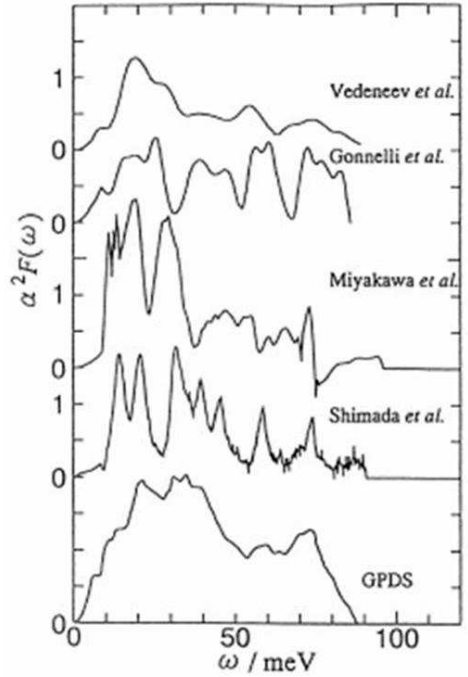}}
\end{center}
\caption{The spectral functions $\protect\alpha ^{2}F(\protect\omega )$ from
measurements of various groups: Vedeneev et al. \protect\cite
{Tunneling-Vedeneev}, Gonnelli et al.\protect\cite{Gonnelli}, Miyakawa et
al. \protect\cite{Tun3}, Shimada et al. \protect\cite{Tun2}. The generalized
density of states GPDS for Bi2212 is at the bottom. From \protect\cite{Tun2}%
. }
\label{Tun-All}
\end{figure}

These results show very similar features to those obtained in \cite
{Tunneling-Vedeneev} on break-junctions. It is worth mentioning that several
groups \cite{Tun3}, \cite{Tun4}, \cite{Gonnelli} have obtained similar
results for the shape of the spectral function $\alpha ^{2}F(\omega )$ from
the $I-V$ measurements on various $HTSC$ cuprates - see the comparison in
Fig.~\ref{Tun-All}. The latter results leave no much doubts about the
importance of the $EPI$ in pairing mechanism of $HTSC$ cuprates. \newline

\begin{figure}[tbp]
\begin{center}
\resizebox{.4 \textwidth}{!}
{\includegraphics*[width=6cm]{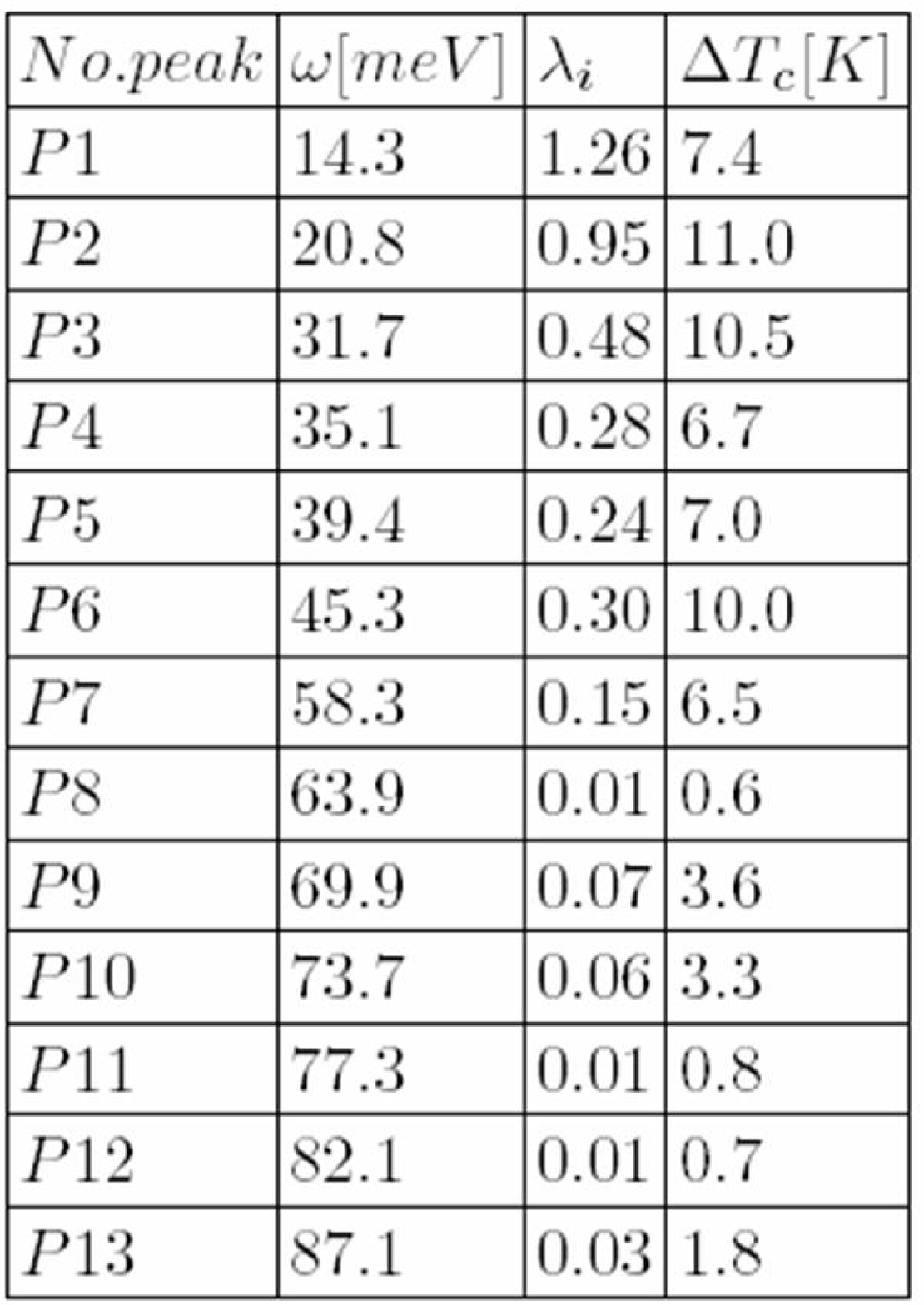}}
\end{center}
\caption{Table I - Phonon frequency $\protect\omega$, EPI coupling constant $%
\protect\lambda_{i}$ of the peaks $P1-P13$ and contribution $\Delta T_{c}$
to $T_{c}$ of each peak in $\protect\alpha ^{2}F(\protect\omega )$-shown in
Fig.\ref{Shimada-Wette-SF}, obtained from the tunnelling conductance of $%
Bi_{2}Sr_{2}CaCu_{2}O_{8}$. $\Delta T_{c}$ is the decrease in $T_{c}$ when
the peak in $\protect\alpha ^{2}F(\protect\omega )$ is eliminated. From
\protect\cite{Tsuda}.}
\label{TableI}
\end{figure}

\begin{figure}[tbp]
\begin{center}
\resizebox{.4 \textwidth}{!}
{\includegraphics*[width=7cm]{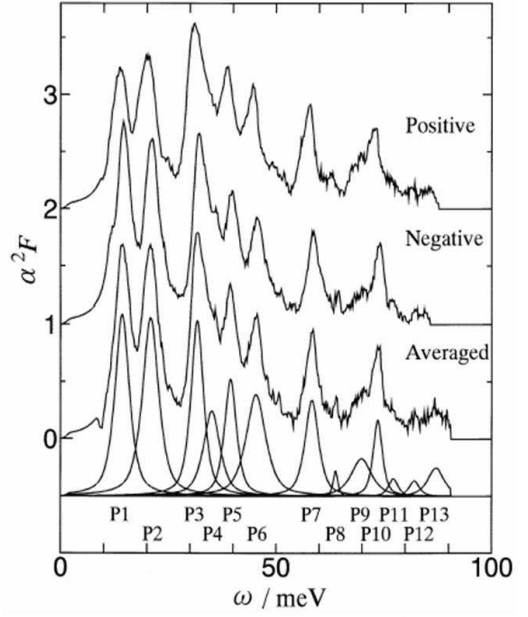}}
\end{center}
\caption{The spectral functions $\protect\alpha ^{2}F(\protect\omega )$ from
the tunnelling conductance of $Bi_{2}Sr_{2}CaCu_{2}O_{8}$ for the positive
and the negative bias voltages, and the averaged one \protect\cite{Tun2}.
The averaged one is divided into $13$ components. The origin of the ordinate
is $2,1,0$ and $-0.5$ from the top down. From \protect\cite{Tsuda},
\protect\cite{Tun2}}
\label{Shimada-Wette-SF}
\end{figure}

In that respect tunnelling measurements on \textit{slightly overdoped} $%
Bi_{2}Sr_{2}CaCu_{2}O_{8}$ \cite{Tun2}, \cite{Tsuda} are impressive, since
the Eliashberg spectral function $\alpha ^{2}F(\omega )$ was extracted from
the measurements of $d^{2}I/dV^{2}$ and by solving the inverse problem. The
extracted $\alpha ^{2}F(\omega )$ has several peaks in broad energy region
up to $80$ $meV$ as it is seen in Fig.~\ref{Tun-Shimada}-~\ref{Tun-All},
which coincide rather well with the peaks in the phonon density of states $%
F_{ph}(\omega )$. In \cite{Tsuda} numerous peaks, from $P1-P13$, in $\alpha
^{2}F(\omega )$ are discerned as shown in Fig.\ref{Shimada-Wette-SF}, which
correspond to various groups of phonon modes - laying in (and around) these
peaks. Moreover, in \cite{Tun2}, \cite{Tsuda} are extracted the coupling
constants for these modes and their contribution ($\Delta T_{c}$) to $T_{c}$
as it is seen in Fig.~\ref{TableI}. Note, due to the nonlinearity of the
problem, the sum of $\Delta T_{c}$ is not equal to $T_{c}$.

The next remarkable result is that the extracted $EPI$ coupling constant is
very large,i.e. $\lambda ^{ep}(=2\int d\omega \alpha ^{2}F(\omega )/\omega
)=\sum_{i}\lambda _{i}\approx 3.5$ - see Fig.~\ref{TableI}. It is obvious
from Figs.~(\ref{TableI}-\ref{Shimada-Wette-SF}) that almost \textit{all
phonon modes contribute} to $\lambda ^{ep}$and $T_{c}$, which means that on
the average, each particular phonon mode is not too strongly coupled to
electrons thus keeping the lattice stable. None of the modes has too large $%
\lambda _{i}<1.3$, thus keeping the lattice stability.

\begin{figure}[tbp]
\begin{center}
\resizebox{.4 \textwidth}{!}
{\includegraphics*[width=7cm]{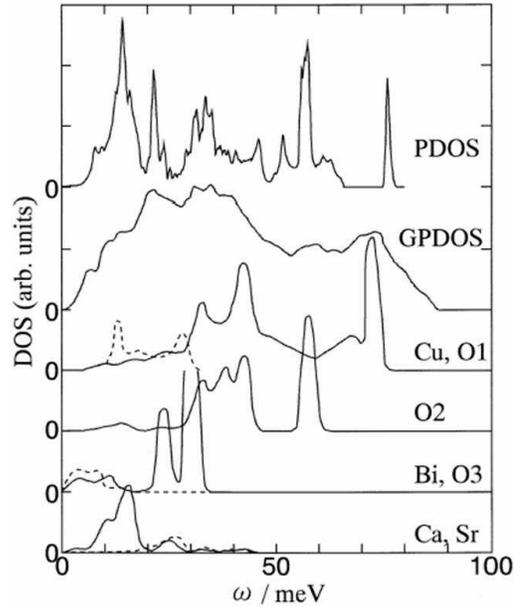}}
\end{center}
\caption{The phonon density of states $F(\protect\omega )$ (PDOS) of $%
Bi_{2}Sr_{2}CaCu_{2}O_{8}$ compared with the generalized density of states
(GPDOS) \protect\cite{Renker}. Atomic vibrations: O1 - O in the $CuO_{2}$
plane; O2 - apical O; O3 - O in the BiO plane. From \protect\cite{Tun2}}
\label{Shimada-Wette-DOS}
\end{figure}

\begin{figure}[tbp]
\begin{center}
\resizebox{.45 \textwidth}{!}
{\includegraphics*[width=6cm]{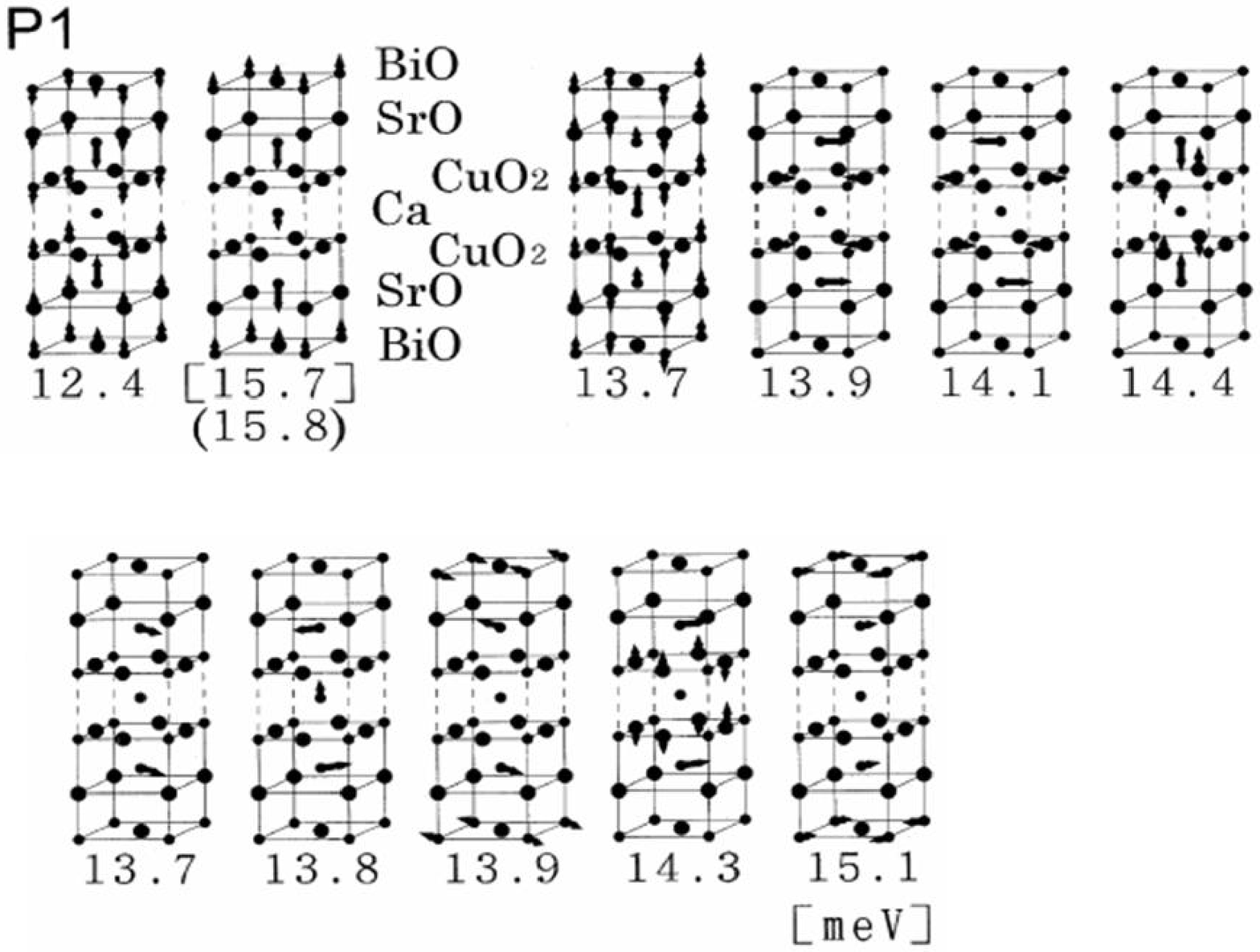}}
{\includegraphics*[width=9cm]{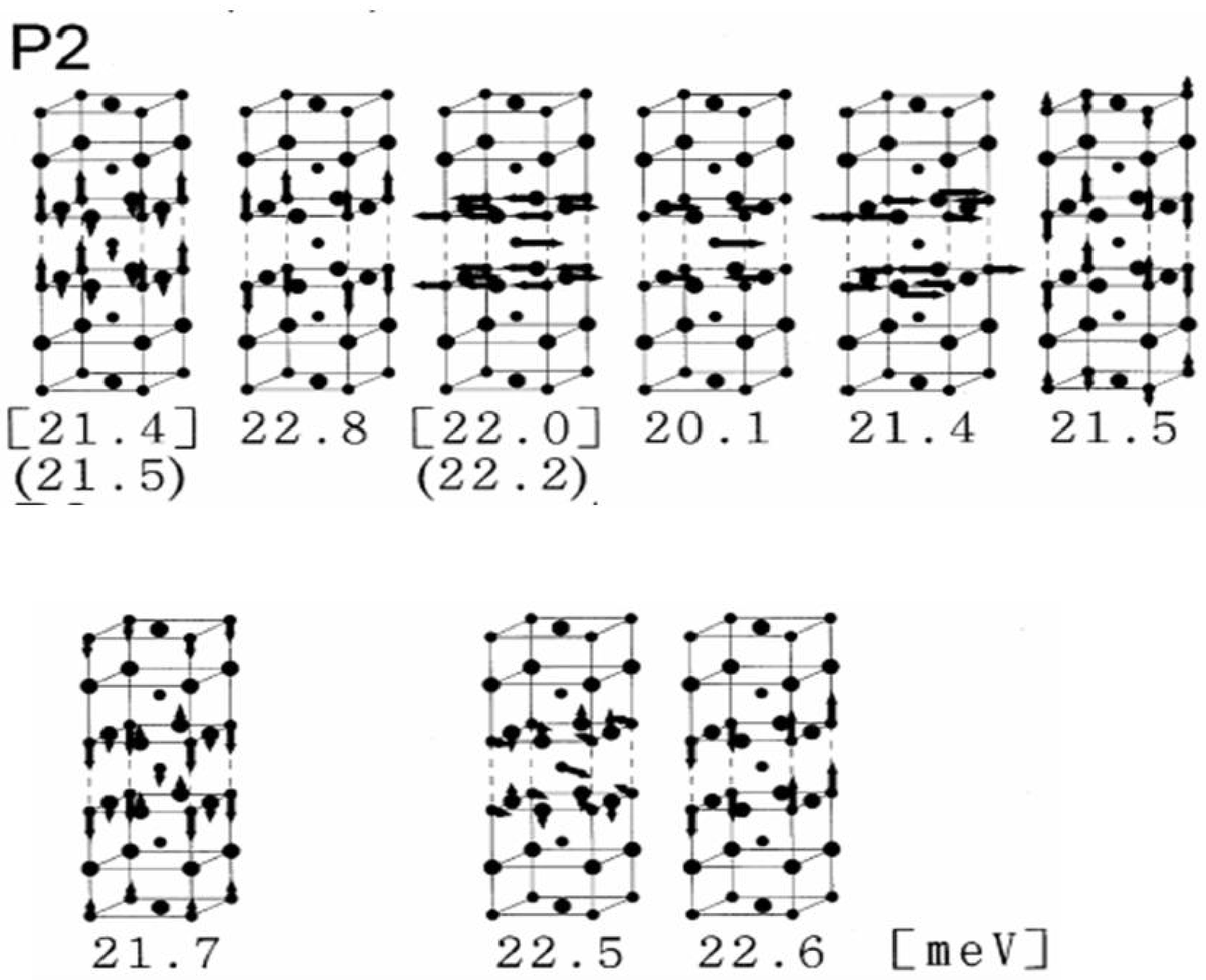}}
\end{center}
\caption{Atomic polarization vectors and their frequencies (in $meV$) at
special points in the Brillouin zone. The larger closed circles in the
lattice are O-ions. $\Gamma - X$ is along the Cu-O-Cu direction. Arrows
indicate displacements. The modes in square and round brackets are the
transverse and longitudinal optical modes respectively. (Top) - modes of the
P1 peak. (Bottom) - modes of the P2 peak. From \protect\cite{Tun2},
\protect\cite{Tsuda}.}
\label{Shimada-Wette-Phonons12}
\end{figure}

\begin{figure}[tbp]
\begin{center}
\resizebox{.45 \textwidth}{!}
{\includegraphics*[width=6cm]{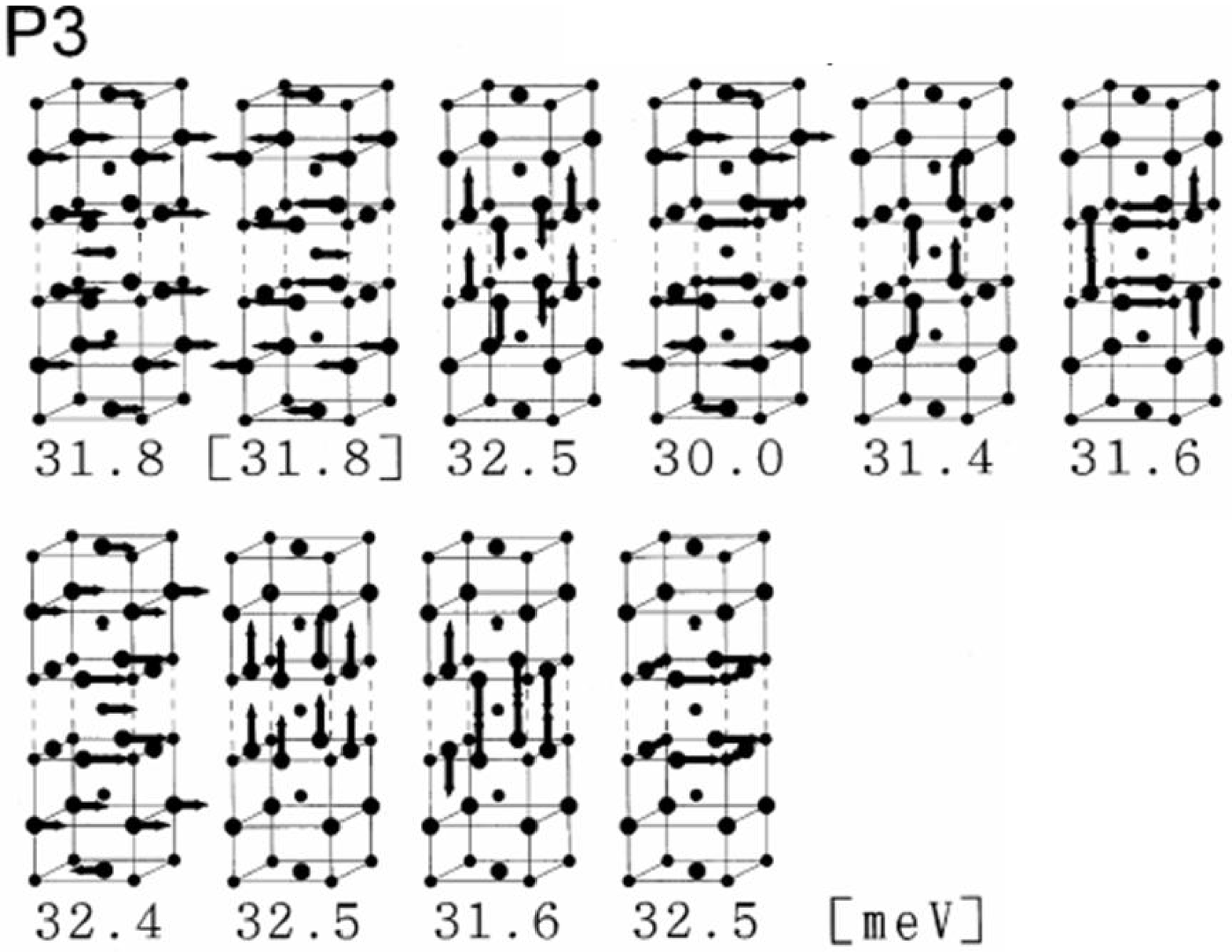}}
{\includegraphics*[width=9cm]{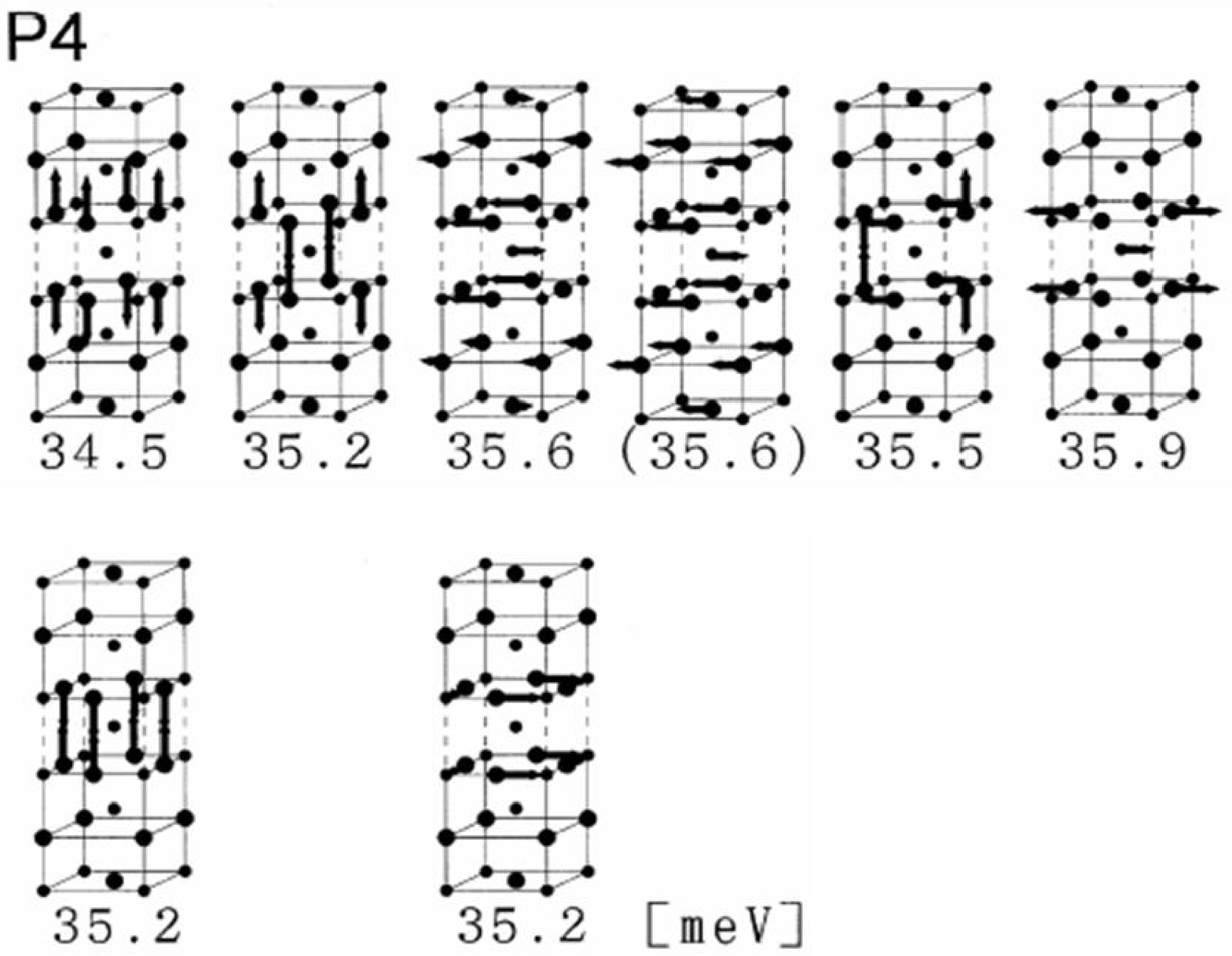}}
\end{center}
\caption{Atomic polarization vectors and their frequencies (in $meV$) at
special points in the Brillouin zone. The larger closed circles in the
lattice are O-ions.$\Gamma - X$ is along the Cu-O-Cu direction. Arrows
indicate displacements. The modes in square and round brackets are the
transverse and longitudinal optical modes, respectively.(Top) - modes of the
P3 peak. (Bottom) - modes of the P4 peak. From \protect\cite{Tun2},
\protect\cite{Tsuda}.}
\label{Shimada-Wette-Phonons34}
\end{figure}

For a better understanding of the the EPI coupling in these systems we show
in Fig. \ref{Shimada-Wette-DOS} the total and partial density of phononic
states.
In Fig.~\ref{TableI} it is seen that lower frequency modes from $P1-P3$,
corresponding to $Cu,Sr$ and $Ca$ vibrations, are rather strongly coupled to
electrons (with $\lambda _{\kappa }\sim 1$) and give appreciable
contributions to $T_{c}$. It is also seen in Fig.~\ref{TableI} that the
coupling constants $\lambda _{i}$ of the high-energy phonons ($P9-P13$ with $%
\omega \geq 70$ $meV$) have $\lambda _{i}\ll 1$ and give moderate
contribution to $T_{c}$ - around 10 \%. These results confirm the \textit{%
importance of modes which cause the change of the Madelung energy} in the
ionic-metallic structure of HTSC cuprates, the idea also conveyed in \cite
{Maksimov-Review}, \cite{Kulic-Review} - see more in \textit{Part II}. If
definitely confirmed, these results are in accordance with the moderate
oxygen isotope effect in cuprates near the optimal doping. We stress that
each peak $P1-P13$ in $\alpha ^{2}F(\omega )$ corresponds to many modes. In
order to get filling on the structure of vibrations possibly strongly
involved in pairing, we show in Figs. \ref{Shimada-Wette-Phonons12}-\ref
{Shimada-Wette-Phonons34} the structure of these vibrations at special
points in the Brillouin zone. It is seen in Fig. \ref
{Shimada-Wette-Phonons12} that the low-frequency phonons $P1-P2$ are
dominated by Cu, Sr, Ca vibrations.

It is seen in Fig.~\ref{TableI} that $P3$ modes are stronger coupled to
electrons than $P4$ ones, although the density of state for the $P4$ modes
is larger. The reason for such an anomalous behavior might be due to
symmetries of corresponding phonons as it is shown in Fig. \ref
{Shimada-Wette-Phonons34}. Namely to the $P3$ peak contribute \textit{axial
vibrations} of O(1) in the Cu$_{2}$ plane which are odd under inversion,
while in the $P4$ peak these modes are even. The in-plane modes of Ca and
O(1) are present in $P3$ which are in-phase and out-of-phase modes, while in
$P4$ they are all out-of-phase modes. For more information on other modes $%
P5-P13$ see \cite{Tsuda}. We stress that the Eliashberg equations based on
the extracted $\alpha ^{2}F(\omega )$ of the slightly overdoped $%
Bi_{2}Sr_{2}CaCu_{2}O_{8}$ with the ratio $(2\Delta /T_{c})\approx 6.5$
describe rather well numerous optical, transport and thermodynamic
properties \cite{Tsuda}. However, in underdoped systems with $(2\Delta
/T_{c})\approx 10$, where the pseudogap phenomena are pronounced, there are
serious disagreements between experiments and the Eliashberg-like theory.

\begin{figure}[tbp]
\begin{center}
\resizebox{.5 \textwidth}{!}
{\includegraphics*[width=7cm]{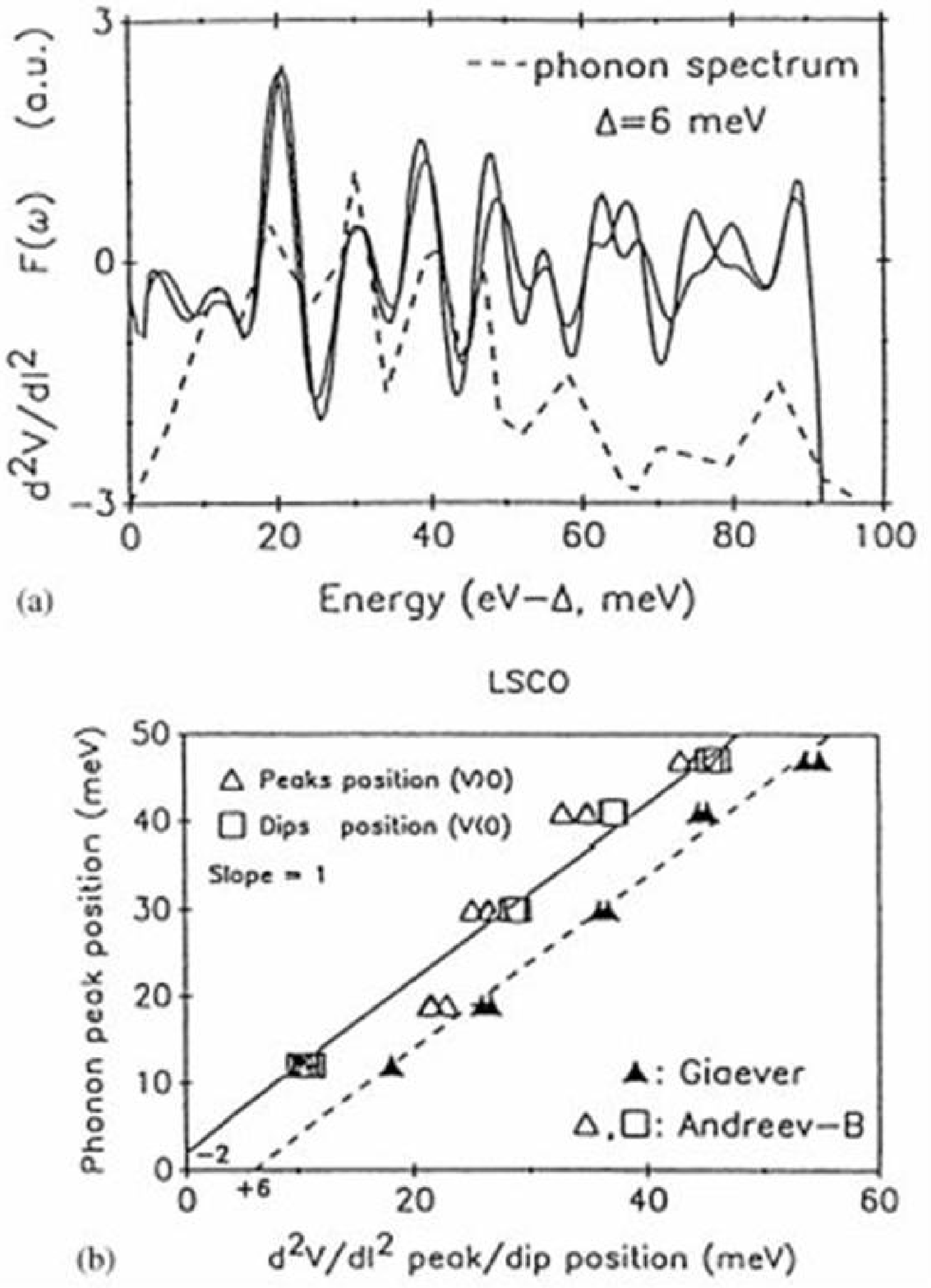}}
\end{center}
\caption{(a) $d^{2}I/dV^{2}$ of a Giaver-like contact in $La$ $%
_{2-x}Sr_{x}CuO_{4}$ - note the large structure below $50meV$; (b) $%
d^{2}I/dV^{2}$ of an Andreev- and Giaver-like contact compared to the peaks
in the phonon density of states. From \protect\cite{Deutscher}}
\label{Tun-Deutsch}
\end{figure}

Similar conclusion, regarding the properties of the $EPI$ spectral function $%
\alpha ^{2}F(\omega )$ in $HTSC$ cuprates, comes out from tunneling
measurements on the Andreev ($Z\ll 1,$ low barrier)- and Giaver ($Z\gg 1$,
high barrier) -type junctions in $La_{2-x}Sr_{x}CuO_{4}$ and $YBCO$
compounds \cite{Deutscher}, where the extracted $\alpha ^{2}F(\omega )$ is
in accordance with the phonon density of states $F_{ph}(\omega )$ see Fig.
\ref{Tun-Deutsch}.

Note that the BTK parameter $Z$ is related to the transmission and
reflection coefficients for the normal metal $(1+Z^{2})^{-1}$ and $%
Z^{2}(1+Z^{2})^{-1}$ respectively.

\begin{figure}[tbp]
\begin{center}
\resizebox{.45 \textwidth}{!}
{\includegraphics*[width=7cm]{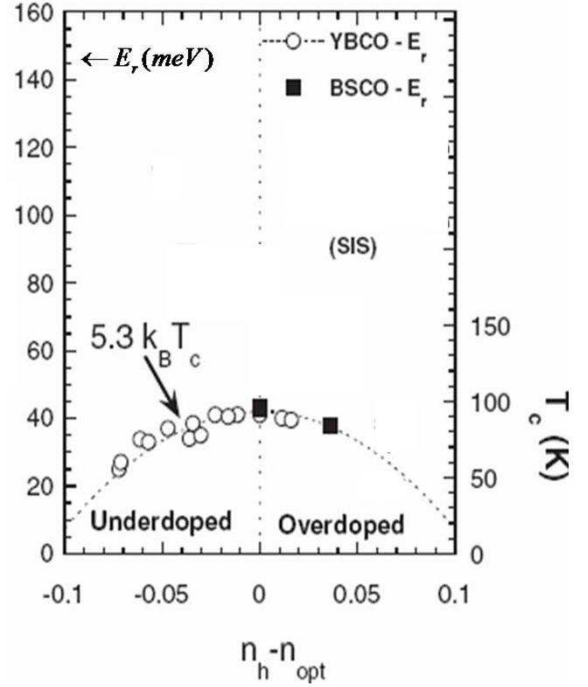}}
\end{center}
\caption{Doping dependence of the energy $E_{r}$ of the magnetic resonance
peak at $\protect\pi ,\protect\pi $ in YBCO and Bi2212 measured at low
temperatures by inelastic neutron scattering. From \protect\cite{Keimer-pss}}
\label{Keimer-reson}
\end{figure}

\begin{figure}[tbp]
\begin{center}
\resizebox{.45 \textwidth}{!}
{\includegraphics*[width=7cm]{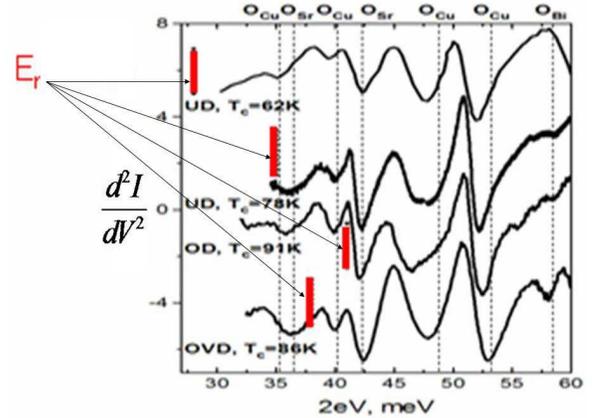}}
\end{center}
\caption{Second derivative of $I(V)$ for a $Bi2212$ tunnelling junctions for
various doping: UD-underdoped; OD-optimally doped; OVD-overdoped system. The
structure of minima in $d^{2}I/dV^{2}$ can be compared with the phonon
density of states $F(\protect\omega )$. The full and vertical lines mark the
positions of the magnetic resonance energy $E_{r}\approx 5.4T_{c}$ for
various doping taken from Fig.\ref{Keimer-reson}. Red tiny arrows mark
positions of the magnetic resonance $E_{r}$ in various doped systems. Dotted
vertical lines mark various phonon modes. From \protect\cite
{Ponomarev-Tunnel}}
\label{Ponomarev-tunnel}
\end{figure}

Although most of the peaks in $\alpha ^{2}F(\omega )$ in $HTSC$ cuprates
coincide with the peaks in the phonon density of states it is legitimate to
put the question - can the magnetic resonance in the superconducting state
give contribution to the $\alpha ^{2}F(\omega )$? In that respect very
important inelastic neutron scattering measurements of the magnetic
resonance as a function of doping \cite{Keimer-pss} give that the resonance
energy $E_{r}$ scales with $T_{c}$, i.e. $E_{r}=(5-6)T_{c}$ as shown in Fig.%
\ref{Keimer-reson}.
This means that if one of the peaks in $\alpha ^{2}F(\omega )$ is due to the
magnetic resonance at $\omega =E_{r}$, then it shifts strongly with doping
as it is observed in \cite{Keimer-pss}. This is contrary to phonon peaks
(energies) whose positions are doping independent. To this end, recent
tunnelling experiments on Bi2212 \cite{Ponomarev-Tunnel} show clear \textit{%
doping independence }of $\alpha ^{2}F(\omega )$ as it is seen in Fig. \ref
{Ponomarev-tunnel}. This remarkable result is an additional and strong
evidence in favor of EPI and against the SFI mechanism of pairing in HTSC
cuprates.

\begin{figure}[tbp]
\begin{center}
\resizebox{.5 \textwidth}{!}
{\includegraphics*[width=7cm]{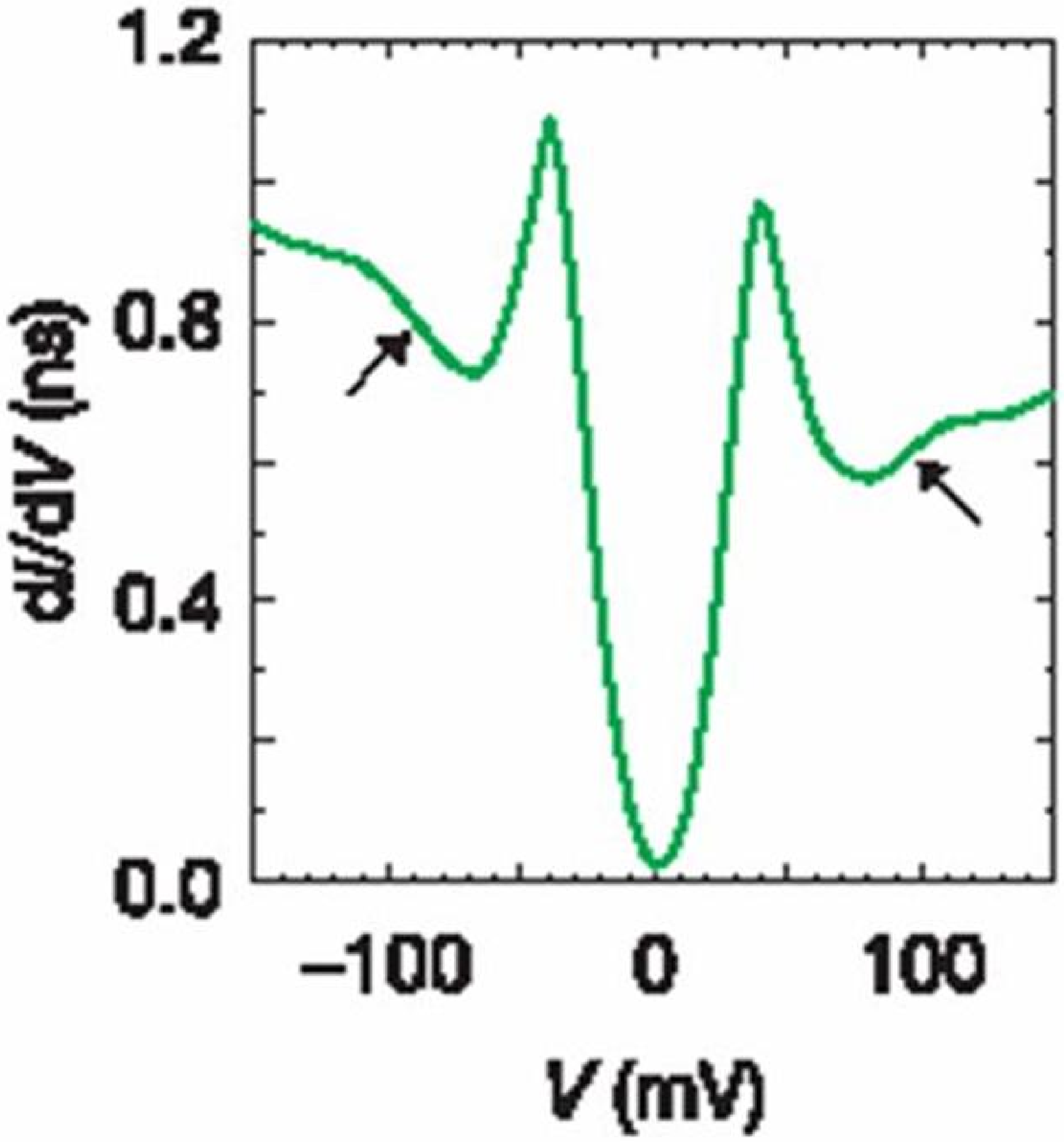}
\includegraphics*[width=7cm]{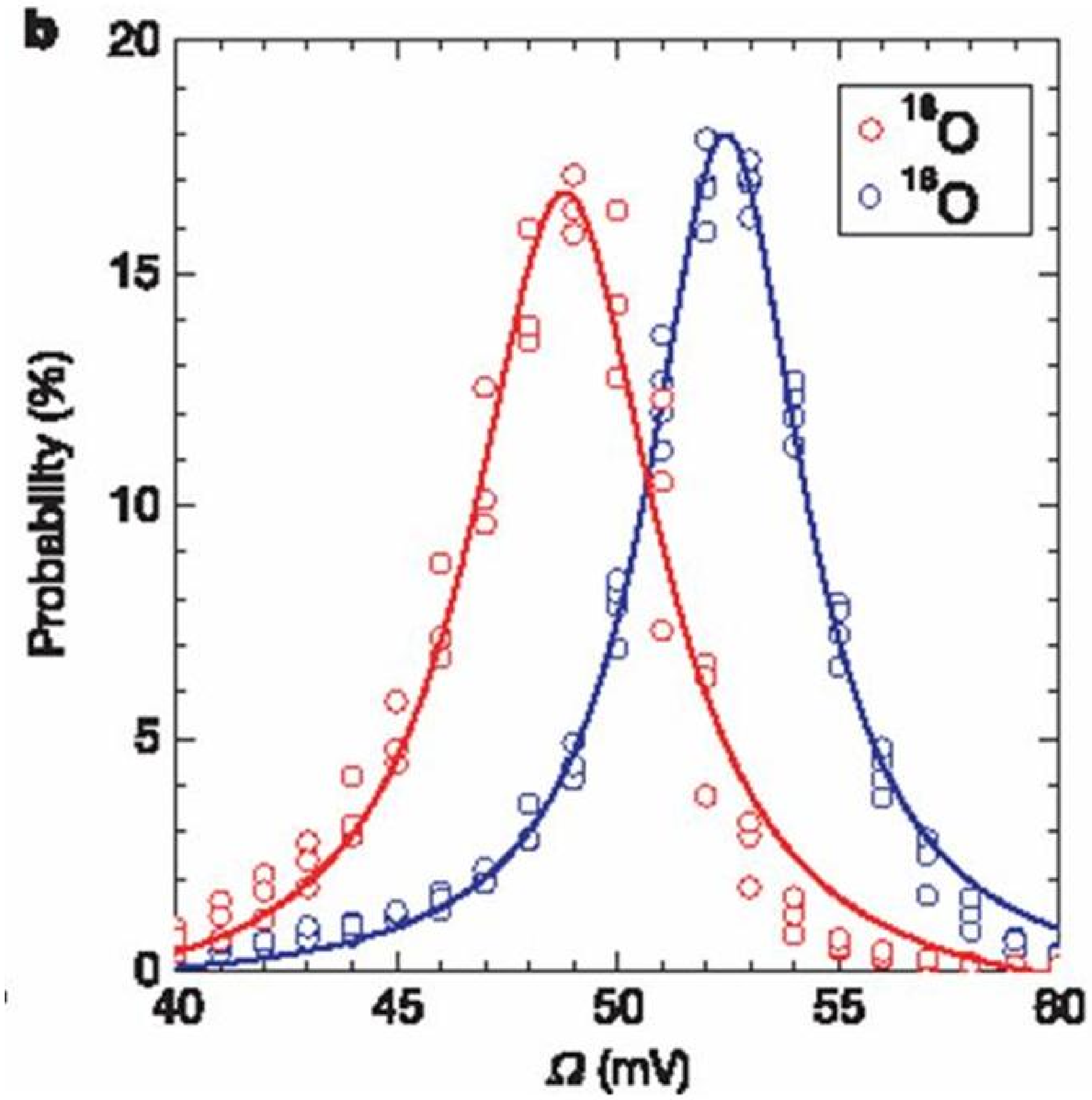}}
\end{center}
\caption{(a) Typical conductance $dI/dV(\mathbf{r},E)$. The ubiquitous
feature at $eV>\Delta$(gap) with maximal slopes, which give peaks in $%
d^{2}I/dV^{2}(\mathbf{r},E)$ are indicated by arrows. (b) The histograms of
all values of $\Omega (\mathbf{r})$ for samples with $O^16$ - right curve
and with $O^18$ - left curve. From \protect\cite{Davis}}
\label{Tun-Davis}
\end{figure}

It is interesting that in the vacuum tunneling $STM$ measurements \cite
{Fischer-Renner-RM} the fine structure in $d^{2}I/dV^{2}$ at $eV>\Delta $
was not seen below $T_{c}$, while the pseudogap structure is observed at
temperatures near and above $T_{c}$. This result could mean that the $STM$
tunnelling is likely dominated by the nontrivial structure of the tunnelling
matrix element (along the $c$-axis), which is derived from the band
structure calculations \cite{AJLM}. However, recent $STM$ experiments on $%
Bi2212$ \cite{Davis} give important information on possible nature of the
bosonic mode which couples with electrons. In \cite{Davis} the local
conductance $dI/dV(\mathbf{r},E)$ is measured where it is found that $%
d^{2}I/dV^{2}(\mathbf{r},E)$ has peak at $E(\mathbf{r})=\Delta (\mathbf{r})$
$+\Omega (\mathbf{r})$ where $dI/dV(\mathbf{r},E)$ has the maximal slope -
see Fig.~ \ref{Tun-Davis}(a).

It turns out that the average phonon energy $\bar{\Omega}$ depends on the
oxygen mass, i.e. $\bar{\Omega}\sim M_{O}^{-1/2}$, with $\bar{\Omega}%
_{16}=52 $ $meV$ and $\bar{\Omega}_{18}\approx 48$ $meV$ - as it is seen in
Fig.~\ref{Tun-Davis}(b). This result is a convincing evidence that phonons
are strongly involved in the quasi-particle scattering. A possible
explanation is put forward in \cite{Davis} by assuming that this isotope
effect is due to the B$_{1g}$ phonon which interacts with anti-nodal
quasi-particles.

In our opinion the important message of tunnelling experiments in $HTSC$
cuprates (by including $Ba_{1-x}K_{x}BiO_{3}$ too \cite{Huang}, \cite{Jensen}%
) is that there is strong evidence for the importance of EPI and that no
particular phonon mode can be singled out in the spectral function $\alpha
^{2}F(\omega )$ as being the only one which dominates in pairing mechanism.
This important result means that the high $T_{c}$ is not attributable to a
particular phonon mode in the $EPI$ mechanism, i.e. all phonon modes
contribute to $\lambda ^{ep}$. Having in mind that the phonon spectrum in $%
HTSC$ cuprates is very broad (up to $80$ $meV$ ), then the large $EPI$
constant ($\lambda ^{ep}\gtrsim 2$) obtained in tunnelling experiments is
not surprising at all.

\subsection{Phonon spectra and EPI}

Although experiments, such as inelastic neutron and Raman scattering,
related to phonon spectra and their renormalization by $EPI$ do not give
directly $\alpha ^{2}F(\omega )$, as the tunnelling and optic spectra do,
they nevertheless give useful information on the strength of EPI for some
particular phonons. We stress in advance that the interpretation of
experimental results in terms of EPI for weakly interacting electrons might
be risky since in strongly correlated systems, such as HTSC cuprates, the
phonon renormalization due to EPI is rather different from that in weakly
correlated metals - see \cite{Gunnarsson-Rosch-epi}. Since these questions
are thoroughly studied in the excellent review \cite{Gunnarsson-Rosch-epi} -
see also \textit{Part II}, we shall briefly enumerate the main points: (1)
In strongly correlated systems, the EPI coupling for a number of phononic
modes can be significantly larger than the LDA and Hartree-Fock methods
predict. This is due to many-body effects \cite{Gunnarsson-Rosch-epi}. (2)
In strongly correlated systems the quasi-particle charge susceptibility,
which enters the phonon self-energy $\Pi \mathbf{(q},\omega )(\sim \chi _{c}%
\mathbf{(q},\omega ))$, is much more suppressed than in weakly correlated
metals and these effects are out of the LDA possibilities \cite{Kulic-Review}%
, \cite{Khaliullin}, \cite{Gunnarsson-Rosch-epi}. This is one of
the reasons that the analysis of experiments on phonon
renormalizations in the framework of \textit{LDA underestimates
the EPI coupling constant significantly} - on all these questions,
see more details in \textit{Part II}.

\subsubsection{\textit{The phonon Raman scattering }}

The \textit{phonon} Raman scattering gives also evidence for appreciable EPI
in cuprates \cite{Cardona1}, \cite{Cardona2}, \cite{Hadjiev}. We enumerate
some of them - see more in \cite{Kulic-Review} and References therein. (i)
There is a pronounced asymmetric line-shape (of the Fano resonance) in the
metallic state. For instance, in $YBa_{2}Cu_{3}O_{7}$ two Raman modes at $115
$ $cm^{-1}$ (Ba dominated mode) and at $340$ $cm^{-1}$ (O dominated mode in
the CuO$_{2}$ planes)\ show pronounced asymmetry which is absent in $%
YBa_{2}Cu_{3}O_{6}$. This result points to appreciable interaction
of Raman active phonons with continuum states (quasi-particles)
\cite{Cardona1}, \cite {Cardona2},. (ii) The phonon frequencies
for some $A_{1g}$ and $B_{1g}$ are strongly renormalized in the
superconducting state, between $(6-10)$ $\%$, pointing again to
large EPI \cite{Hadjiev} - see also in \cite{Kulic-Review},
\cite{Kulic-AIP}. To this point we mention that the
\textit{electronic Raman scattering} in cuprates show a remarkable
correlation between the Raman cross-section $\tilde{S}_{\exp
}(\omega )$ and the optical conductivity in the a-b plane $\sigma
(\omega )$, i.e. $\tilde{S}_{\exp }(\omega )\sim \sigma (\omega )$
\cite{Kulic-Review}. It was argued above that EPI with the very
broad spectral function $\alpha ^{2}F(\omega )$ explains in a
natural way the $\omega $ and $T$ dependence of $\sigma (\omega
)$. This means that the electronic Raman spectra in cuprates can
be explained by EPI in conjunction with strong correlations. This
conclusion is supported by calculations of the Raman cross-section
\cite{Rashkeev} which take into account EPI with $\alpha
^{2}F(\omega )$ extracted from tunnelling measurements in
$YBa_{2}Cu_{3}O_{6+x}$ and $Bi_{2}Sr_{2}CaCu_{2}O_{8+x}$
\cite{Kulic-Review}, \cite{Tunneling-Vedeneev}-\cite{Tsuda}. Quite
similar
properties (to cuprates) of the electronic Raman scattering, as well as of $%
\sigma (\omega )$, $R(\omega )$ and $\rho (T)$, were observed in
experiments \cite{Bozovic} on isotropic 3D metallic oxides
$La_{0.5}Sr_{0.5}CoO_{3}$ and $Ca_{0.5}Sr_{0.5}RuO_{3}$ where
there are no signs of antiferromagnetic fluctuations. This means
that low-dimensionality and antiferromagnetic spin fluctuations
cannot be a prerequisite for anomalous scattering of
quasi-particles and EPI must be inevitably taken into account
since it is present in all these compounds.

\subsubsection{\textit{Neutron scattering, phonon spectra and EPI }}

The softening of numerous phonon modes has been observed in the normal state
of cuprates giving important evidence for pronounced EPI. There are several
important reviews on this subject \cite{Pintschovius} and here we discuss
briefly two important examples which demonstrate in an impressive way the
inefficiency of the LDA band structure calculations to treat quantitatively
and qualitatively EPI in HTSC cuprates. Namely, by doping the Cu-O
bond-stretching phonon mode shows a \textit{substantial softening} at $%
\mathbf{q}_{hb}=(0.5,0,0)$ - called the \textit{half-breathing phonon}, and
a \textit{large broadening} by 5 meV at 15\% doping. While the softening can
be partly described by the LDA method \cite{Bohnen-Heid-2003}, it\textit{\
predicts an order of magnitude smaller broadening} than the experimental
one. The reason for this failure lies in strong correlations, which are not
included in the LDA method as explained in \cite{Kulic-Review}, \cite
{Gunnarsson-review-2008}, \cite{Khaliullin}, \cite{Gunnarsson-Rosch-epi}.
They give rise to an increase of the EPI coupling and to strong suppression
of the charge fluctuations which enter the phonon self-energy via the charge
susceptibility - see below and in \textit{Part II}. The neutron scattering
in $La_{1.85}Sr_{0.15}CuO_{4}$ gives evidence for large (30\%) softening of
the O$_{Z}^{Z}$ with $\Lambda _{1}$ symmetry with the energy $\omega \approx
60$ $meV$, which is theoretically predicted in \cite{Falter-O-phonon}, and
for the large line-width about 17 meV which also suggest strong EPI.

\begin{figure}[tbp]
\begin{center}
\resizebox{.55 \textwidth}{!}
{\includegraphics*[width=8cm]{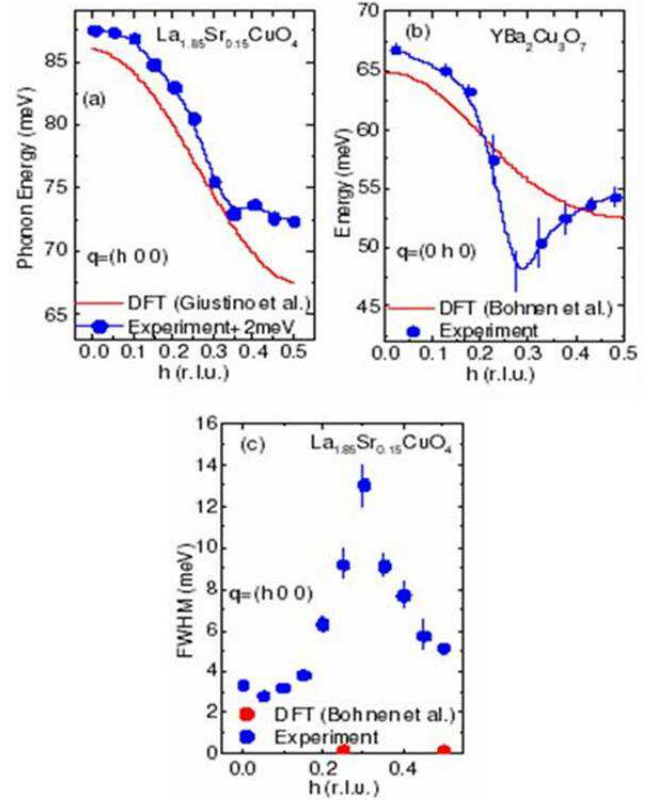}}
\end{center}
\caption{Comparison of DFT calculations with experimental results:(a) in $%
La_{1.85}Sr_{0.15}CuO_{4}$; (b) in $Y_{B}a_{2}Cu_{3}O_{7}$. (c) Phonon line
widths in $La_{1.85}Sr_{0.15}CuO_{4}$. DFT calculations \protect\cite
{Bohnen-Cohen} gives much smaller width than experiments \protect\cite
{Phonon-Exp}. From \protect\cite{ReSaGuDe}.}
\label{LDAvsExp}
\end{figure}

As it is discussed in the Introduction, there are recently several
calculations of the EPI coupling constant $\lambda ^{ep}$ in the framework
of DFT (or LDA), where very small $\lambda ^{ep}\approx 0.3$ was obtained
\cite{Bohnen-Cohen}, \cite{Giuistino}. However, the LDA method is inadequate
for strongly correlated systems as it does not correctly take into account
exchange-correlations and many-body effects, and therefore overestimates the
screening in cuprates. If DFT is able to describe EPI correctly, it must
also be able to calculate phonon renormalization, such as softening and
broadening of the spectrum. In fact DFT completely fails to describe this
renormalization for some important phonon modes and therefore fails to
describe the effect of EPI on the electronic spectrum. The critique of LDA
(DFT) results in HTSC cuprates is done in \cite{Kulic-Review} and recently
strongly argued in \cite{Gunnarsson-review-2008}, \cite{ReSaGuDe} by its
disagreement with neutron scattering measurements as it is shown in Fig.~\ref
{LDAvsExp}.

The point is that DFT (LDA) can reproduce the phonon softening in $%
La_{1.85}Sr_{0.15}CuO_{4}$ and $Y_{B}a_{2}Cu_{3}O_{7}$ rather good at low
momenta $\mathbf{q}=(h,0,0)$ but predicts smooth softening at higher $%
\mathbf{q}$, while experiments show pronounced features for $h=0.3$. At the
same time DFT predicts an order of magnitude smaller line width than
experiments \cite{Phonon-Exp}.

In \textit{Part II} we shall discuss some theoretical approaches related to
EPI renormalization of phonons in strongly correlated systems. Here, we
point out two results. \textit{First}, there is an appreciable difference in
the \textit{phonon renormalization} in strongly and weakly correlated
systems. Namely, the change of phonon frequencies in the presence of
conduction electrons is proportional to the coupling constant $\left| g_{%
\mathbf{q}}\right| $ and charge susceptibility $\chi _{c}$, i.e. $\delta
\omega (\mathbf{q})\sim \left| g_{\mathbf{q}}\right| ^{2}\func{Re}\chi _{c}$%
, while the line width is given by $\Gamma _{\omega (\mathbf{q})}\sim \left|
g_{\mathbf{q}}\right| ^{2}\left| \func{Im}\chi _{c}\right| $. It turns out
that in strongly correlated systems doped with hole concentration $\delta
\ll 1$ the charge fluctuations are suppressed in which case the following
sum-rule holds \cite{Gunnarsson-review-2008}, \cite{Khaliullin}
\[
\frac{1}{\pi N}\sum_{\mathbf{q}\neq 0}\int_{-\infty }^{\infty }\left| \func{%
Im}\chi _{c}(\mathbf{q})\right| d\omega =2\delta (1-\delta )N,
\]
while in the LDA method one has
\[
\frac{1}{\pi N}\sum_{\mathbf{q}\neq 0}\int_{-\infty }^{\infty }\left| \func{%
Im}\chi _{c}(\mathbf{q})\right| ^{LDA}d\omega =(1-\delta )N.
\]
This means that for low doping $\delta \ll 1$ (note $n=1-\delta $), one has $%
\left| \func{Im}\chi _{c}\right| \ll \left| \func{Im}\chi _{c}^{LDA}\right| $
and LDA \textit{strongly underestimates the coupling constant}, i.e. $\left|
g_{\mathbf{q}}\right| _{LDA}\ll \left| g_{\mathbf{q}}\right| $. We stress
that there is no such strong suppression in the quasi-particle self-energy
\cite{Gunnarsson-review-2008}.

\textit{Second}, the theory gives that the coupling constant $\left| g_{%
\mathbf{q}}\right| $ in HTSC cuprates can be significantly larger than LDA
predicts, which is due to some many-body effects not present in the latter
\cite{Gunnarsson-review-2008}, \cite{Khaliullin}. It can be shown that for
some phonon modes one has $\left| g_{\mathbf{q}}\right| ^{2}\gg $ $\left| g_{%
\mathbf{q}}\right| _{LDA}^{2}$. For instance, for the half-breathing mode,
one has $\left| g_{\mathbf{q}}\right| ^{2}\approx 3\left| g_{\mathbf{q}%
}\right| _{LDA}^{2}$ that is first calculated in \cite{Khaliullin}. These
results point to inadequacy of LDA in calculations of EPI effects in HTSC
cuprates.

\subsection{Isotope effect for various doping}

The isotope effect $\alpha _{T_{C}}$ in the critical temperature $T_{c}$ was
one of the very important proof for the EPI pairing in low-temperature
superconductors (LTSC). As a curiosity the isotope effect in $LTSC$ systems
was measured almost exclusively in monoatomic systems and in few polyatomic
systems: the hydrogen isotope effect in $PdH$, the $Mo$ and $Se$ isotope
shift of $T_{c}$ in $Mo_{6}Se_{8}$, and the isotope effect in $Nb_{3}Sn$ and
$MgB_{2}$. We point out that very small ($\alpha _{T_{C}}\approx 0$ in $Zr$
and $Ru$) and even negative (in $PdH$) isotope effect in some polyatomic
systems of $LTSC$ materials are compatible with the $EPI$ pairing mechanism
but in the presence of substantial Coulomb interaction or lattice
anharmonicity. The isotope effect $\alpha _{T_{C}}$cannot be considered as
the smoking gun since it is sensitive to numerous influences. For instance,
in $MgB_{2}$ it is with certainty proved that the pairing is due to EPI and
strongly dominated by the boron vibrations, but the boron isotope effect is
significantly reduced, i.e. $\alpha _{T_{C}}\approx 0.3$. It is still
unexplained. The situation in HTSC cuprates is much more complicated because
they contain many-atoms in unit cell. Additionally, the situation is
complicated with the presence of intrinsic and extrinsic inhomogeneities
which can mask real effects. On the other hand new techniques such as ARPES,
STM, $\mu SR$ allow studies of the isotope effects in quasi-particle
self-energies, i.e. $\alpha _{\Sigma }$, which will be discussed below.

\subsubsection{Isotope effect $\protect\alpha _{T_{C}}$ in $T_{c}$}

This problem will be discussed only briefly since more extensive discussion
can be found in \cite{Kulic-Review}. It is well known that in the pure EPI
pairing mechanism, the total isotope coefficient $\alpha $ is given by
\begin{equation}
\alpha _{T_{C}}=\sum_{i,p}\alpha _{i}^{(p)}=-\sum_{i,p}\frac{d\ln T_{c}}{%
d\ln M_{i}^{(p)}},  \label{isotope}
\end{equation}
where $M_{i}^{(p)}$ is the mass of the i-th element in the p-th
crystallographic position. Note that in the case when the screened Coulomb
interaction is negligible, i.e. $\mu _{c}^{\ast }=0$, one has $\alpha
_{T_{C}}=1/2$. From this formula one can deduce that the relative change of $%
T_{c}$, $\delta T_{c}/T_{c}$, for heavier elements is rather small - for
instance it is 0.02 for $^{135}Ba\rightarrow ^{138}Ba$, 0.03 for $%
^{63}Cu\rightarrow ^{65}Cu$ and 0.07 for $^{138}La\rightarrow ^{139}La$.
This means that measurements of $\alpha _{i}$ for heavier elements are at/or
beyond the ability of present day experimental techniques. Therefore most
isotope effect measurements were done by substituting light atoms $^{16}O$
by $^{18}O$ only. It turns out that in most optimally doped HTSC cuprates $%
\alpha _{O}$ is small. For instance $\alpha _{O}\approx 0.02-0.05$ in $%
YBa_{2}Cu_{3}O_{7}$ with $T_{c,\max }\approx 91$ $K$, but it is appreciable
in $La_{1.85}Sr_{0.15}CuO_{4}$ with $T_{c,\max }\approx 35$ $K$ where$\
\alpha _{O}\approx 0.1-0.2$. In $Bi_{2}Sr_{2}CaCu_{2}O_{8}$ with $T_{c,\max
}\approx 76$ $K$ one has $\alpha _{O}\approx 0.03-0.05$ while $\alpha
_{O}\approx 0.03$ and even negative ($-0.013$) in $%
Bi_{2}Sr_{2}Ca_{2}Cu_{2}O_{10}$ with $T_{c,\max }\approx 110$ $K$. The
experiments on $Tl_{2}Ca_{n-1}BaCu_{n}O_{2n+4}$ ($n=2,3$) with $T_{c,\max
}\approx 121$ $K$ are still unreliable and $\alpha _{O}$ is unknown: In the
electron-doped $(Nd_{1-x}Ce_{x})_{2}CuO_{4}$ with $T_{c,\max }\approx 24$ $K$
one has $\alpha _{O}<0.05$ while in the underdoped materials $\alpha _{O}$
increases. The largest $\alpha _{O}$ is obtained even in the optimally doped
compounds like in systems with substitution, such as $%
La_{1.85}Sr_{0.15}Cu_{1-x}M_{x}O_{4}$, $M=Fe,Co$, where $\alpha _{O}\approx
1.3$ for $x\approx 0.4$ $\%$. In $La_{2-x}M_{x}CuO_{4}$ there is a $Cu$
isotope effect which is of the order of the oxygen one, i.e. $\alpha
_{Cu}\approx \alpha _{O}$ giving $\alpha _{Cu}+\alpha _{O}\approx 0.25-0.35$
for optimally doped systems ($x=0.15$). In case when $x=0.125$ with $%
T_{c}\ll T_{c,\max }$ one has$\ \alpha _{Cu}\approx 0.8-1$ with $\alpha
_{Cu}+\alpha _{O}\approx 1.8$ \cite{Franck}. The appreciation of copper
isotope effect in $La_{2-x}M_{x}CuO_{4}$ tells us that vibrations other than
oxygen ions are important in giving high T$_{c}$. In that sense one should
have in mind the tunnelling experiments discussed above, which tell us that
all phonons contribute to the Eliashberg pairing function $\alpha ^{2}F(%
\mathbf{k},\omega )$ and according to these results the oxygen modes give
moderate contribution to $T_{c}$ \cite{Tsuda}. Having these facts in mind,
then the small oxygen isotope effect $\alpha _{T_{c}}^{(O)}$ in optimally
doped cuprates, if it is intrinsic property, does not exclude the EPI
mechanism of pairing.

\subsubsection{Isotope effect $\protect\alpha_{\Sigma }$ in the self-energy}

\begin{figure}[tbp]
\begin{center}
\resizebox{.45 \textwidth}{!}
{\includegraphics*[width=8cm]{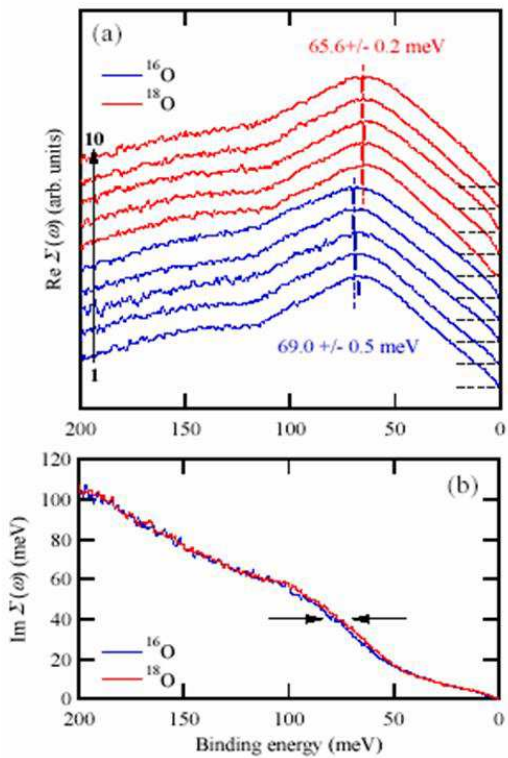}}
\end{center}
\caption{(a) Effective $Re\Sigma $ for five samples for $O^{16}$ (blue) and $%
O^{18}$ (red) along the nodal direction. (b) Effective $Im\Sigma $
determined from MDC full widths. An impurity term is subtracted at $\protect%
\omega =0$. From \protect\cite{Iwasawa-isotop-2}.}
\label{Iwasawa-isotop}
\end{figure}

The fine structure of the quasi-particle self-energy $\Sigma (\mathbf{k}%
,\omega )$, such as kinks, can be resolved in ARPES measurements and in some
respect in STM. It turns out that there is an isotope effect in the
self-energy in the optimally doped $Bi2212$ samples \cite{Lanzara-isotope},
\cite{Douglas-isotop}, \cite{Iwasawa-isotop}. In the first paper on this
subject \cite{Lanzara-isotope}, there is a red shift $\delta \omega
_{k,70}\sim -(10-15)$ $meV$ of the nodal kink at $\omega _{k,70}\simeq 70$ $%
meV$ for the $^{16}O\rightarrow ^{18}O$ substitution. This isotope shift of
the self-energy $\delta \Sigma =\Sigma _{16}-\Sigma _{18}\sim 10$ $meV$ is
more pronounced at large energies $\omega =100-300$ $meV$ . However, there
is a dispute on the latter result which is not confirmed experimentally \cite
{Douglas-isotop}, \cite{Iwasawa-isotop}. The isotope effect in $\func{Re}%
\Sigma (\mathbf{k},\omega )$ \cite{Douglas-isotop}, \cite{Iwasawa-isotop}
can be well described in the framework of the Migdal-Eliashberg theory for
EPI \cite{Ma-Ku-Do} which is in accordance with the recent ARPES
measurements with low-energy photons $\sim 7$ $eV$ \cite{Iwasawa-isotop-2}.
The latter allowed very good precision in measuring the isotope effect in
the nodal point of Bi2212 with $T_{c}^{16}=92.1$ $K$ and $T_{c}^{18}=91.1$ $%
K $ \cite{Iwasawa-isotop-2}. They observed shift in the maximum of Re$\Sigma
(\mathbf{k}_{N},\omega )$ - at $\omega _{k,70}\approx 70$ $meV$ which
corresponds to the half-breathing or breathing phonon, by $\delta \omega
_{k,70}\approx 3.4\pm 0.5$ $meV$ as shown in Fig. \ref{Iwasawa-isotop}.

By analyzing the shift in Im$\Sigma (\mathbf{k}_{N},\omega )$ - shown in
Fig. \ref{Iwasawa-isotop}, one finds similar result $\delta \omega
_{k,70}\approx 3.2\pm 0.6$ $meV$. The similar shift was obtained in STM
measurements \cite{Davis} which is shown in Fig.~\ref{Tun-Davis}(b) and can
have its origin in different phonons. We would like to stress two points:
(i) in compounds with $T_{c}\sim 100$ $K$ the isotope effect in $T_{c}$ is
moderate, i.e. $\alpha _{T_{c}}^{(O)}\lesssim 0.1$, \cite{Iwasawa-isotop-2}.
If we consider this value to be intrinsic then it is not in conflict with
the tunnelling experiments in \cite{Tsuda} which give evidence that
vibrations of heavier ions contribute significantly to $T_{c}$ - see the
above discussion on the tunnelling spectroscopy; (ii) the extracted value in
\cite{Iwasawa-isotop-2} of the effective EPI coupling constant $\lambda
_{eff}^{ep}$ $\sim 0.6$, which is smaller than the real $\lambda ^{ep}$ -
see above the discussion on ARPES, is significantly larger than the LDA
theory predicts $\lambda _{LDA}^{ep}<0.3$ \cite{Bohnen-Cohen}, \cite
{Giuistino}. This again points that the LDA method does not pick up the
many-body effects due to strong correlations- see Part II.

\section{Summary of \textit{Part I}}

The analysis of experimental data in HTSC cuprates which are related to
optics, tunnelling and ARPES \textit{measurements near the optimal doping}
give evidence for the large electron-phonon interaction with the coupling
constant $1<\lambda ^{ep}<3.5$. The analysis is done in the framework of the
Migdal-Eliashberg theory which is a reliable approach for systems near the
optimal doping. The spectral function averaged over the Fermi surface $%
\alpha ^{2}F(\omega )$ is extracted from various tunnelling measurements and
it contains peaks at the same positions as the phonon density of states. The
energy width of $\alpha ^{2}F(\omega )$ is the same as the width of the
phonon density of states $F(\omega )$. This is an unambiguous proof for the
important role which EPI plays in the pairing mechanism of cuprates. The
optical IR reflectivity data provide additional support for this finding
since the transport spectral function has the width and global properties
similar to $F(\omega )$. These findings are additionally and strongly
supported by ARPES measurements on BISCO compounds. The ARPES kinks in the
quasi-particle self-energy can be explained exclusively by EPI and there is
no much room for spin fluctuations (SFI) effects. The weakness of SFI is
unambiguously proved in magnetic neutron scattering on YBCO where the
imaginary part of the susceptibility is drastically reduced in the low
energy region by going from slightly underdoped toward optimally doped
systems, while $T_{c}$ is practically unchanged. This implies that the SFI
coupling constant is limited to the value $\lambda ^{sf}\lesssim 0.3$. All
these results do not leave doubts on the significance of EPI and weakness of
SFI. The obtained total EPI coupling constant is rather large, i.e. $%
1<\lambda ^{ep}<3.5$, while the transport coupling constant is $\lambda
_{tr}\sim \lambda /3$. The different renormalization of the quasi-particle
and transport self-energies by the Coulomb interaction and strong
correlations points to dominance of the small-momentum scattering in EPI.
This will be discussed in \textit{Part II}.

Inelastic neutron scattering measurements in cuprates show that the
broadening of phonon lines is by an order of magnitude larger than the LDA
(DFA) method predicts. Since the phonon line-widths depend on the EPI
coupling and the charge susceptibility it is evident that calculations of
both quantities are beyond the range of applicability of LDA. As a
consequence, LDA overestimates electronic screening and thus underestimates
the EPI coupling. This means that LDA is suitable only for weakly correlated
systems, while many-body effects due to strong correlations are not
contained in this mean-field type theory.

In spite of the very promising and encouraging results about the dominance
of EPI in cuprates the theory is still confronted with the task of obtaining
sufficiently large coupling constant in the d-channel in order that EPI
conforms with d-wave pairing. At present we do not have such a detailed
microscopic theory although some concepts, such as the the dominant EPI
scattering at small transfer momenta, are understood at least qualitatively.
These set of problems and questions will be discussed in \textit{Part II}.

\textbf{Acknowledgement}. We are thankful to Vitalii Lazarevich Ginzburg for
his permanent interest in our work and for support in many respects over
many years. One of us (M. L. K.) is thankful to Karl Bennemann for inspiring
discussions on many subjects related to physics of HTSC cuprates. We thank
Godfrey Akpojotor for careful reading of the manuscript. M. L. K. is
thankful to the Max-Born-Institut f\"{u}r Nichtlineare Optik und
Kurzzeitspektroskopie, Berlin for the hospitality and financial support
during his stay where part of this work has been done.

\section{Appendix - Spectral functions}

\subsection{Spectral functions $\protect\alpha ^{2}F(\mathbf{k},\mathbf{k}%
^{\prime },\protect\omega )$ and $\protect\alpha ^{2}F(\protect\omega )$}

Before discussing experimental results in cuprate superconductors which are
related to various spectral functions, we introduce the reader briefly into
the subject. The quasiparticle-bosonic \ (Eliashberg) spectral function $%
\alpha ^{2}F(\mathbf{k},\mathbf{k}^{\prime },\omega )$ and its Fermi surface
average $\alpha ^{2}F(\omega )=\left\langle \left\langle \alpha ^{2}F(%
\mathbf{k},\mathbf{k}^{\prime },\omega )\right\rangle \right\rangle _{%
\mathbf{k},\mathbf{k}^{\prime }}$ enter the quasi-particle self-energy $%
\Sigma (\mathbf{k},\omega )$, while the transport spectral function $\alpha
^{2}F_{tr}(\omega )$ enters the transport self-energy $\Sigma _{tr}(\mathbf{k%
},\omega )$ and dynamical conductivity $\sigma (\omega )$. Since the
Migdal-Eliashberg theory for EPI is well defined we define the spectral
functions for this case and the generalization to other electron-boson
interaction is straightforward. In the superconducting state the Matsubara
Green's function $\hat{G}(\mathbf{k},\omega _{n})$ and $\hat{\Sigma}(\mathbf{%
k},\omega _{n})$ are $2\times 2$ matrices with the diagonal elements $%
G_{11}\equiv G(\mathbf{k},\omega _{n}),G_{22},\Sigma _{11}\equiv \Sigma (%
\mathbf{k},\omega _{n}),\Sigma _{22}$ and off-diagonal elements $%
G_{12}\equiv F(\mathbf{k},\omega _{n}),G_{21},\Sigma _{12}\equiv \Phi (%
\mathbf{k},\omega _{n}),\Sigma _{21}$ which describe superconducting
pairing. By defining $i\omega _{n}\left[ 1-Z(\mathbf{k},\omega _{n})\right] =%
\left[ \Sigma (\mathbf{k},\omega _{n})-\Sigma (\mathbf{k},-\omega _{n})%
\right] /2$ and $\chi (\mathbf{k},\omega _{n})=\left[ \Sigma (\mathbf{k}%
,\omega _{n})+\Sigma (\mathbf{k},-\omega _{n})\right] /2$, the Eliashberg
functions for EPI in the presence of the Coulomb interaction (in the pairing
channel) read \cite{Allen-Mitrovic}, \cite{Maksimov-Eliashberg}, \cite
{Marsiglio-Carbotte-Book}
\begin{equation}
Z(\mathbf{k},\omega _{n})=1+\frac{T}{N}\sum_{\mathbf{p},m}\frac{\lambda _{%
\mathbf{kp}}^{Z}(\omega _{n}-\omega _{m})\omega _{m}}{N(\mu )\omega _{n}}%
\frac{Z(\mathbf{p},\omega _{m})}{D(\mathbf{p},\omega _{m})},  \label{Z-Eli}
\end{equation}
\begin{equation}
\chi (\mathbf{k},\omega _{n})=-\frac{T}{N}\sum_{\mathbf{p},m}\frac{\lambda _{%
\mathbf{kp}}^{Z}(\omega _{n}-\omega _{m})}{N(\mu )}\frac{\epsilon (\mathbf{p}%
)-\mu +\chi (\mathbf{p},\omega _{m})}{D(\mathbf{p},\omega _{m})},
\label{chi-Eli}
\end{equation}
\begin{equation}
\Phi (\mathbf{k},\omega _{n})=\frac{T}{N}\sum_{\mathbf{p},m}\left[ \frac{%
\lambda _{\mathbf{kp}}^{\Delta }(\omega _{n}-\omega _{m})}{N(\mu )}-V_{%
\mathbf{kp}}\right] \frac{\Phi (\mathbf{p},\omega _{m})}{D(\mathbf{p},\omega
_{m})},  \label{Fi-Eli}
\end{equation}
where $N(\mu )$ is the density of states at the Fermi surface, $\omega
_{n}=\pi T(2n+1)$, $\Phi (\mathbf{k},\omega _{n})\equiv Z(\mathbf{k},\omega
_{n})\Delta (\mathbf{k},\omega _{n})$ and $D=\omega _{m}^{2}Z^{2}+\left(
\epsilon -\mu +\chi \right) ^{2}+\Phi ^{2}$. (For studies of optical
properties - see below, it is useful to introduce the renormalized frequency
$i\tilde{\omega}_{n}(i\omega _{n})(\equiv i\omega _{n}Z(\omega _{n}))=\omega
_{n}-\Sigma (\omega _{n})$ (or its analytical continuation $\tilde{\omega}%
(\omega )=Z(\omega )\omega =\omega -\Sigma (\omega )$). These equations are
supplemented with the electron number equation $n(\mu )$ ($\mu $ is the
chemical potential)
\[
n(\mu )=\frac{2T}{N}\sum_{\mathbf{p},m}G(\mathbf{p},\omega _{m})e^{i\omega
_{m}0^{+}}
\]
\begin{equation}
=1-\frac{2T}{N}\sum_{\mathbf{p},m}\frac{\epsilon (\mathbf{p})-\mu +\chi (%
\mathbf{p},\omega _{m})}{D(\mathbf{p},\omega _{m})}.  \label{n-mju}
\end{equation}
Note that in the case of EPI one has $\lambda _{\mathbf{kp}}^{\Delta }(\nu
_{n})=\lambda _{\mathbf{kp}}^{Z}(\nu _{n})(\equiv \lambda _{\mathbf{kp}}(\nu
_{n}))$ (with $\nu _{n}=\pi Tn$) where $\lambda _{\mathbf{kp}}(\nu _{n})$ is
defined by
\begin{equation}
\lambda _{\mathbf{kp}}(\nu _{n})=2\int_{0}^{\infty }\frac{\nu \alpha _{%
\mathbf{kp}}^{2}F(\nu )d\nu }{\nu ^{2}+\nu _{n}^{2}}  \label{lambda-kp}
\end{equation}

\begin{equation}
\alpha _{\mathbf{kp}}^{2}F(\nu )=N(\mu )\sum_{\kappa }\left| g_{\kappa ,%
\mathbf{kp}}^{ren}\right| ^{2}B_{\kappa }(\mathbf{k}-\mathbf{p,}\nu )
\label{alpha-kp}
\end{equation}
where $B_{\kappa }(\mathbf{k}-\mathbf{p;}\nu )$ is the phonon spectral
function of the $\kappa $-th phonon mode related to the phonon propagator
\begin{equation}
D_{\kappa }(\mathbf{q,}i\nu _{n})=-\int_{0}^{\infty }\frac{\nu }{\nu
^{2}+\nu _{n}^{2}}B_{\kappa }(\mathbf{q,}\nu )d\nu .  \label{D-phonon}
\end{equation}
The renormalized coupling constant $g_{\kappa ,\mathbf{kp}}^{ren}(\approx
g_{\kappa ,\mathbf{kp}}^{0}\varepsilon ^{-1}\gamma )$ comprises the
screening effect due to long-range Coulomb interaction ($\sim \varepsilon
^{-1}$ - the inverse electronic dielectric function) and short-range strong
correlations ($\sim \gamma $ - the vertex function) - see more in \textit{%
Part II.} Usually in the case of low-temperature superconductors (LTS) with
s-wave pairing the anisotropy is rather small (or in the presence of
impurities it is averaged out) which allows an averaging of the Eliashberg
equations \cite{Allen-Mitrovic}, \cite{Maksimov-Eliashberg}, \cite
{Marsiglio-Carbotte-Book}
\begin{equation}
Z(\omega _{n})=1+\frac{\pi T}{\omega _{n}}\sum_{m}\frac{\lambda (\omega
_{n}-\omega _{m})\omega _{m}}{\sqrt{\omega _{m}^{2}+\Delta ^{2}(\omega _{m})}%
},  \label{Z-iso}
\end{equation}
\[
Z(\omega _{n})\Delta (\omega _{n})=\pi T\sum_{m}[\lambda (\omega _{n}-\omega
_{m})
\]
\begin{equation}
-\mu (\omega _{c})\theta (\omega _{c}-\left| \omega _{m}\right| )]\frac{%
\Delta (\omega _{m})}{\sqrt{\omega _{m}^{2}+\Delta ^{2}(\omega _{m})}},
\label{Gap-iso}
\end{equation}
\begin{equation}
\lambda (\omega _{n}-\omega _{m})=\int_{0}^{\infty }d\nu \frac{2\nu \alpha
^{2}F(\nu )}{\nu ^{2}+(\omega _{n}-\omega _{m})^{2}}.  \label{lambda}
\end{equation}
Here $\alpha ^{2}F(\omega )=\left\langle \left\langle \alpha ^{2}F(\mathbf{k}%
,\mathbf{k}^{\prime },\omega )\right\rangle \right\rangle _{\mathbf{k},%
\mathbf{k}^{\prime }}$where $\left\langle \left\langle ...\right\rangle
\right\rangle _{\mathbf{k},\mathbf{k}^{\prime }}$ is the average over the
Fermi surface. The above equations can be written on the real axis by the
analytical continuation $i\omega _{m}\rightarrow \omega +i\delta $ where the
gap function is complex i.e. $\Delta (\omega )=\Delta _{R}(\omega )+i\Delta
_{I}(\omega )$. The solution for $\Delta (\omega )$ allows the calculation
of the current-voltage characteristic $I(V)$ and \textit{tunnelling
conductance} $G_{NS}(V)=dI_{NS}/dV$ in the superconducting state of the $NIS$
tunneling junction where $I_{NS}(V)$ is given by

\[
I_{NS}(V)=2e\sum_{\mathbf{k},\mathbf{p}}\mid T_{\mathbf{k},\mathbf{p}}\mid
^{2}\int_{-\infty }^{\infty }\frac{d\omega }{2\pi }
\]
\begin{equation}
A_{N}(\mathbf{k},\omega )A_{S}(\mathbf{p},\omega +eV)[f(\omega )-f(\omega
+eV)].  \label{I(V)}
\end{equation}
Here, $A_{N,S}(\mathbf{k},\omega )=-2ImG_{N,S}(\mathbf{k},\omega )$ are the
spectral functions of the normal metal and superconductor, respectively and $%
f(\omega )$ is the Fermi distribution function. Since the angular and energy
dependence of the tunnelling matrix elements $\mid T_{\mathbf{k},\mathbf{p}%
}\mid ^{2}$ is practically unimportant for $s-wave$ superconductors, then in
that case the relative conductance $\sigma _{NS}(V)\equiv
G_{NS}(V)/G_{NN}(V) $ is proportional to the tunnelling density of states $%
N_{T}(\omega )=\int A_{S}(\mathbf{k},\omega )d^{3}k/(2\pi )^{3}$, i.e. $%
\sigma _{NS}(\omega )\approx N_{T}(\omega )$ where
\begin{equation}
N_{T}(\omega )=Re\left\{ \frac{\omega +i\tilde{\gamma}(\omega )}{\sqrt{%
(\omega +i\tilde{\gamma}(\omega ))^{2}-\tilde{Z}^{2}(\omega )\Delta (\omega
)^{2}}}\right\} .  \label{Sigma-NS}
\end{equation}
Here, $\tilde{Z}(\omega )=Z(\omega )/ReZ(\omega )$, $\tilde{\gamma}(\omega
)=\gamma (\omega )/ReZ(\omega )$, $Z(\omega )=ReZ(\omega )+i\gamma (\omega
)/\omega $ and the \textit{quasi-particle scattering rate }in the
superconducting state $\gamma _{s}(\omega ,T)=-2Im\Sigma (\omega ,T)$ is
given by
\[
\gamma _{s}(\omega ,T)=2\pi \int\limits_{0}^{\infty }d\nu \alpha ^{2}F(\nu
)N_{s}(\nu +\omega )\{2n_{B}(\nu )
\]
\begin{equation}
+n_{F}(\nu +\omega )+n_{F}(\nu -\omega )\}+\gamma ^{imp},  \label{Gamma-T}
\end{equation}
where $N_{s}(\omega )=\func{Re}\{\omega /(\omega ^{2}-\Delta ^{2})^{1/2}$ is
the quasi-particle density of states in the superconducting state, $%
n_{B,F}(\nu )$ are Bose and Fermi distribution function respectively. Since
the structure of phononic spectrum is contained in $\alpha ^{2}F(\omega )$,
it is reflected on $\Delta (\omega )$ for $\omega >\Delta _{0}$ (the real
gap obtained from $\Delta _{0}=Re\Delta (\omega =\Delta _{0})$) which gives
the structure in $G_{S}(V)$ at $V=\Delta _{0}+\omega _{ph}$. On the contrary
one can extract the spectral function $\alpha ^{2}F(\omega )$ from $%
G_{NS}(V) $ by the inversion procedure proposed by McMillan and Rowell \cite
{McMillan-Rowell}, \cite{Kulic-Review}. It turns out that in low-temperature
superconductors, negative peaks of $d^{2}I/dV^{2}$ at $eV_{i}=\Delta +\omega
_{ph,i}$, correspond to the peak positions of $\alpha ^{2}F(\omega )$ and $%
F(\omega )$. However, we would like to point out that in HTSC cuprates the
gap function is unconventional and very anisotropic, i.e. $\Delta (\mathbf{k}%
,i\omega _{n})\sim \cos k_{x}a-\cos k_{y}a$. Since in this case the
extraction of $\alpha ^{2}F(\mathbf{k},\mathbf{k}^{\prime },\omega )$ is
extremely difficult and at present rather unrealistic task, then an
''average'' $\alpha ^{2}F(\omega )$ is extracted from the experimental curve
$G_{S}(V)$. There is belief that it gives relevant information on the real
spectral function such as the energy width of the bosonic spectrum ($%
0<\omega <\omega _{\max }$) and positions and distributions of peaks due to
bosons. It turns out that even such an approximate procedure gives valuable
information in HTSC cuprates - see discussion in \textit{Section III D}.

Note that in the case of spin-fluctuation interaction (the SFI model) one
should make difference between $\lambda _{\mathbf{kp}}^{Z}(i\nu _{n})$ and $%
\lambda _{\mathbf{kp}}^{\Delta }(i\nu _{n})$ since they differ by sign i.e. $%
\lambda _{\mathbf{kp}}^{Z}(i\nu _{n})=-\lambda _{\mathbf{kp}}^{\Delta }(i\nu
_{n})>0$ since SFI is repulsive in the pairing-channel - see Eqs. (\ref
{lambdaZ}-\ref{lambdaD}).

\subsubsection{Inversion of tunnelling data}

Phonon features in the conductance $\sigma _{NS}(V)$ at $eV=\Delta
_{0}+\omega _{ph}$ makes the tunnelling spectroscopy a powerful method in
obtaining the Eliashberg spectral function $\alpha ^{2}F(\omega )$. Two
methods were used in the past for extracting $\alpha ^{2}F(\omega )$.

The \textit{first} \textit{method} is based on solving the \textit{inverse
problem} of the nonlinear Eliashberg equations. Namely, by measuring $\sigma
_{NS}(V)$, one obtains the tunnelling density of states $N_{T}(\omega )(\sim
\sigma _{NS}(\omega ))$ and by the inversion procedure one gets $\alpha
^{2}F(\omega )$ \cite{McMillan-Rowell}. In reality the method is based on
the iteration procedure - the McMillan-Rowell ($MR$) inversion, where in the
first step an initial $\alpha ^{2}F_{ini}(\omega )$, $\mu _{ini}^{\ast }$
and $\Delta _{ini}(\omega )$ are inserted into Eliashberg equations (for
instance $\Delta _{ini}(\omega )=\Delta _{0}$ for $\omega <\omega _{0}$ and $%
\Delta _{ini}(\omega )=0$ for $\omega >\omega _{0}$) and then $\sigma
_{ini}(V)$ is calculated. In the second step the functional derivative $%
\delta \sigma (\omega )/\delta \alpha ^{2}F(\omega )$ ($\omega \equiv eV$)
is found in the presence of a small change of $\alpha ^{2}F_{ini}(\omega )$
and then the iterated solution $\alpha ^{2}F_{\mathbf{(1)}}(\omega )=\alpha
^{2}F_{ini}(\omega )+$ $\delta \alpha ^{2}F(\omega )$ is obtained, where the
correction $\delta \alpha ^{2}F(\omega )$ is given by
\begin{equation}
\delta \alpha ^{2}F(\omega )=\int d\nu \lbrack \frac{\delta \sigma _{ini}(V)%
}{\delta \alpha ^{2}F(\nu )}]^{-1}[\sigma _{exp}(\nu )-\sigma _{ini}(\nu )].
\label{Inv-dat}
\end{equation}
The procedure is iterated until $\alpha ^{2}F_{(n)}(\omega )$ and $\mu
_{(n)}^{\ast }$ converge to $\alpha ^{2}F(\omega )$ and $\mu ^{\ast }$which
reproduce the experimentally obtained conductance $\sigma _{NS}^{\exp }(V)$.
In such a way the obtained $\alpha ^{2}F(\omega )$ for $Pb$ resembles the
phonon density of states $F_{Pb}(\omega )$, that is obtained from neutron
scattering measurements. Note that the method depends explicitly on $\mu
^{\ast }$ but on the contary it requires only data on $\sigma _{NS}(V)$ up
to the voltage $V_{\max }=\omega _{ph}^{\max }+\Delta _{0}$ where $\omega
_{ph}^{\max }$ is the maximum phonon energy ($\alpha ^{2}F(\omega )=0$ for $%
\omega >\omega _{ph}^{\max }$) and $\Delta _{0}$ is the zero-temperature
superconducting gap. One pragmatical feature for the interpretation of
tunnelling spectra (and for obtaining the spectral pairing function $\alpha
^{2}F(\omega )$) in $LTS$ and $HTSC$ cuprates is that the negative peaks of $%
d^{2}I/dV^{2}$ are at the peak positions of $\alpha ^{2}F(\omega )$ and $%
F(\omega )$. This feature will be discussed later on in relation with
experimental situation in cuprates.

The \textit{second method }has been invented in \cite{GDS-method} and it is
based on the combination of the Eliashberg equations and dispersion
relations for the Greens functions - we call it GDS method. First, the
tunnelling density of states is extracted from the tunneling conductance in
a more rigorous way \cite{Ivanshenko}
\[
N_{T}(V)=\frac{\sigma _{NS}(V)}{\sigma _{NN}(V)}-\frac{1}{\sigma ^{\ast }(V)}%
\int_{0}^{V}du
\]
\begin{equation}
\times \frac{d\sigma ^{\ast }(u)}{du}\left[ N_{T}(V-u)-N_{T}(V)\right]
\label{NT(V)}
\end{equation}
where $\sigma ^{\ast }(V)=\exp \{-\beta V\}\sigma _{NN}(V)$ and the constant
$\beta $ is obtained from $\sigma _{NN}(V)$ at large biases - see \cite
{GDS-method}. $N_{T}(V)$ under the integral can be replaced by the BCS
density of states. Since the second method is used in extracting $\alpha
^{2}F(\omega )$ in a number of LTSC as well as in HTSC cuprates - see below,
we describe it briefly for the case of isotropic EPI at T=0 K. In that case
the Eliashberg equations are \cite{Allen-Mitrovic}, \cite
{Maksimov-Eliashberg}, \cite{Marsiglio-Carbotte-Book}, \cite{GDS-method}
\[
Z(\omega )\Delta (\omega )=\int_{\Delta _{0}}^{\infty }d\omega ^{\prime }%
\func{Re}\left[ \frac{\Delta (\omega ^{\prime })}{\left[ \omega ^{\prime
2}-\Delta ^{2}(\omega ^{\prime })\right] ^{1/2}}\right]
\]
\begin{equation}
\times \left[ K_{+}(\omega ,\omega ^{\prime })\right] -\mu ^{\ast }\theta
(\omega _{c}-\omega )  \label{Gap-iso2}
\end{equation}
\begin{equation}
Z(\omega )=\frac{1}{\omega }\int_{\Delta _{0}}^{\infty }d\omega ^{\prime }%
\func{Re}\left[ \frac{\omega ^{\prime }}{\left[ \omega ^{\prime 2}-\Delta
^{2}(\omega ^{\prime })\right] ^{1/2}}\right] K_{-}(\omega ,\omega ^{\prime
})  \label{Z-iso2}
\end{equation}
where
\[
K_{\pm }(\omega ,\omega ^{\prime })=\int_{\Delta _{0}}^{\omega _{ph}^{\max
}}d\nu \alpha ^{2}F(\nu )(\frac{1}{\omega ^{\prime }+\omega +\nu +i0^{+}}
\]
\begin{equation}
\pm \frac{1}{\omega ^{\prime }-\omega +\nu -i0^{+}}).  \label{Kern}
\end{equation}
Here $\mu ^{\ast }$ is the Coulomb pseudopotential, the cutoff $\omega _{c}$
is approximately $(5-10)$ $\omega _{ph}^{\max }$, $\Delta _{0}=\Delta
(\Delta _{0})$ is the energy gap. Now by using the dispersion relation for
the matrix Greens functions $\hat{G}(\mathbf{k},\omega _{n})$ one obtains
\cite{GDS-method}
\begin{equation}
\func{Im}S(\omega )=\frac{2\omega }{\pi }\int_{\Delta _{0}}^{\infty }d\omega
^{\prime }\frac{N_{T}(\omega ^{\prime })-N_{BCS}(\omega ^{\prime })}{\omega
^{2}-\omega ^{\prime 2}}  \label{ImS}
\end{equation}
where $S(\omega )=\omega /\left[ \omega ^{2}-\Delta ^{2}(\omega )\right]
^{1/2}$. From Eqs. (\ref{Gap-iso2}-\ref{Z-iso2}) one obtains
\[
\int_{0}^{\omega -\Delta _{0}}d\nu \alpha ^{2}F(\omega -\nu )\func{Re}%
\left\{ \Delta (\nu )\left[ \nu ^{2}-\Delta ^{2}(\nu )\right] ^{1/2}\right\}
\]
\[
=\frac{\func{Re}\Delta (\omega )}{\omega }\int_{0}^{\omega -\Delta _{0}}d\nu
\alpha ^{2}F(\nu )N_{T}(\omega -\nu )+\frac{\func{Im}\Delta (\omega )}{\pi }
\]
\begin{equation}
+\frac{\func{Im}\Delta (\omega )}{\pi }\int_{0}^{\infty }d\omega ^{\prime
}N_{T}(\omega ^{\prime })\int_{0}^{\omega _{ph}^{\max }}d\nu \frac{2\alpha
^{2}F(\nu )}{(\omega ^{\prime }+\nu )^{2}-\omega ^{2}}.  \label{IntEq}
\end{equation}

Based on Eqs. (\ref{NT(V)}-\ref{IntEq}) one obtains the scheme for
extracting $\alpha ^{2}F(\omega )$%
\[
\sigma _{NS}(V),\sigma _{NN}(V)\rightarrow N_{T}(V)
\]
\[
\rightarrow \func{Im}S(\omega )\rightarrow \Delta (\omega )\rightarrow
\alpha ^{2}F(\omega ).
\]
The advantage in this method is that the explicit knowledge of $\mu ^{\ast }$
is not required \cite{GDS-method}. However, the integral equation for $%
\alpha ^{2}F(\omega )$ is linear Fredholm equation of the first kind which
is ill-defined - see the duscussion in Section II.B.2.

\subsubsection{Phonon effects in $N_{T}(\protect\omega )$}

We briefly discuss the physical origin for the phonon effects in $%
N_{T}(\omega )$ by considering a model with only one peak, at $\omega _{0}$,
in the phonon density of states $F(\omega )$ by assuming for simplicity $\mu
^{\ast }=0$ and neglecting the weak structure in $N_{T}(\omega )$ at $%
n\omega _{0}+\Delta _{0}$, which is due to the nonlinear structure of the
Eliashberg equations \cite{SSW}.
\begin{figure}[tbp]
\begin{center}
\resizebox{.45 \textwidth}{!}
{\includegraphics*[width=8cm]{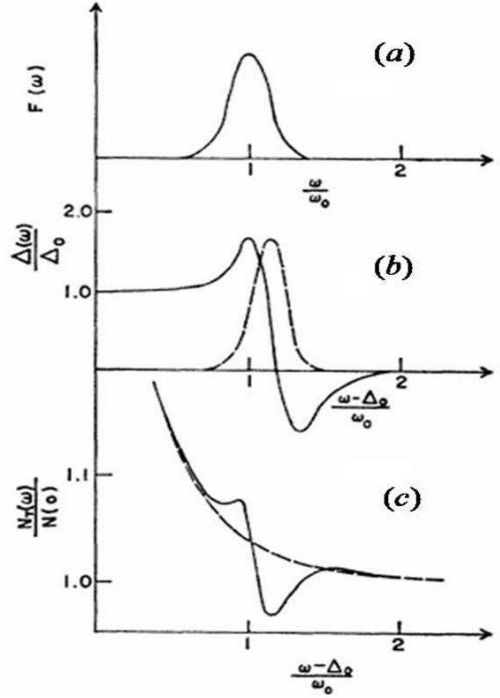}}
\end{center}
\caption{(a) Model phonon density of states $F(\protect\omega )$ with the
peak at $\protect\omega _{0}$. (b) The real (solid) $\Delta _{R}$ and
imaginary part $\Delta _{I}$ (dashed) of the gap $\Delta (\protect\omega )$.
(c) The normalized tunnelling density of states $N_{T}(\protect\omega )/N(0)$
(solid) compared with the BCS density of states (dashed). From \protect\cite
{SSW}.}
\label{Tunnel-density}
\end{figure}
In Fig.\ref{Tunnel-density} it is seen that the real gap $\Delta _{R}(\omega
)$ reaches a maximum at $\omega _{0}+\Delta _{0}$ then decreases, becomes
negative and zero, while $\Delta _{I}(\omega )$ is peaked slightly beyond $%
\omega _{0}+\Delta _{0}$ that is the consequence of the effective
electron-electron interaction via phonons.

It follows that for $\omega <\omega _{0}+\Delta _{0}$ most phonons have
higher energies than the energy $\omega $ of electronic charge fluctuations
and there is over-screening of this charge by ions giving rise to
attraction. For $\omega \approx \omega _{0}+\Delta _{0}$ charge fluctuations
are in resonance with ion vibrations giving rise to the peak in $\Delta
_{R}(\omega )$. For $\omega _{0}+\Delta _{0}<\omega $ the ions move out of
phase with respect to charge fluctuations giving rise to repulsion and
negative $\Delta _{R}(\omega )$. This is shown in Fig. \ref{Tunnel-density}%
(b). The structure in $\Delta (\omega )$ is reflected on $N_{T}(\omega )$ as
shown in Fig. \ref{Tunnel-density}(c) which can be reconstructed from the
approximate formula for $N_{T}(\omega )$ expanded in powers of $\Delta
/\omega $%
\[
\frac{N_{T}(\omega )}{N(0)}\approx 1+\frac{1}{2}\left[ \left( \frac{\Delta
_{R}(\omega )}{\omega }\right) ^{2}-\left( \frac{\Delta _{I}(\omega )}{%
\omega }\right) ^{2}\right] .
\]
As $\Delta _{R}(\omega )$ increases above $\Delta _{0}$ this gives $%
N_{T}(\omega )>N_{BCS}(\omega )$, while for $\omega \gtrsim \omega
_{0}+\Delta _{0}$ the real value $\Delta _{R}(\omega )$ decreases while $%
\Delta _{I}(\omega )$ rises and $N_{T}(\omega )$ decreases giving rise for $%
N_{T}(\omega )<N_{BCS}(\omega )$.

\subsection{Transport spectral function $\protect\alpha _{tr}^{2}F(\protect%
\omega )$}

The spectral function $\alpha _{tr}^{2}F(\omega )$ enters the dynamical
conductivity $\sigma _{ij}(\omega )$ ($i,j=a,b,c$ axis in $HTS$ systems)
which generally speaking is a tensor quantity given by the following formula
\[
\sigma _{ij}(\omega )=-\frac{e^{2}}{\omega }\int \frac{d^{4}q}{(2\pi )^{4}}%
\gamma _{i}(q,k+q)
\]
\begin{equation}
\times G(k+q)\Gamma _{j}(q,k+q)G(q),  \label{Sigma-gen}
\end{equation}
where $q=(\mathbf{q},\nu )$ and $k=(\mathbf{k}=0,\omega )$ and the bare
current vertex $\gamma _{i}(q,k+q;\mathbf{k}=0)$ is related to the Fermi
velocity $v_{F,i}$, i.e. $\gamma _{i}(q,k+q;\mathbf{k}=0)=v_{F,i}.$ The
vertex function $\Gamma _{j}(q,k+q)$ takes into account the renormalization
due to all scattering processes responsible for finite conductivity \cite
{SchrieffLTS}. In the following we study only the in-plane conductivity at $%
\mathbf{k}=0$. The latter case is realized due to the long penetration depth
in HTSC cuprates and the skin depth in the normal state are very large. In
the $EPI$ theory, $\Gamma _{j}(q,k+q)\equiv \Gamma _{j}(\mathbf{q},i\omega
_{n},i\omega _{n}+i\omega _{m})$ is a solution of an approximative integral
equation written in the symbolic form \cite{KMS}
\begin{equation}
\Gamma _{j}=v_{j}+V_{eff}GG\Gamma _{j}  \label{Gamma-vertex}
\end{equation}
where the effective potential $V_{eff}$ (due to EPI) is given by $%
V_{eff}=\sum_{\kappa }\mid g_{\kappa }^{ren}\mid ^{2}D_{\kappa }$, where $%
D_{\kappa }$ is the phonon Green's function. In such a case the Kubo theory
predicts $\sigma _{ii}^{intra}(\omega )$ ($i=x,y,z$)

\[
\sigma _{ii}(\omega )=\frac{\omega _{p,ii}^{2}}{4i\pi \omega }%
\{\int_{-\omega }^{0}d\nu th(\frac{\omega +\nu }{2T})S^{-1}(\omega ,\nu )
\]
\begin{equation}
+\int_{0}^{\infty }d\nu \lbrack th(\frac{\omega +\nu }{2T})-th(\frac{\nu }{2T%
})]S^{-1}(\omega ,\nu )\},  \label{Sigma-tr}
\end{equation}
where $S(\omega ,\nu )=\omega +\Sigma _{tr}^{\ast }(\omega +\nu )-\Sigma
_{tr}(\nu )+i\gamma _{tr}^{imp}$, and $\gamma _{tr}^{imp}$ is the impurity
contribution$.$ In the following we omit the tensor index $ii$ in $\sigma
_{ii}(\omega )$. In the presence of several bosonic scattering processes the
transport self-energy $\Sigma _{tr}(\omega )=Re\Sigma _{tr}(\omega
)+iIm\Sigma _{tr}(\omega )$ is given by
\begin{equation}
\Sigma _{tr}(\omega )=-\sum_{l}\int_{0}^{\infty }d\nu \alpha
_{tr,l}^{2}F_{l}(\nu )[K_{1}(\omega ,\nu )+iK_{2}(\omega ,\nu )],
\label{Self-tr}
\end{equation}
\begin{equation}
K_{1}(\omega ,\nu )=Re[\Psi (\frac{1}{2}+i\frac{\omega +\nu }{2\pi T})-\Psi (%
\frac{1}{2}+i\frac{\omega -\nu }{2\pi T})],  \label{K1}
\end{equation}
\begin{equation}
K_{2}(\omega ,\nu =\frac{\pi }{2}[2cth(\frac{\nu }{2T})-th(\frac{\omega +\nu
}{2T})+th(\frac{\omega -\nu }{2T})]  \label{K2}
\end{equation}
Here $\alpha _{tr,l}^{2}F_{l}(\nu )$ is the \textit{transport spectral
function }which measures the strength of the $l$-th (bosonic) scattering
process and $\Psi $ is the digamma function. The index $l$ enumerates EPI,
charge and spin-fluctuation scattering processes. Like in the case of EPI,
the transport bosonic spectral function $\alpha _{tr,l}^{2}F(\Omega )$ is
given explicitly by
\[
\alpha _{tr,l}^{2}F(\omega )=\frac{1}{N^{2}(\mu )}\int \frac{dS_{\mathbf{k}}%
}{v_{F,\mathbf{k}}}\int \frac{dS_{\mathbf{k}^{\prime }}}{v_{F,\mathbf{k}%
^{\prime }}}\times
\]
\begin{equation}
\left[ 1-\frac{v_{F,\mathbf{k}}^{i}v_{F,\mathbf{k}}^{i}}{(v_{F,\mathbf{k}%
}^{i})^{2}}\right] \alpha _{\mathbf{kk}^{\prime },l}^{2}F(\omega ).
\label{Tr-spec-fun}
\end{equation}
We stress that in the phenomenological SFI theory \cite{Pines} one assumes $%
\alpha _{\mathbf{kk}^{\prime }}^{2}F(\omega )\approx N(\mu )g_{sf}^{2}$Im$%
\chi (\mathbf{k}-\mathbf{p},\omega )$, which, as we have repeated several
times, can be justified only for small $g_{sf}$, i.e. $g_{sf}\ll W_{b}$ (the
band width).

In case of weak coupling ($\lambda <1$), $\sigma (\omega )$ can be written
in the generalized (extended) Drude form as discussed in Section III.B.


\begin{references}

\bibitem{Maksimov-Review} E. G. Maksimov,
Uspekhi Fiz. Nauk {\bf 170}, 1033 (2000); V. L. Ginzburg, E,\ G.
Maksimov, Physica {\bf C 235-240}, 193 (1994); Superconductivity
(in Russian) {\bf 5}, 1505 (1992)

\bibitem{Kulic-Review} M. L. Kuli\'{c}, Phys. Reports
{\bf 338}, 1-264 (2000)

\bibitem{Falter} C. Falter, phys. stat. solidi
(b) {\bf 242}, 118 (2005)

\bibitem{Alexandrov} A. S. Alexandrov,
arXiv:0808.3520; T. M. Hardy, J. P. Hague, J. H. Samson, A. S.
Alexandrov, arXiv:0806.2810

\bibitem{Gunnarsson-review-2008} O.
Gunnarsson, O. R\"{o}sch, J. Phys: Condens. Matter {\bf 20},
043201 (2008); O. R\"{o}sch, J. E. Hahn, O. Gunnarsson, V. H. Crespi,
phys. stat. solidi (b) {\bf 242}, 78 (2005)

\bibitem{Pines} A. J. Millis, H. Monien, D. Pines,
Phys. Rev. {\bf B 42}, 167 (1990); P. Monthoux and D. Pines, Phys.
Rev. Lett. {\bf 69}, 961 (1992); Phys. Rev. B {\textbf{4}7}, 6069
(1993); B. P. Stojkovi\'{c}, D. Pines, Phys. Rev. {\bf B56}, 11931
(1997); D. Pines, preprint CNSL Newsletter, LALP-97-010, No. 138,
June 1997; Physica {\bf B163}, 78 (1990)

\bibitem{Patrick-Lee} P. A. Lee, N. Nagaosa, C.-G. Wen,
Rev. Mod. Phys. {\bf 78}, 17 (2006)

\bibitem{Imada-MC} T. Ami, M. Imada,
J. Phys. Soc. Jpn. {\bf 76}, 113708 (2007)

\bibitem{Scalapino-Drude-weight} D. J.
Scalapino, S. R. White, S. C. Zhang, Phys. Rev. Lett. {\bf 68},
2830 (1992)

\bibitem{Pryadko} L. Pryadko, S. Kivelson, O.
Zachar, Phys. Rev. Lett. {\bf 92}, 067002 (2004)

\bibitem{Shen-review}
A. Damascelli, Z. Hussain, Z. -X. Shen, Rev. Mod. Phys. {\bf 75},
473 (2003); J. C. Campuzano, M. R. Norman, M. Randeria,
cond-mat/0209476 (2002)

\bibitem{Hwang-Timusk-1} J. Hwang et al., Phys. Rev. Lett.
{\bf 100}, 137005 (2008)

\bibitem{Bohnen-Cohen} R. Heid et al., Phys.
Rev. Lett. {\bf 100}, 137001 (2008)

\bibitem{Giuistino} F. Giuistino et
al., Nature {\bf 452}, 975 (2008)

\bibitem{MaKuDoAk} E. G. Maksimov, M.
L. Kuli\'{c}, O. V. Dolgov, G. Akpojotor, in preparation

\bibitem{Bourges}
Ph. Bourges, in {\it The Gap Symmetry and Fluctuations in High
Temperature Superconductors}, J. Bok, G. Deutscher, D. Pavuna, S.
A. Wolf, Eds. (Plenum, New York, 1998), pp. 349-371; preprint
cond-mat/9901333 (1999)

\bibitem{Allen} P. B. Allen, Phys. Rev. {\bf B3}, 305 (1971)

\bibitem{Dolgov-Shulga} O. V. Dolgov, S. V. Shulga, J. of
Superconductivity {\bf 8}, 611 (1995); S. V. Shulga, O. V. Dolgov,
E. G. Maksimov, Physica {\bf C178}, 266 (1991), O. V. Dolgov, E.
G. Maksimov, S. V. Shulga, in {\it Electron-Phonon Interaction in
Oxide Superconductors}, ed. R. Baquero, World Scien., p.30 (1991)

\bibitem{Shulga} S. V. Shulga, in {\it High-$T_{c}$
Superconductors and Related Materials}, p. 323 (2001), eds. S.-L.
Drechsler, T. Mishonov, Kluwer Academic Publishers

\bibitem{Maks-Karakoz-1} A. E. Karakozov, E. G. Maksimov, O. V.
Dolgov, Sol. St. Comm. {\bf 124}, 119 (2002)

\bibitem{Maks-Karakoz-2} A. E. Karakozov, E.
G. Maksimov, Solid St. Commun. {\bf 139}, 80 (2006)

\bibitem{Kusar-2008}
P. Kusar et al., arXiv: 0805.0536v2

\bibitem{%
Tunneling-Vedeneev} L. N. Bulaevskii, O. V. Dolgov, I.P. Kazakov,
S.N. Maksimovskii, M.O. Ptitsyn, V.A. Stepanov, S.I. Vedeneev,
Supercond. Science $\And$ Technology, {\textbf 1}, 205 (1988); S.
I. Vedeneev, A. G. M. Jensen, P. Samuely, V. A. Stepanov, A. A.
Tsvetkov and P. Wyder, Phys. Rev. {\textbf 49}, 9823 (1994); S. I.
Vedeneev, A. G. M. Jansen, A. A. Tsvetkov, P. Wyder, Phys. Rev.
{\bf B 51}, 16380 (1995); S. I. Vedeneev, A. G. M. Jensen and P.
Wyder, Physica {\textbf B 218}, 213 (1996)

\bibitem{Tun2} D.
Shimada, Y. Shiina, A. Mottate, Y. Ohyagi and N. Tsuda, Phys. Rev.
{\textbf B} {\textbf 51}, 16495 (1995); N. Miyakawa, A. Nakamura,
Y. Fujino, T. Kaneko, D. Shimada, Y. Shiina and N. Tsuda, Physica
{\textbf C 282-287}, 1519 (1997);

\bibitem{Tun3}  N. Miyakawa, Y. Shiina, T. Kaneko and N.
Tsuda, J. Phys. Soc. Jpn. {\textbf 62}, 2445 (1993); N. Miyakawa,
Y. Shiina, T. Kido and N. Tsuda, J. Phys. Soc. Jpn. {\textbf 58},
383 (1989)

\bibitem{Tun4}  Y. Shiina, D. Shimada, A. Mottate, Y. Ohyagi and
N. Tsuda, J. Phys. Soc. Jpn. {\textbf 64}, 2577 (1995); Y. Ohyagi,
D. Shimada, N. Miyakawa, A. Mottate, M. Ishinabe, K. Yamauchi and
N. Tsuda, J. Phys. Soc. Jpn. {\textbf 64}, 3376 (1995)

\bibitem{Gonnelli} R. S. Gonnelli, F. Asdente and D. Andoreone, Phys. Rev.
{\textbf B 49}, 1480 (1994)

\bibitem{Tsuda}
D. Shimada et al., Physica {\bf C 298}, 195 (1998);
 N. Tsuda et al., in {\it New Research on
Superconductivity}, Ed. Barry P. Martinis, pp. 105-141, 2007 Nova
Science Publishers, Inc.

\bibitem{Ponomarev-Tunnel} Y. G. Ponomarev et al.,
in Proc. of the Intern. Conf. on Strongly Correlated Systems,
Houston, Texas, USA, May 13-17, 2007, p. 190; Physica {\bf C}
(2008)

\bibitem{Davis} K. McElroy, R. W.
Simmonds, J. E. Hoffman, D. H. Lee, K. M. Lang, J. Orenstein, H.
Eisaki, S. Uchida, J. C. Davis, Nature {\bf 422}, 592 (2003); J.
Lee et al., Bull. Am. Phys. Soc. {\bf 50}, 299 (2005); J. C. Davis
et al., Bull. Am. Phys. Soc. {\bf 50}, 1223 (2005)

\bibitem{Marsiglio-1994} F. Marsiglio et al.,
Phys. Rev. {\bf B50}, 7023 (1994)

\bibitem{Klein-1994} O. Klein et al.,
Phys. Rev. {\bf B50}, 6307 (1994)

\bibitem{MgB2-isotop} D. G. Hinks, J.
D. Jorgensen, Physica {\bf C385}, 98 (2003)

\bibitem{Phillips} J. C.
Phillips, phys. stat. solidi (b) {\bf 242}, 51 (2005)

\bibitem{Cohen} M.
L. Cohen, P. W. Anderson, {\it Superconductivity in} $d $ {\it and
}$f$ {\it band metals}, AIP Conference Proceedings (ed. D. H.
Douglass) New York, p.17 (1972); reprint in P.W. Anderson
A Career in Theoretical Physics (World Scientific, 1994) p. 288

\bibitem{Maksimov-Dolgov-2007} O. V. Dolgov, E. G.
Maksimov, Uspekhi Fiz. Nauk {\bf 177},
983 (2007) ( Phys.-Uspekhi, 50, 933 (2007))

\bibitem{Kirzhnitz} D. A.
Kirzhnitz, in {\it High Temperature Superconductivity}, eds. V. L.
Ginzburg and D. Kirzhnitz, Consultant Bureau New York, London, 1982;
D. A. Kirzhnitz, Chapter 2 in: "Dielectric Function of Condensed Systems",
eds. L.V.Keldysh, D.A.Kirzhnits, A.A.Maradudin,
Elsevier Publ., Amsterdam (1989)

\bibitem{DKM}
O. V. Dolgov, D. A. Kirzhnitz, E. G. Maksimov, Rev. Mod. Phys.
{\bf 53}, 81 (1981);
O.V. Dolgov, D.A. Kirzhnits, E.G. Maksimov,
Chapter 2 in: "Superconductivity, Superdiamagnetism and Superfluidity",
ed. by V.L.Ginzburg, MIR Publ., Moscow (1987) (in English);
O.V. Dolgov, E.G.Maksimov,  Chapter 4 in: "Dielectric Function
of Condensed Systems", eds. L.V.Keldysh, D.A.Kirzhnits, A.A.Maradudin,
Elsevier Publ., Amsterdam (1989)

\bibitem{Cohen2} J. E. Moussa, M. L. Cohen,
Phys. Rev. {\bf B 74}, 094520 (2006)

\bibitem{Anderson2} P. W. Anderson, Science {\bf 316}, 1705 (2007)

\bibitem{Tsui-Kirtley} C. C. Tsuei, J. R. Kirtley, Rev. Mod. Phys.
{\bf 72}, 969 (2000)

\bibitem{Licht}  A. I. Lichtenstein, M.
L. Kuli\'{c}, Physica C {\textbf 245}, 186 (1995)

\bibitem{Scalapino-Review}
D. J. Scalapino, Physics Reports {\bf 250}, 329(1995)


\bibitem{Kulic-AIP} M. L. Kuli\'{c}, {\it AIP
Conference Proceedings Volume }{\bf 715, }75 (2004), Lectures On
The Physics Of Highly Correlated Electron Systems; M. L.
Kuli\'{c}, O. V. Dolgov, phys. stat. solidi (b) {\bf 242}, 151
(2005)

\bibitem{Kulic1}
M. L. Kuli\'{c}, R. Zeyher, Phys. Rev. {\bf B 49},
4395 (1994); Physica {\bf C 199-200}, 358 (1994); Physica {\bf C %
235-240}, 358 (1994)

\bibitem{Kivelson} H. Y. Kee, S. A. Kivelson, G. Aeppli,
Phys. Rev. Lett. {\bf 88}, 257002 (2002)

\bibitem{Kulic-Kulic} M. L. Kuli\'{c}, I. M.
Kuli\'{c}, Physica {\bf 391}, 42 (2003)

\bibitem{Lanzara} A. Lanzara et
al., Nature {\bf 412}, 510 (2001)

\bibitem{Valla} T. Valla et al.,
Science {\bf 285}, 2110 (1999)

\bibitem{Bozovic-Plasma} I. Bo\v{z}%
ovi\'{c}, Phys. Rev. {\bf B 42}, 1969 (1990)

\bibitem{Schlesinger} Z.
Schlesinger et al., Phys. Rev. Lett. {\bf 65}, 801 (1990)

\bibitem{Romero92} D. B. Romero et al.,
Phys. Rev. Lett. {\bf 68}, 1590 (1992); D. B. Romero et al., Sol.
St. Comm. {\bf 82}, 183 (1992)

\bibitem{Puchkov}
A. V. Puchkov, D. N. Basov, T. Timusk, J. Phys.: Condens. Matter
{\bf 8}, 10049 (1996); J. Hwang, T. Timusk, G. D. Gu, Nature {\bf
427}, 714 (2004); M. Norman, Nature {\bf 427}, 692 (2004).

\bibitem{Timusk-old} F. Gao et
all., Phys. Rev. {\bf B47}, 1036 (1993)

\bibitem{BorisMPI} A. V. Boris
et al., Science {\bf 304}, 708 (2004).

\bibitem{Kaufmann} H. J.
Kaufmann, Ph. D. Thesis, Uni. Cambridge, February 1999

\bibitem{Schutzmann} J. Schutzmann et al.,
Phys. Rev. {\bf B46}, 512 (1992)

\bibitem{Kamaras} K. Kamaras et al.,
Phys. Rev. Lett. {\bf 64}, 84 (1990)

\bibitem{Carbotte} E. Schachinger, J. P. Carbotte, Phys. Rev. {\bf B
64}, 094501 (2001)

\bibitem{Hwang-Timusk-2} J. Hwang et al., Phys. Rev. {\bf
B75}, 144508 (2007); J. Hwang et al., Nature (London) {\bf 427},
714 (2004)

\bibitem{Vignolle} B. Vignolle et al., Nature Phys. {\bf 3}, 163
(2007)

\bibitem{Bozovic} I. Bo\v{z}ovi\'{c}, J. H.
Kim, J. S. Harris, Jr., C. B. Eom, J. M. Phillips, J. T. Cheung,
Phys. Rev. Lett., {\bf 73}, 1436 (1995)

\bibitem{Hirsch} J. E. Hirsch, Physica {\bf
201}, 347 (1992)

\bibitem{Maldague} P. F. Maldague, Phys. Rev. {\bf
B16}, 2437 (1977)

\bibitem{Carbone} F. Carbone et al., Phys. Rev. {\bf
B74}, 024502 (2006); F. Carbone et al., Phys. Rev. {\bf B74},
064510 (2006)

\bibitem{Molegraaf} H. J. Molegraaf et al., Science {\bf
295}, 2239 (2002)

\bibitem{Deutscher-optics}
G. Deutscher et al., Phys. Rev. {\bf B72}, 095504 (2005)

\bibitem{KMS} H. J. Kaufmann, E. G.
Maksimov, E. K. H. Salje, J. of Supercond. {\bf 11}, 755 (1998)

\bibitem{%
Mazin-Dolgov} I. I. Mazin, O. V. Dolgov, Phys. Rev. {\bf B45},
2509 (1992)

\bibitem{Friedman} T. A. Friedman et al.,
Phys. Rev. {\bf B42}, 6217 (1990)

\bibitem{Allen-kinky} P. B. Allen,
Nature {\bf 412}, 494 (2001)

\bibitem{Santander-2003}
A. F. Santander-Syro eta al., Europhys. Lett. {\bf 62}, 568 (2003)

\bibitem{Kulic2} R. Zeyher, M. L. Kuli\'{c},
Phys. Rev. {\bf B53}, 2850 (1996); R. Zeyher, M. L. Kuli\'{c},
Phys. Rev. {\bf B54}, 8985 (1996); M. L. Kuli\'{c} and R. Zeyher,
Mod. Phys. Lett. {\bf B 11}, 333 (1997)

\bibitem{Mihailovic-Kabanov} D. Mihailovi\'{c}, V. V. Kabanov,
Struct Bond (2005) {\bf 114}, p. 331-364, Springer-Verlag Berlin-Heidelberg 2005

\bibitem{Kabanov-PRL}
V. V. Kabanov et al., Phys. Rev. Lett. {\bf 95}, 147002 (2005)

\bibitem{Bansil} A. Bansil, M. Lindros,
Phys. Rev. Lett. {\bf 83}, 5154 (1999)

\bibitem{Cuk} T. Cuk et al., Phys. Rev.
Lett. {\bf 93}, 117003 (2004)

\bibitem{Zhou-PRL} X. J. Zhou et
al., Phys. Rev. Lett. {\bf 95}, 117001 (2005)

\bibitem{%
Kulic-Dolgov-lambda} M. L. Kuli\'{c}, O. V. Dolgov, Phys. Rev.
{\bf B 76}, 132511 (2007)

\bibitem{Shen-Cuk-review} T. Cuk et al.,
phys. stat. solidi (b) {\bf 242}, 11 (2005)

\bibitem{Lanzara-isotope} G.-H. Gweon et
al.,  Nature {\bf 430}, 187 (2004)

\bibitem{Ma-Ku-Do} E. G. Maksimov, O.
V. Dolgov, M. L. Kulic, Phys. Rev. {\bf B72}, 212505 (2005)

\bibitem{Valla-2006}
T. Valla, T. E. Kidd, Z.-H. Pan, A. V. Fedorov, W.-G. Yin, G. D.
Gu, P. D. Johnson, Phys. Rev. Lett. {\bf 98}, 167003 (2007)

\bibitem{%
Shen-polarons} K. M. Shen et al., Phys. Rev. Lett. {\bf 93},
267002 (2004)

\bibitem{Gunnar-Nagaosa-Ciuchi} O. R\"{o}sch et al.,
Phys. Rev. Lett. {\bf 95}, 227002 (2005); A. S. Mishenko, N.
Nagaosa, Phys. Rev. Lett. {\bf 93}, 036402 (2004); S. Ciuchi et
al., Phys. Rev. {\bf B 56}, 4494 (1997)

\bibitem{Tsunekawa} M. Tsunekawa, New Journal of Physics {\bf
10}, 073005 (2008)

\bibitem{Eisaki} S. R. Park et al.,
Phys. Rev. Lett. {\bf 101}, 117006 (2008)

\bibitem{Kulic-Dolgov-shift} M. L. Kuli\'{c}, O.
V. Dolgov, Phys. Rev. {\bf B 71}, 092505 (2005)

\bibitem{Chen-Shen-4-layered} Y. Chen et al.,
Phys. Rev. Lett. {\bf 97}, 23640 (2006)

\bibitem{Shen-Anders-4layer} W. Xie et al.,
Phys. Rev. Lett. {\bf 98}, 047001 (2007)

\bibitem{Zhu} L. Zhu, P. J. Hirschfeld, D. J.
Scalapino, Phys. Rev. {\bf B70}, 214503 (2004)

\bibitem{KuOudo}  M. L.
Kuli\'{c}, V. Oudovenko, Solid State Comm. {\textbf 104}, 731
(1997)

\bibitem{Kulic-Dolgov-imp} M. L. Kuli\'{c}, O. V. Dolgov, Phys.
Rev.{\textbf B60}, 13062(1999)

\bibitem{Kee-Tc} H. -Y. Kee, Phys.
Rev.{\textbf B64}, 012506(2001)

\bibitem{Kirtley} J.
Kirtley et al.,  in Superconductivity I, eds. J. Kettersson, K. H.
Bennemann, 2008 Springer Verlag, Berlin

\bibitem{Renker} B. Renker et al., Z. Phys. {\bf B77}, 65 (1989)

\bibitem{Deutscher}
G. Deutscher et al., J. Supercond. {\bf 7}, 371 (1994)

\bibitem{Keimer-pss}
Y. Sidis et al., phys. stat. sol. (b) {\bf 241}, 1204 (2004) and
References therein

\bibitem{Fischer-Renner-RM}
O. Fischer et al., Rev. Mod. Phys. {\bf 79}, 353 (2007)

\bibitem{AJLM} O. K. Andersen, O. Jepsen, A. I. Lichtenstein,
I. I. Mazin, Phys. Rev. {\bf B49}, 4145 (1994)

\bibitem{Huang} Q. Huang,
J. F. Zasadinski, N. Tralshawala, K. E. Gray, D. Hinks, J. L. Peng
and R. L. Greene, Nature (London) {\bf 347}, 389 (1990)

\bibitem{Jensen} P.
Samuely, N. L Bobrov, A. G. N. Jansen, P. Wyder, S. N. Barilo, S.
V. Shiryaev, Phys. Rev. {\bf B48}, 13904 (1993); P. Samuely, P.
Szabo, A. G. N. Jansen, P. Wyder, J. Marcus, C. Escribe-Filippini,
M. Afronte, Physica {\bf B 194-196}, 1747 (1994)

\bibitem{Gunnarsson-Rosch-epi} O.
R\"{o}sch, O. Gunnarsson, Phys. Rev. Lett. {\bf 92}, 146403 (2004)

\bibitem{Khaliullin} K. J. Szczepanski, K. W. Becker,
Z. Phys. {\bf B89}, 327 (1992); G. Khaliullin, P. Horsch, Phys.
Rev. {\bf B54}, R9600 (1996)

\bibitem{Rashkeev} S. N. Rashkeev and G. Wendin, Phys. Rev. {\bf
B47}, 11603 (1993)

\bibitem{Falter-O-phonon} T. Bauer, C.
Falter, arXiv:0808.2765

\bibitem{Franck} J. F. Franck, in {\it Physical
Properties of High Temperature Supercoductors V }, ed. D. M.
Ginsberg, (World Scientific, Singapore, 1994); Physica {\bf
C282-287}, 198 (1997)

\bibitem{Pintschovius} L. Pintschovius,
phys. stat. solidi (b) {\bf 242}, 30 (2005)

\bibitem{Bohnen-Heid-2003} K. -P. Bohnen, R. Heid, M.
Kraus, Europhys. Lett. {\bf 64}, 104 (2003)

\bibitem{ReSaGuDe} D. Reznik
et al., Nature {\bf 455}, E6 (2008)

\bibitem{Cardona1}
C. Thomsen, M. Cardona, B. Gegenheimer, R. Liu and A. Simon,
Phys. Rev. {\bf B 37}, 9860 (1988)

\bibitem{Cardona2}
C. Thomsen and M. Cardona, in Physical Properties of High
Temperature Superconductors I, ed. by D. M. Ginzberg (World
Scientific, Singapore, 1989), pp. 409; R. Feile, Physica {\bf C
159}, 1 (1989); C. Thomsen, in Light Scattering in Solids VI, ed.
by M. Cardona and G. Guentherodt (Berlin, Heidelberg, New York,
Springer, 1991), pp. 285

\bibitem{Hadjiev}
V. G. Hadjiev, X. Zhou, T. Strohm, M. Cardona, Q. M. Lin, C. W.
Chu, Phys. Rev. {\bf B 58}, 1043 (1998)


\bibitem{Phonon-Exp} D. Reznik et al.,
Nature {\bf 440}, 1170 (2006); D. Reznik et al., J. Low Temp.
Phys. {\bf 147}, 353 (2007); L. Pintschovius et al., Phys. Rev.
{\bf B69}, 214506 (2004)

\bibitem{Douglas-isotop} J. F. Douglas et al., Nature {\bf
446}, E5 (2007)

\bibitem{Iwasawa-isotop} H. Iwasawa et al., Physica {\bf
C463-465}, 52 (2007)

\bibitem{Iwasawa-isotop-2} H. Iwasawa et al., arXiv:
0808.1323

\bibitem{Allen-Mitrovic} P.B. Allen, B. Mitrovi\'{c}, Solid
State Physics, ed. H. Ehrenreich, F. Seitz, D. Turnbull, Academic,
New York, {\bf V37}, p. 1, (1982)

\bibitem{Maksimov-Eliashberg} O. V.
Dolgov, E. G. Maksimov, Uspekhi Fiz. Nauk {\bf 138}, 95 (1982);
O.V. Dolgov, E.G.Maksimov, Chapter 1 in: "Thermodynamics and
Electrodynamics of Superconductors", ed. by V.L.Ginzburg,
Nova Science Publ., N.Y.(1987)

\bibitem{Marsiglio-Carbotte-Book} F. Marsiglio, J. P. Carbotte, in
Superconductivity I, eds. J. Kettersson, K. H. Bennemann, 2008
Springer Verlag, Berlin

\bibitem{McMillan-Rowell} W. L. McMillan and J. M. Rowell,
Phys. Rev. Lett. {\bf 14}, 108 (1965)

\bibitem{GDS-method} A. A. Galkin, A. I. Dyashenko,
V. M. Svistunov, Zh. Eksp. Teor. Fiz., {\bf 66}, 2262 (1974); V.
M. Svistunov et al, J. Low Temp. Phys. {\bf 31}, 339 (1978)

\bibitem{Ivanshenko} Yu. M. Ivanshenko, Yu. V. Medvedev,
Fiz. Nizkih Temp., {\bf 2}, 143 (1976)

\bibitem{SchrieffLTS} J. R.Schrieffer, {\it Theory of
Superconductivity}, New York, (1964)

\bibitem{SSW} D. J. Scalapino, J. R.
Schrieffer, J. W. Wilkins, Phys. Rev. {\bf 148}, 263 (1966)


\end{references}
\end{document}